\documentstyle[prd,aps,eqsecnum]{revtex}             

\newcommand{\be}{\begin{equation}}
\newcommand{\ee}{\end{equation}}
\newcommand{\bea}{\begin{eqnarray}}
\newcommand{\eea}{\end{eqnarray}}

\def\thefootnote{\fnsymbol{footnote}}

\newcommand{\artref}[5]{{\rm #1}, {\rm #2} {\bf #3}, {\rm #4 (#5)}}

\newcommand{\bookref}[3]{{\rm #1}, {\it #2} {\rm (#3)}}

\def\B{B}
\def\Z{{\cal Z}}
\def\W{{\cal W}}
\def\gp{{g^\prime}}
\def\gs{g^2}
\def\gps{\gp^2}
\def\p{\partial}
\def\jmunut{(\p_\mu J_{\nu}^\Z - \p_\nu J_{\mu}^\Z)}

\def\lag{{\cal L}}

\def\polemr{\Lambda_\varepsilon(2\mrs)}
\def\lr{\lambda_r}
\def\gr{g_r}
\def\gpr{{g_r^\prime}}
\def\grs{g_r^2}
\def\gprs{\gpr^2}
\def\mrs{m_r^2}

\def\Y{{\cal Y}}

\def\A{{\cal A}}
\def\PT{{\rm PT}}
\def\PL{{\rm PL}}
\def\PTT{\widetilde{\rm PT}}

\def\g{g}

\def\tq{q}
\def\tvarD{\widetilde\varD}
\def\hvarD{\widehat\varD}
\def\u{\underline}

\def\varD{{\cal D}}

\def\half{{1\over 2}}

\def\y{\Y^\Z}
\def\yv{\Y}

\def\V{{\cal V}}        
\def\UoneParam{\omega}  
\def\Bphi{\Phi}
\def\tBphi{\widetilde\Phi}
\def\polem{\Lambda_\varepsilon(2m^2)}
\def\tr{{\rm tr}}

\def\order{{\cal O}}

\def\G{{\cal G}}  

\def\eres{{e_{\mathrm{res}}}}
\def\ers{{e_{\mathrm{res}}^2}}

\def\poleMHbare{\Lambda_\varepsilon(M_H^2)}

\def\d{{\rm d}}   
\def\intdx{\int\d^d x\/}

\def\Tr{{\rm Tr\/}}

\def\Uoneem{U(1)_{\mathrm{em}}}
\def\phase{{\varphi}}

\def\SM{{\mathrm SM}}

\def\Allfields{{\cal F}}
\def\cAllfields{{{\cal F}^{cl}}}
\def\fluct{{y}}
\def\gaugeparam{{\omega}}
\def\tildeD{{\widetilde D}}
\def\Trafo{{\cal T}}

\def\Umat{{\hat U}}
\def\covDmat{{\hat D}}
\def\Wmat{{\hat W}}
\def\Bmat{{\hat B}}
\def\Vmat{{\hat V}}
\def\Tmat{{\hat T}}
\def\op{L}  
\def\bD{\bar D}
\def\bvarD{\bar{\varD}}
\def\bP{\bar P}

\def\DeltaEoM{{T}}  

\def\low{l}

\def\llambda{{\lambda}}

\def\fourth{{1\over4}}

\def\w{w} 
\def\b{b} 

\def\ourop{{\cal O}}  
\def\ourlow{l} 

\def\tY{{\widetilde \Y}}

\def\MHp{M_{H,\mathrm{pole}}}  
\def\MWp{M_{W,\mathrm{pole}}}
\def\MZp{M_{Z,\mathrm{pole}}}

\def\MHps{M_{H,\mathrm{pole}}^2}  
\def\MWps{M_{W,\mathrm{pole}}^2}
\def\MZps{M_{Z,\mathrm{pole}}^2}

\def\root{\sigma}  

\def\Dfull{{\tildeD + P P^T + \delta_P}}

\def\bU{{\bar U}}
\def\bB{{\bar B}}
\def\bW{{\bar \W}} 
\def\bZ{{\bar \Z}}
\def\bA{{\bar \A}}
\def\bh{{\bar h}}
\def\bJ{{\bar J}}
\def\bj{{\bar j}}
\def\bK{{\bar K}}
\def\bd{{\bar d}}

\def\bY{{\bar \Y}}

\def\trho{{\bar \rho}}

\def\pole{\Lambda_\varepsilon}

\def\bv{{\bar v}}
\def\bg{{\bar g}}
\def\bgp{{\bar {g^\prime}}}
\def\bgs{{{\bar g}^2}}
\def\bgps{{{\bg^\prime}{}^2}}
\def\bc{{\bar c}}
\def\bs{{\bar s}}
\def\bes{{{\bar e}^2}}

\def\Dslash{{\not\!\! D}}

\def\bu{u}
\def\bdf{d}
\def\bPsi{{\Psi}}
\def\tbU{{\widetilde{\bU}}}
\def\brho{{\bar\rho}}

\begin{document}

\preprint{$\begin{array}{r} \mbox{DESY 99--088} \\
\mbox{hep-ph/9907294} \\ \\ \end{array}$} 

\draft

\title{The electroweak chiral Lagrangian reanalyzed}

\author{Andreas Nyf\/feler\footnote{Present address: Centre de Physique
    Th\'{e}orique, CNRS-Luminy, Case 907, F-13288 Marseille Cedex 9,
    France. Email address: nyf\/feler@cpt.univ-mrs.fr}}
\address{DESY, Platanenallee 6, D-15738 Zeuthen, Germany} 

\author{Andreas Schenk} 
\address{Hambacher Stra\ss e 14, D-64625 Bensheim-Gronau, Germany} 
    
\date{26 October 2000} 

\maketitle

\begin{abstract}
In this paper we reanalyze the electroweak chiral Lagrangian with
particular focus on two issues related to gauge invariance. Our
analysis is based on a manifestly gauge-invariant approach that we
introduced recently. It deals with gauge-invariant Green's functions
and provides a method to evaluate the corresponding generating
functional without fixing the gauge.  First we show, for the case
where no fermions are included in the effective Lagrangian, that the
set of low-energy constants currently used in the literature is
redundant. In particular, by employing the equations of motion for the
gauge fields one can choose to remove two low-energy constants which
contribute to the self-energies of the gauge bosons. If fermions are
included in the effective field theory analysis the situation is more
involved. Even in this case, however, these contributions to the
self-energies of the gauge bosons can be removed.  The relation of
this result to the experimentally determined values for the oblique
parameters~$S, T,$ and~$U$ is discussed. In the second part of the
paper we consider the matching relation between a full and an
effective theory. We show how the low-energy constants of the
effective Lagrangian can be determined by matching gauge-invariant
Green's functions in both theories. As an application we explicitly
evaluate the low-energy constants for the standard model with a heavy
Higgs boson. The matching at the one-loop level and at next-to-leading
order in the low-energy expansion is performed employing functional
methods.
\end{abstract}


\renewcommand{\thefootnote}{\arabic{footnote}}
\setcounter{footnote}{0}



\section{Introduction}
\label{sec:intro}

The symmetry breaking sector of the standard model is still poorly
understood from a theoretical point of view.  Furthermore no direct
experimental evidence of the Higgs boson has been found so far.  In
this situation the method of effective field theory has repeatedly
been used in recent years to analyze the symmetry breaking
sector~\cite{L_eff_SM}. It provides a convenient and model independent
parametrization of various scenarios which are discussed in the
literature, regarding the nature of the spontaneous breaking of the
electroweak symmetry. In this approach, the unknown physics is hidden
in the low-energy constants of an effective Lagrangian, which
describes the effective field theory. Effective Lagrangians thereby
allow a unified treatment of different parametrizations of new physics
effects, such as oblique corrections to gauge bosons
self-energies~\cite{STU,epsilon_i} and anomalous triple~\cite{TGV} and
quartic~\cite{QGV} vertices of the gauge bosons.

The low-energy structure of a theory containing light and heavy
particle species which are separated by a mass gap can adequately be
described by an effective field theory which contains only the light
fields. In the case of the standard model one can construct effective
Lagrangians by introducing higher dimensional operators that preserve
the $SU(2)_L \times U(1)_Y$ gauge symmetry. In the presence of a light
Higgs boson, i.e.\ in the decoupling case~\cite{decoupling_theorem},
the symmetry is linearly realized and the corresponding effective
Lagrangian, which contains the Higgs field, was presented in
Ref.~\cite{Buchmueller_Wyler}. For a strongly interacting symmetry
breaking sector, i.e.\ in the non-decoupling case, the effective
Lagrangian can be
built~\cite{heavy_Higgs,Longhitano,Appelquist_Wu,Feruglio} in analogy
to the chiral Lagrangian~\cite{Weinberg,GL84_85} for QCD and it is
therefore called electroweak chiral Lagrangian. The use of effective
Lagrangians might in fact be the only way, apart from lattice
calculations, to gain insight into strongly interacting theories for
the electroweak symmetry breaking sector, similarly to the situation
with QCD at low energies. We note that by employing the electroweak
chiral Lagrangian it was shown recently~\cite{BaggerFalkSwartz} that
present electroweak precision data are still compatible with a
strongly interacting model of symmetry breaking with a scale of new
physics as high as $3$~TeV.

The purpose of this paper is to take another look at the electroweak
chiral Lagrangian and to investigate two issues related to gauge
invariance where there are some subtleties involved, because one has
to deal with off-shell quantities.  According to
Refs.~\cite{Weinberg,GL84_85} the effective field theory should
describe the physics of the underlying full theory at low
energies. Symmetry principles thereby play a crucial role for the
construction of the effective field theory and, apart from the
occurrence of anomalies, the effective field theory can be described
by an effective Lagrangian which respects these (possibly broken)
symmetries~\cite{Leutwyler_94}. In order to preserve the gauge
symmetry even when dealing with off-shell quantities we employ a
manifestly gauge-invariant approach that was introduced
recently~\cite{SM_gaugeinv}. It deals with gauge-invariant Green's
functions and provides a method to evaluate the corresponding
generating functional without fixing the gauge.

The first topic is the analysis of the general effective field theory
which describes a strongly interacting electroweak symmetry breaking
sector. We are particularly interested in the question of how many
independent, physically relevant parameters are contained in the
effective Lagrangian. It is well known from chiral perturbation
theory~\cite{GL84_85,Bij_Col_Eck} that one can use the equations of
motion which are derived from the lowest order effective Lagrangian to
remove redundant terms that appear at higher orders in the low-energy
expansion. This procedure is well defined within a functional approach
where one performs an expansion around the solutions of the classical
equations of motion in the path-integral representation of the
generating functional of suitably chosen Green's functions. It is only
in this framework where we will use the equations of motion later on.
Equivalently, one can also remove terms in the effective Lagrangian by
performing appropriate reparametrizations of the fields and external
sources in the path integral~\cite{Fearing_Scherer}.

In the usual gauge-dependent framework the equations of motion for the
gauge fields are gauge-dependent. For instance, contributions from the
gauge-fixing terms and from the non-gauge-invariant source terms would
appear in Eqs.~(\ref{eomWpm_eff})--(\ref{eomB_eff_2}) below. It is
doubtful whether these equations can then be used to eliminate
redundant gauge-invariant terms from the effective Lagrangian. As a
matter of fact, we do not know of any reference where this has been
tried. The equations of motion in our approach are
gauge-invariant. Employing them we first show for a purely bosonic
effective field theory, i.e.\ when no fermions are included in the
effective Lagrangian, that the set of parameters currently used in the
literature contains two redundant low-energy constants which can be
removed.  In particular, one can choose to remove two low-energy
constants which contribute to the self-energies of the gauge bosons
which are not observable anyway.  If fermions are present, the
situation is more involved. We will show that these two parameters
renormalize the coupling of the massive gauge bosons to charged and
neutral currents and, thus, have no physical meaning in a full
effective Lagrangian analysis. The relation of this result to the
experimentally determined values for the oblique parameters~$S, T,$
and $U$~\cite{STU} as quoted by the particle data group will be
discussed. 

The second topic of this paper is to study the evaluation of the
low-energy constants in the effective Lagrangian for a given
underlying theory.  Comparing the theoretical predictions for the
low-energy constants for different models with experimental
constraints might help to rule out some of the underlying theories
under consideration before direct effects become visible. This point
motivates to determine the values of the low-energy constants in the
effective theory for various models.  At low energies, the standard
model with a heavy Higgs boson in the spontaneously broken phase can
adequately be described by such an effective field theory.  In order
to determine the effective Lagrangian one can require, for instance,
that corresponding Green's functions in both theories have the same
low-energy structure. One can take this matching condition as the
definition of the effective field theory.  At this point the issue of
gauge invariance is crucial. If gauge-dependent Green's functions are
used in this matching procedure one has to make sure that no gauge
artifacts enter the low-energy constants of the effective Lagrangian.

Several groups~\cite{H_M,E_M,D_GK} have performed such a matching
calculation for the standard model with a heavy Higgs boson in recent
years, thereby extending the results which were obtained long time
ago~\cite{heavy_Higgs,Longhitano}. Gradually the importance to
maintain gauge invariance in the matching procedure was recognized.
Whereas the matching was performed with gauge-dependent Green's
functions in Ref.~\cite{H_M}, the authors of Refs.~\cite{E_M,D_GK}
proposed new methods to overcome these gauge artifacts. See
Ref.~\cite{Abelian_Higgs} for a more detailed account of the
development. The extension of the method proposed in Ref.~\cite{E_M}
to the two-loop level was discussed in Ref.~\cite{Matias}.
Nevertheless, the problems with gauge dependencies have not yet been
fully resolved. In the meantime similar matching calculations have
been performed for various
models~\cite{twoloop_heavyHiggs,twoHDM,Decoupling_SUSY} without
considering the issue of gauge invariance any further.

To avoid any problems with gauge dependencies one should in fact match
only gauge-invariant quantities, such as $S$-matrix
elements~\cite{E_M}. As it turns out, however, matching $S$-matrix
elements is quite cumbersome because one has to deal with the whole
infrared physics. Techniques which involve Green's functions are much
easier to use. We therefore propose to match Green's functions of
gauge-invariant fields in order to determine the effective
Lagrangian. In this way no gauge artifacts can appear through the
matching procedure and one can employ functional
methods~\cite{LSM}. For the Abelian Higgs model such a manifestly
gauge-invariant matching calculation has been performed in
Ref.~\cite{Abelian_Higgs}. In the present paper we show how one can
determine the effective Lagrangian for the standard model with a heavy
Higgs boson by matching gauge-invariant Green's functions in the full
and the effective theory at low energies at the one-loop level.  For
this purpose we can use a generating functional of gauge-invariant
Green's functions for the bosonic sector of the standard model which
was discussed in a recent paper~\cite{SM_gaugeinv}.  In this way the
starting point of the matching procedure is well defined and gauge
invariance is manifestly preserved throughout the whole calculation.

In view of the fact that all fits to electroweak
precision data over the last couple of years tend to prefer a light
Higgs boson\footnote{For instance, at the Moriond 2000 meeting the
value $M_H = \left( 67 {+60 \atop -33} \right)$~GeV was
presented~\cite{Higgs_fit}. 
}, we will regard the standard model with a heavy Higgs boson merely
as a model of a strongly interacting symmetry breaking sector, where,
however, perturbation theory can still be applied if the coupling
constant is not too strong. Thus, it serves as a testing ground for
our gauge-invariant method of matching. The corresponding values for
the low-energy constants will also represent a reference point for
other strongly interacting models. As pointed out in Ref.~\cite{Lane},
it is very difficult to get any reliable estimate for the low-energy
parameters for genuinely strongly interacting models of the
electroweak symmetry breaking sector.

This paper is organized as follows: In the next section we introduce
the general effective field theory for a strongly interacting
electroweak symmetry breaking sector within the gauge-invariant
functional framework presented in Ref.~\cite{SM_gaugeinv}. We discuss
our choice of gauge-invariant operators and the corresponding source
terms which emit one-particle states of the gauge bosons. We then
determine the number of independent low-energy constants by employing
the equations of motion to remove redundant terms from the effective
Lagrangian. We sketch the inclusion of fermions in the effective field
theory and relate our findings to the experimentally determined
oblique parameters $S, T,$ and $U$. In order to prepare the matching
calculation in the second part of this paper we briefly recapitulate
in Sec.~\ref{sec:SM_gauge_inv} the main results from our manifestly
gauge-invariant approach to the standard model~\cite{SM_gaugeinv}. We
calculate the generating functional for the gauge-invariant Green's
functions in the bosonic sector up to the one-loop level. In this
section we also present the renormalization prescriptions for the
fields, the mass parameter and the coupling constants of the model. In
Sec.~\ref{sec:matching} we evaluate the matching condition between
gauge-invariant Green's functions in the full and the effective theory
at low energies at the one-loop level for the case of the standard
model with a heavy Higgs boson. The effective Lagrangian for the
bosonic sector is determined up to order $p^4$ in the low-energy
expansion.  In Sec.~\ref{sec:renorm} we express the result for the
effective Lagrangian in terms of the physical masses of the Higgs and
the gauge bosons and the electric charge. Finally, we compare our
results with those obtained by other groups. We summarize our findings
in Sec.~\ref{sec:summary}.  The source terms which appear in the
general effective Lagrangian at order $p^4$ are listed in
Appendix~\ref{app:sources_p4}. The relations between our set of
operators for the electroweak chiral Lagrangian and the basis which is
usually used in the literature can be found in
Appendix~\ref{app:EW_chiral_Lag}. Some technical details needed for
the calculation of the one-loop generating functional in the standard
model are presented in Appendix~\ref{app:diffop_HH}.


\section{Effective Field Theory}
\label{sec:EFT}

\subsection{The general effective Lagrangian}
\label{sec:def_EFT}

In this section we will discuss the general effective field theory for
the bosonic\footnote{The electroweak chiral Lagrangian including
matter fields was presented in Ref.~\cite{EW_Leff_matter}.} part of a
strongly interacting electroweak symmetry breaking sector, closely
following the functional approach to the standard model introduced in
Ref.~\cite{SM_gaugeinv}. The relation of our approach to the one that
is usually adopted in the
literature~\cite{heavy_Higgs,Longhitano,Appelquist_Wu,Feruglio} will
be discussed below. According to Refs.~\cite{Weinberg,GL84_85} the
effective field theory should describe the physics of the underlying
full theory at low energies. We assume that
\begin{equation} \label{efflagcondition1}
        p^2, M_W^2, M_Z^2 \ll M^2 \, , 
\end{equation}
where $p$ is a typical momentum and $M$ is the mass scale for heavy
particles in the underlying theory, e.g.\ a heavy Higgs boson in the
standard model or a technirho in some technicolor
model~\cite{Technicolor}.  In general, symmetry principles are crucial
for the construction of the effective field theory and, apart from the
occurrence of anomalies, the effective field theory can be described
by an effective Lagrangian which respects these (possibly broken)
symmetries~\cite{Leutwyler_94}. In our case this Lagrangian is
gauge-invariant and depends on the Goldstone boson field~$\bU$,
confined to the sphere~$\bU^\dagger \bU = 1$, the $SU(2)_L$ gauge
fields $\bar W_\mu^a \, (a=1,2,3)$, the $U(1)_Y$ gauge field
$\bB_\mu$, and external sources $\bK_{\mu\nu}, \bJ_\mu^a, (a=1,2,3)$
\begin{equation} \label{lageff_1}
  \lag_{eff} = \lag_{eff}\left(\bar W_{\mu\nu}^a, \bB_{\mu\nu}, \bU,
  \bD_\mu \bU, \bD_\mu \bD_\nu \bU, \ldots; \bK_{\mu\nu},
  \bJ_\mu^a \right) \, , 
\end{equation}
where the Goldstone boson doublet $\bU$ is coupled to the gauge  
fields through the covariant derivative
\be
\label{cov_deriv_EFT}
\bD_\mu \bU = \left( \p_\mu - i {\tau^a\over 2} \bar W_\mu^a
                          - i {1\over 2} \bB_\mu \right) \bU \, .
\ee
Note that we have absorbed the coupling constants $\bg$ and $\bgp$ into
the gauge fields $\bar W_\mu^a$ and $\bB_\mu$, respectively. The field
strengths are given by
\bea
\bar W_{\mu\nu}^a  & = & \p_\mu \bar W_\nu^a - \p_\nu \bar W_\mu^a
                        + \varepsilon^{abc} \bar W_\mu^b \bar W_\nu^c
\, , \label{Wmunu} \\
\bB_{\mu\nu}    & = & \p_\mu \bB_\nu - \p_\nu \bB_\mu \,
. \label{Bmunu} 
\eea
The fields and the sources in the effective theory have been denoted
with a bar in order to distinguish them from those occurring in the
standard model which will be discussed below.  The Goldstone boson
field $\bU$ and the gauge fields $\bar W_\mu^a, \bB_\mu$ transform
under $SU(2)_L$ gauge transformations in the following way: 
\bea 
\bU & \to & \V \bU \, , \quad \V \in SU(2) \, , \nonumber \\
\bar W_\mu & \to & \V \bar W_\mu \V^\dagger - i (\partial_\mu \V)
\V^\dagger  \, , \quad \bar W_\mu \equiv \bar W_\mu^a {\tau^a\over 2}
\, , 
\label{SUtwo_trafo_EFT}
\eea
and under $U(1)_Y$ gauge transformations as follows: 
\bea
\bU   & \to & e^{- i \UoneParam / 2} \ \bU \, , \nonumber \\
\bB_\mu & \to & \bB_\mu - \partial_\mu \UoneParam \ .
\label{Uone_trafo_EFT}
\eea

The effective Lagrangian in Eq.~(\ref{lageff_1}) describes the
dynamics of the massive gauge bosons $\bar W_\mu^\pm, \bar Z_\mu,$ and
the massless photon $\bar A_\mu$. In order to have nontrivial
solutions of the equations of motion, we furthermore couple external
sources, denoted by $\bK_{\mu\nu}$ and $\bJ_\mu^a$ in
Eq.~(\ref{lageff_1}), to the gauge fields. In the applications that we
will discuss below we will be forced to deal with off-shell
quantities. Therefore, we want to preserve the gauge symmetry, which
is imposed in the construction of the effective Lagrangian, even in
the presence of these external sources.

As discussed in detail for the Abelian Higgs model in
Ref.~\cite{Abelian_Higgs}, for QED in Ref.~\cite{QED_gaugeinv}, and
for the standard model in Ref.~\cite{SM_gaugeinv}, the appropriate
choice of the source terms is crucial for a manifestly
gauge-invariant analysis. The sources will only respect the gauge
symmetry, if they do not couple to the gauge degrees of
freedom. Otherwise, one has to impose constraints on the fields in
order to solve the equations of motion.  Usually, this problem is
cured by fixing a gauge. However, one can also turn the argument
around and consider only those external sources which couple to
gauge-invariant operators. As we will see below, such a manifestly
gauge-invariant treatment is in fact possible at the classical level
as well as when quantum corrections are taken into account.

In this respect our approach to the effective field theory description
of a strongly interacting electroweak symmetry breaking sector differs
from the one that is usually adopted in the
literature~\cite{heavy_Higgs,Longhitano,Appelquist_Wu,Feruglio}.
Although the authors of these references also start with a
gauge-invariant effective Lagrangian they then add gauge-fixing and
Faddeev-Popov terms. Since these terms break the gauge symmetry these
authors, as well as those of
Refs.~\cite{H_M,E_M,D_GK,Matias,twoloop_heavyHiggs,twoHDM,Decoupling_SUSY},
are then working with gauge-dependent Green's functions.

In order to write down appropriate source terms we will first introduce
fields for the dynamical degrees of freedom which are already
invariant under the non-Abelian group $SU(2)_L$ and, in parts, under
the Abelian group $U(1)_Y$ as well. It has been known for a long time
\cite{EW_confinement,tHooft,gauge_inv_fields} that all fields in the
standard model Lagrangian can be written, in the spontaneously broken
phase, in a gauge-invariant way up to the unbroken $\Uoneem$. A
similar approach can be employed for the effective field theory
description. Defining the $Y$-charge conjugate doublet by
\be
\tilde \bU = i \tau_2 \bU^* \, , 
\ee
we can introduce the following fields, see also Ref.~\cite{SM_gaugeinv}: 
\begin{eqnarray}\label{bfieldsf_EFT}
\bW^+_\mu &=& {i\over2} \left(\tilde \bU^\dagger (\bD_\mu \bU) -
(\bD_\mu\tilde \bU)^ \dagger \bU\right) \, , \\
\bW^-_\mu &=& {i\over2} \left( \bU^\dagger (\bD_\mu \tilde \bU) -
(\bD_\mu \bU)^\dagger \tilde \bU\right) \, , \\
\bZ_\mu &=& i \left( \tilde \bU^\dagger (\bD_\mu \tilde \bU) -
\bU^\dagger (\bD_\mu \bU) \right) \, , \label{defZ_EFT} \\
\bA_\mu &=& \bB_\mu + \bs^2 \bZ_\mu \, , \label{bfieldsl_EFT} \\
\bW^\pm_\mu &=& {1\over2} (\bW^1_\mu \mp i \bW^2_\mu) \, , 
\label{defWpm_EFT}
\end{eqnarray}
which are invariant under the $SU(2)_L$ gauge transformations from
Eq.~(\ref{SUtwo_trafo_EFT}). In Eq.~(\ref{bfieldsl_EFT}) we used the
following definition of the weak mixing angle: 
\be \label{cos_theta_EFT}
\bc^2 \equiv \cos^2 \bar \theta_W = M_W^2 / M_Z^2 \, , \,
\bs^2 \equiv 1 - \bc^2 \ .
\ee

In order to calculate Green's functions from which we then can extract
physical masses, coupling constants and $S$-matrix elements, we have
to introduce external sources which emit one-particle states of the
gauge bosons. In analogy to our effective field theory analysis of the
Abelian Higgs model~\cite{Abelian_Higgs} we couple a source to the
field strength $\bB_{\mu\nu}$. For the massive gauge bosons the
situation is more involved. Whereas the field $\bZ_\mu$ is fully
$SU(2)_L \times U(1)_Y$ gauge-invariant, the charged gauge fields
$\bW_\mu^\pm$ have a residual gauge dependence under the $U(1)_Y$ gauge
transformations from Eq.~(\ref{Uone_trafo})\footnote{Note that the
$SU(2)_L$ invariant field $\bA_\mu$ from Eq.~(\ref{bfieldsl_EFT})
transforms under $U(1)_Y$ as $\bA_\mu \to \bA_\mu - \partial_\mu
\UoneParam$, i.e.\ like an Abelian gauge field.}:
\be \label{calW_trafo_EFT}
\bW_\mu^{\pm} \to e^{\mp i \UoneParam} \bW_\mu^{\pm} \ . 
\ee
We can, however, compensate this gauge dependence by multiplying the
charged fields $\bW_\mu^\pm$ by a phase
factor~\cite{Steinmann,Horan_Lavelle_McMullan,QED_gaugeinv,SM_gaugeinv}.
Appropriate $SU(2)_L \times U(1)_Y$ gauge-invariant
source terms for all the fields can then be written in the following
way:  
\be
\label{basic_sources_EFT}
\bK_{\mu\nu} \bB_{\mu\nu} \, , \quad 
\bJ_\mu^+ \bar \phase^+ \bW_\mu^- + \bJ_\mu^- \bar
\phase^- \bW_\mu^+ \, , \quad
\bJ_\mu^\Z \bZ_\mu \, , 
\ee
with external sources $\bK_{\mu\nu}, \bJ_\mu^\pm,$ and $\bJ_\mu^\Z$. The 
phase factor in Eq.~(\ref{basic_sources_EFT}) is defined by
\be \label{phase_complex_EFT}
\bar \phase^\pm(x)      =  \exp\left({\mp i \int d^dy  \, \G_0(x-y) \,
\partial_\mu \bB_\mu}(y) \right)   .
\ee
with
\be \label{G_zero} 
\G_0(x-y) = \langle x | {1 \over - \Box} | y \rangle .
\ee
For computational convenience we are working in Euclidean space-time. 

Using identities of the form
\begin{eqnarray}
 \bD_\mu \bU & = & {i\over 2} \bZ_\mu \bU - i \bW_\mu^+ \tilde{\bU} \, 
, \nonumber \\
 \bD_\mu \tilde{\bU} & = & - i \bW_\mu^- \bU - {i\over 2} \bZ_\mu
 \tilde{\bU} \, , \nonumber \\
 \bD_\mu \bD_\nu \bU & = & \left( {i\over 2} (\p_\mu \bZ_\nu) - {1\over
   4} \bZ_\mu \bZ_\nu - \bW_\mu^- \bW_\nu^+ \right) \bU 
+ \left( - i \bd_\mu \bW_\nu^+ + {1\over 2}
   \bW_\mu^+ \bZ_\nu - {1\over 2} \bZ_\mu \bW_\nu^+ \right)
\tilde{\bU} \, , \nonumber \\
 \bD_\mu \bD_\nu \tilde{\bU} & = & \left( - i \bd_\mu \bW_\nu^- +
 {1\over 2} \bZ_\mu \bW_\nu^- - {1\over 2} \bW_\mu^- \bZ_\nu \right)
 \bU  
+  \left( - {i\over 2} (\p_\mu \bZ_\nu) - {1\over
   4} \bZ_\mu \bZ_\nu - \bW_\mu^+ \bW_\nu^- \right) \tilde{\bU} \, , 
\end{eqnarray}
where 
\be \label{d_Wpm_EFT} 
\bd_\mu \bW_\nu^\pm = \left( \p_\mu \mp i \bB_\mu \right) \bW_\nu^\pm \, , 
\ee
one can express the Lagrangian in terms of the fields~$\bW_\mu^\pm,
\bZ_\mu, \bB_\mu$, and covariant derivatives thereof
\begin{equation} \label{L_eff_prelim} 
  \lag_{eff} = \lag_{eff}\left(\bW_\mu^\pm, \bZ_\mu, \bB_\mu, \ldots;
  \bK_{\mu\nu}, \bJ_\mu^\pm, \bJ_\mu^\Z, \right) \ .
\end{equation}
As a matter of convenience we write the field $\bB_\mu$
in Eq.~(\ref{L_eff_prelim}) instead of the photon field
$\bA_\mu$. 

The generating functional in the effective field theory is given by
the path integral
\begin{equation} \label{genfunc_pathint_EFT} 
  e^{-W_{eff}[\bK_{\mu\nu}, \bJ_\mu^\pm, \bJ_\mu^\Z]} = 
\int \d\mu[\bU, \bar W_\mu^a, \bB_\mu] e^{-\intdx \lag_{eff}} \ .
\end{equation}
Note that we still integrate over the original fields $\bU, \bar
W_\mu^a,$ and $\bB_\mu$ in
Eq.~(\ref{genfunc_pathint_EFT}). Furthermore, we have absorbed an
appropriate normalization factor into the measure $\d\mu[\bU,\bar
W_\mu^a,\bB_\mu]$. Derivatives of this functional with respect to the
source $\bK_{\mu\nu}$ generate Green's functions of the field strength
$\bB_{\mu\nu}$, while derivatives with respect to $\bJ_\mu^\pm$ and
$\bJ_\mu^\Z$ generate Green's functions for the gauge-invariant fields
$\bar \phase^\mp \bW_\mu^\pm$ and $\bZ_\mu$, respectively. As was
pointed out in Refs.~\cite{Abelian_Higgs,QED_gaugeinv,SM_gaugeinv} it
is possible to evaluate the path integral in
Eq.~(\ref{genfunc_pathint_EFT}) without the need to fix a gauge as
will be shown below.

The effective Lagrangian~$\lag_{eff}$ in Eq.~(\ref{L_eff_prelim}) is a sum
of terms with an increasing number of derivatives, mass factors, and
powers of external sources, corresponding to an expansion in powers of
the momenta and the masses,
\begin{equation} \label{Lag_eff_series}
        \lag_{eff} = \lag_2 + \lag_4 + \lag_6 + \cdots \ ,
\end{equation}
where $\lag_k$ is of order $p^k$ and has the general form
\be \label{lag_k_exp}
\lag_k = \sum_i \low_i^{(k)} {\cal O}_i^{(k)} \ .
\ee
The coefficients $\low_i^{(k)}$ in Eq.~(\ref{lag_k_exp}) represent the
low-energy constants of the effective theory and count as order $p^0$.
The operators ${\cal O}_i^{(k)}$ involve the light fields and the
sources in such a way that they respect the $SU(2)_L \times U(1)_Y$
gauge symmetry.

In order to evaluate the low-energy expansion up to a given order, we
follow the counting rules usually adopted in chiral perturbation
theory~\cite{Weinberg,GL84_85} for the bookkeeping of the terms in the
effective Lagrangian. These rules are necessary for the internal
consistency of the effective field theory. We note that they are
formulated completely within the framework of the effective field
theory. In particular, there is no expansion with respect to some
heavy mass scale in the underlying theory involved. We thus treat the
covariant derivative $\bar{D}_\mu$, the gauge boson masses $M_W$ and
$M_Z$ and the momenta as quantities of order $p$, while the Goldstone
boson field $\bU$ is of order $p^0$.  In counting the masses~$M_W$ and
$M_Z$ as order~$p$, the low-energy expansion is carried out at a fixed
ratio~$p^2/M_W^2$ and $p^2/M_Z^2$, and correctly reproduces all
singularities associated with the gauge bosons. The consistency of
these rules requires that the coupling constants~$\bg, \bgp$ and
therefore the electromagnetic coupling constant $\bar e$, defined in
Eq.~(\ref{bare_quantities_eff}) below, are also treated as quantities
of order~$p$. Note that this is different from the usual dimensional
analysis: the coupling constants have dimension $(\mbox{mass})^0$, yet
they count as order~$p$ in the low-energy expansion. This is similar
to chiral perturbation theory where the quark masses $m_q$ are
quantities of order $p^2$~\cite{GL84_85} and where the electromagnetic
coupling constant~$e$ is counted as order~$p$ if virtual photons are
included~\cite{Urech}. Our counting rules furthermore imply that
$\cos\bar\theta_W$ and $\sin\bar\theta_W$ are treated as quantities of
order $p^0$, whereas the gauge fields $\bar W_\mu^a, \bB_\mu$ and
therefore also $\bW_\mu^\pm, \bZ_\mu,$ and $\bA_\mu$ count as
quantities of order $p$. Finally, the external sources~$\bJ_\mu^\pm$
and~$\bJ_\mu^\Z$ count as quantities of order~$p$, while the
source~$\bK_{\mu\nu}$ and the phase factor~$\bar
\phase^\pm$ are of order $p^0$. 
 
In general, there are two different kinds of contributions to the
generating functional. On the one hand, one has tree-level
contributions given by the integral~$\int d^dx\lag_{eff}$, which has
to be evaluated at the stationary point, i.e., with the solutions of
the equations of motion. On the other hand there are contributions
from loops, which ensure unitarity. General power counting arguments
show that $n$-loop corrections are suppressed by at least~${2n}$
powers of the momentum~\cite{Weinberg}. For instance, tree-level
contributions with one vertex from~$\lag_k$ and any number of vertices
from $\lag_2$ are of order~$p^k$, while one-loop corrections with one
vertex from $\lag_k$ and any number of vertices from $\lag_2$ are of
order~$p^{k+2}$. On the other hand, graphs with more vertices from
$\lag_{k^\prime}$ where $k^\prime > 2$ or with more loops are
suppressed by additional powers of the momentum.  The corresponding
expansion of the generating functional is denoted by
\begin{equation}
        W_{eff} = W_2 + W_4 + W_6 + \cdots \ ,
\end{equation}
where $W_k$ is of order~$p^k$.

\subsubsection{The generating functional at order~$p^2$}

At order $p^2$ the effective Lagrangian can be written in the
form
\be
\lag_2 = \lag_2^0 + \lag_2^{s} \, ,  \label{gen_lageff_2}
\ee
with
\bea
\lag_2^0 & = & {\bv^2\over 2} \left( \bW_\mu^+ \bW_\mu^-
+ \trho {1\over 4} \bZ_\mu \bZ_\mu \right)
+ {1\over 4\bgs}  \bW_{\mu\nu}^a \bW_{\mu\nu}^a
+ {1\over 4\bgps} \bB_{\mu\nu} \bB_{\mu\nu} \, ,
\label{gen_lageff_2_0} 
\eea
and
\be
\lag_2^{s}  = -{1\over2} \bK_{\mu\nu} \bB_{\mu\nu}
+ 2 \bv^2 (\bj_\mu^+ \bW_\mu^- + \bj_\mu^- \bW_\mu^+)
+ \bv^2 \bJ_\mu^\Z \bZ_\mu + 4 \bar c_\W \bv^2 \bJ_\mu^+ \bJ_\mu^-
+ \bar c_\Z \bv^2 \bJ_\mu^\Z \bJ_\mu^\Z \, ,  \label{gen_lageff_2_s}
\ee
where
\bea
\bW_{\mu\nu}^a   & = & \p_\mu \bW_\nu^a - \p_\nu \bW_\mu^a
                        + \varepsilon^{abc} \bW_\mu^b \bW_\nu^c \ ,
                        \ a = 1,2,3 \, , 
\label{defcalWmunu} \\
\bW_\mu^3        & = & \bZ_\mu + \bB_\mu \ , \\
\bj_\mu^\pm 	& = & \bar \phase^\pm \bJ_\mu^\pm \, . 
\label{def_jpm} 
\eea
The Lagrangian $\lag_2^0$ contains only the mass terms and the kinetic
terms of the gauge bosons in the effective theory. Note that in the
general effective Lagrangian $\lag_2^s$ in Eq.~(\ref{gen_lageff_2_s})
there appear additional contact terms involving the external
sources only. The masses of the gauge bosons, the weak mixing angle and the
electric charge can be expressed through the quantities
$\bv,\trho,\bg$, and $\bg^\prime$ as follows: 
\be \label{bare_quantities_eff}
M_W^2 = {\bv^2 \bes \over 4 \bs^2} \ , \
M_Z^2 = \trho {\bv^2 \bes \over 4 \bs^2 \bc^2} \ , \
\bc^2 = {\bgs \over \bgs + \bgps} \ ,
\bes = {\bgs \bgps \over \bgs + \bgps} \ .
\ee
The expression for the weak mixing angle $\bc^2$ follows from the
requirement that the field $\bZ_\mu = \bW^3_\mu - \bB_\mu$ is
invariant under gauge transformations. Similarly, the electric charge
$\bar e$ is determined by the coupling of the charged gauge boson
$\bW_\mu^\pm$ to the photon field $\bA_\mu$. The low-energy constants
$\bv$ and $\trho-1$ are of order $p^0$. Note that $\trho \equiv M_Z^2
\bc^2 / M_W^2$ is the inverse of the usual $\rho$-parameter.  In
allowing~$\trho \ne 1$ we do not assume that custodial symmetry
breaking effects vanish at leading order in the low-energy
expansion. Hence, we follow the first paper of Ref.~\cite{Longhitano}
and Ref.~\cite{Appelquist_Wu}. Note that in the recent literature it
became customary to include such a custodial symmetry breaking term
only at order $p^4$, following the conventions used in the second
paper of Ref.~\cite{Longhitano} and the second paper of
Ref.~\cite{H_M}. Since $\trho - 1$ is very small~\cite{PDG_98} this
might indeed be justified, if the low-energy expansion is carried out
up to order $p^4$ or higher.

At order~$p^2$, the generating functional of the effective field theory
is given by
\begin{equation} \label{W2_eff} 
	W_2[\bK_{\mu\nu}, \bJ_\mu^\pm, \bJ_\mu^\Z] = \int d^dx \lag_2
  \left(\bW_{\mu}^a, \bB_{\mu}; \bK_{\mu\nu}, \bJ_\mu^\pm, \bJ_\mu^\Z 
\right) \ ,
\end{equation}
where the gauge fields satisfy the equations of motion
\begin{eqnarray}
- \bd_\mu \bW^\pm_{\mu\nu} & = & - M_W^2 \bY^\pm_\nu
\pm i ( \bZ_{\mu\nu} + \bB_{\mu\nu}) \bW^\pm_\mu \mp i
\bW^\pm_{\mu\nu} \bZ_\mu  
\mp i (\p_\mu \bZ_\mu) \bW^\pm_\nu \pm i (\p_\mu \bZ_\nu)
\bW^\pm_\mu \nonumber \\
& &\pm i \bZ_\nu \bd_\mu \bW^\pm_\mu \mp i \bZ_\mu \bd_\mu
\bW^\pm_\nu - (\bZ_\mu \bZ_\mu) \bW^\pm_\nu + (\bZ_\mu \bZ_\nu)
\bW^\pm_\mu \pm 2 \bW^\pm_\mu (\bW_\mu^+ \bW_\nu^- - \bW_\nu^+
\bW_\mu^-) \, , \label{eomWpm_eff} \\ 
- \p_\mu (\bZ_{\mu\nu} + \bB_{\mu\nu})  & = & - \bc^2 M_Z^2
  \bY^\Z_\nu + 2 \bZ_\mu ( \bW_\mu^+ \bW_\nu^- + \bW_\nu^+
\bW_\mu^- )  
- 4 \bZ_\nu \bW_\mu^+ \bW_\mu^- + 2 i ( \bW^+_{\mu\nu}
\bW^-_\mu  - \bW^-_{\mu\nu} \bW^+_\mu ) \nonumber \\
& &- 2 i ( \bd_\mu\bW_\mu^+\bW_\nu^-
- \bd_\mu\bW_\mu^-\bW_\nu^+
- \bd_\mu\bW_\nu^+\bW_\mu^- + \bd_\mu\bW_\nu^-\bW_\mu^+) \, ,
\label{equmow3_eff} \\ 
- \p_\mu \bB_{\mu\nu} & = & \bs^2 M_Z^2 \PT_{\nu\mu} \bY^\Z_\mu
- {\bes\over \bc^2} \p_\mu \bK_{\mu\nu} \, . \label{eomB_eff_2}
\eea
Using relation~(\ref{bfieldsl_EFT}) the equations of motion for the massive
gauge field $\bZ_\mu$ and the photon field $\bA_\mu$ can be obtained.
The constraints are given by
\begin{eqnarray}
\bd_\mu\bY^\pm_\mu &=& \pm i \bZ_\mu \bY^\pm_\mu \mp i \trho
\bY^\Z_\mu\bW^\pm_\mu \, , \label{eomU_pm_eff} \\
\p_\mu\bY^\Z_\mu &=& 8 i {1\over \trho} (\bW_\mu^+ \bj_\mu^- - \bW_\mu^-
\bj^+_\mu) \ . \label{eomU_Z_eff}
\end{eqnarray}
They are obtained by varying the effective Lagrangian~$\lag_{2}$ with
respect to the Goldstone boson field~$\bU$. In
Eqs.~(\ref{eomWpm_eff})--(\ref{eomU_Z_eff}) we have introduced the
quantities 
\bea
\bW_{\mu\nu}^\pm & = & \bd_\mu \bW_\nu^\pm - \bd_\nu \bW_\mu^\pm \, ,
\label{def_caldWmunu} \\
\bZ_{\mu\nu} & = & \p_\mu \bZ_\nu - \p_\nu \bZ_\mu \ ,
\label{def_Zmunu} \\  
\bY_\mu^\pm & = & \bW_\mu^\pm + 4 \bj_\mu^\pm \ , \ 
\bY_\mu^\Z = \bZ_\mu + 4 {1\over \trho} \bJ_\mu^\Z \, ,  
\label{defYZ}  \\
\PT_{\mu\nu} & = & \delta_{\mu\nu} - {\p_\mu \p_\nu \over \Box} \,
. \label{def_PT}  
\eea
The covariant derivatives in $\bd_\mu \bW_{\mu\nu}^\pm$ and $\bd_\mu
\bY_\mu^\pm$ are defined in the same way as in Eq.~(\ref{d_Wpm_EFT}). 

Several things about the equations of motion
(\ref{eomWpm_eff})--(\ref{eomU_Z_eff}) are worth notice.  As discussed
in Ref.~\cite{SM_gaugeinv} the equations of motion uniquely determine
only the physical degrees of freedom since we did not fix a gauge. The
equations of motion can be rewritten in a form which only involves
fully $SU(2)_L \times U(1)_Y$ gauge-invariant fields. Solutions for
the massive gauge boson fields $\bar \phase^\mp \bW_\mu^\pm$ follow
from Eq.~(\ref{eomWpm_eff}). Suitable linear combinations of
Eqs.~(\ref{equmow3_eff}) and (\ref{eomB_eff_2}) determine the gauge
boson field $\bZ_\mu$ and the transverse component of the massless
photon field $\bA_\mu^T = \PT_{\mu\nu} \bA_\nu$.  Note that the
equations of motion do not determine the longitudinal component of the
photon field and the phase of the gauge boson fields $\bW_\mu^\pm$
which correspond to the $U(1)_Y$ gauge degree of freedom.  Even more
they do not determine the classical Goldstone boson field $\bU$
either, since it corresponds to the $SU(2)_L$ gauge degrees of
freedom. Thus, gauge invariance implies that these equations have a
whole class of solutions in terms of the original fields $\bU, \bar
W_\mu^a, \bB_\mu$.  Every two representatives are related to each
other by a gauge transformation.  Nevertheless, the physical degrees
of freedom are uniquely determined by these equations of motion.
Moreover, since the action is gauge-invariant, the generating
functional in Eq.~(\ref{W2_eff}) is uniquely determined for the given
set of source terms.

The most important point is the fact that the classical Goldstone
boson field $\bU$ represents the $SU(2)_L$ gauge degrees of freedom.
Thus, no Goldstone bosons are propagating at the classical level of
the theory. All gauge-invariant sources emit physical modes only.
Moreover, Eqs.~(\ref{eomU_pm_eff}) and (\ref{eomU_Z_eff}), which
follow from the requirement that the variation of the Lagrangian with
respect to the Goldstone boson field $\bU$ vanishes, are not equations
of motion, but constraints expressing the fact that the gauge fields
$\bar \phase^\mp \bW_\mu^\pm, \bZ_\mu,$ and $\bA_\mu$ couple to
conserved currents.  They can also be obtained by taking the
derivative of the equations of motion for the gauge fields. Note that
we have already used the constraints to bring these equations of
motion into the form given in
Eqs.~(\ref{eomWpm_eff})--(\ref{eomB_eff_2}).

We note that the equations of motion can be solved in powers of the
external sources, see Ref.~\cite{SM_gaugeinv}.

\subsubsection{The generating functional at order~$p^4$}
\label{subsec:EFT_W_4}

The one-loop contribution to the generating functional can be
evaluated with the saddle-point method.  If we write the fluctuations
$\bar \fluct$ around the classical fields $\bar \Allfields^{cl}$ as
$\bar \Allfields = \bar \cAllfields + \bar \fluct$, we obtain the
following representation for the one-loop approximation to the
generating functional: 
\begin{equation} \label{genfunc_saddlepoint_EFT}
  e^{-W_{eff}[\bK_{\mu\nu},\bJ_\mu^\pm, \bJ_\mu^\Z]} =
e^{-\intdx\lag_{eff}^{cl}} \int\d\mu[\bar \fluct] e^{- (1/2) \intdx
\bar \fluct^T \widetilde{\bD} \bar \fluct} \ . 
\end{equation}
Gauge invariance implies that the operator $\widetilde{\bD}$ has zero
eigenvalues corresponding to fluctuations $\bar \fluct$ which are
equivalent to infinitesimal gauge transformations. Indeed, if $\bar
\Allfields^{cl,i}$ is a solution of the equation of motion, i.e., a
stationary point of the classical action,
\begin{equation} \label{geqmo_EFT}
        \left.{\delta S_{eff}\over\delta\bar\Allfields^i}\right|_{\bar 
	\Allfields=\bar\cAllfields} = 0 \ ,
\end{equation}
then any gauge transformation yields another equivalent solution. The
index $i$ in $\bar \Allfields^{cl,i}$ labels the different fields.
Thus, differentiating equation~(\ref{geqmo_EFT}) with respect to the
gauge parameters $\gaugeparam^A$ one obtains
\begin{equation} \label{zero1_EFT}
        \left.
        {\delta^2 S_{eff} \over\delta\bar\Allfields^i\delta\bar\Allfields^j}
        {\delta\bar\Allfields^j\over\delta\gaugeparam^A}
        \right|_{\bar\Allfields=\bar\cAllfields} = 0 \ .
\end{equation}
The quadratic form which appears in Eq.~(\ref{zero1_EFT}) is identical
to the differential operator $\widetilde{\bD}$.  If these zero modes
are treated properly~\cite{Abelian_Higgs,SM_gaugeinv}, one can
evaluate the path-integral representation for the generating
functional at the one-loop level without the need to fix a gauge and
without introducing ghost fields. Up to an irrelevant infinite
constant one obtains the following result for the generating
functional of the effective field theory at order~$p^4$:
\begin{equation}	\label{actionatorder4}
 \left(W_2 + W_4\right)[\bK_{\mu\nu}, \bJ_\mu^\pm, \bJ_\mu^\Z] =
	\int d^dx \left(\lag_2 + \lag_4\right) +
   	{1\over2}\ln\det{}^\prime\widetilde{\bD} -
	{1\over2} \ln\det \bP^T\bP \, , 
\end{equation}
where~$\lag_4$ is the effective Lagrangian of order~$p^4$.  The first
term on the right-hand side represents the classical action which
describes the tree-level contributions of order $p^2$ and $p^4$ to the
generating functional. The two determinants on the right-hand side of
this equation represent one-loop contributions to the generating
functional. The first determinant describes all one-loop contributions
with vertices from the Lagrangian~$\lag_2$ where
${\det}^\prime\widetilde{\bD}$ is defined as the product of all
non-zero eigenvalues of the operator $\widetilde{\bD}$.  The second
determinant originates from the path integral measure.  The operator
$\bP$ satisfies the relation $ \bar P^T \widetilde{\bD} = 
\widetilde{\bD} \bP = 0$.  The fields in Eq.~(\ref{actionatorder4})
satisfy the equations of motion. At order~$p^4$ the contributions
from~$\lag_4$ to these equations of motion are not relevant. Hence,
they are given by Eqs.~(\ref{eomWpm_eff})--(\ref{eomU_Z_eff}). The
explicit form of the differential operators $\widetilde{\bD}$ and
$\bP$ for the case $\trho \neq 1$ is very complicated and we will not
write it down here. We note that the results for $\widetilde{\bD}$ and
$\bP$ for $\trho = 1$ can be inferred from the corresponding
differential operators in the standard model, see the discussion after
Eq.~(\ref{match2}) below.

The most general effective Lagrangian at order $p^4$ is given by
\be
\lag_4 = \lag_4^0 + \lag_4^{s} \, .   \label{gen_lageff_4}
\ee
The first term can be written in the form
\be \label{lag_4_0}
\lag_4^0 = \sum_{i=1}^{18} \ourlow_i \ourop_i \, , 
\ee
where the operators~$\ourop_i$ are given by
\bea
\ourop_{1} & = & (\bW_\mu^+ \bW_\mu^-) (\bW_\nu^+ \bW_\nu^-) \, ,
\nonumber \\ 
\ourop_{2} & = & (\bW_\mu^+ \bW_\nu^-) (\bW_\mu^+ \bW_\nu^-) \, ,
\nonumber \\ 
\ourop_{3} & = & (\bZ_\mu \bZ_\mu) (\bW_\nu^+ \bW_\nu^-) \, ,
\nonumber \\ 
\ourop_{4} & = & (\bZ_\mu \bZ_\nu) (\bW_\mu^+ \bW_\nu^-) \, ,
\nonumber \\ 
\ourop_{5} & = & (\bZ_\mu \bZ_\mu)(\bZ_\nu \bZ_\nu) \, , \nonumber \\
\ourop_{6} & = & \epsilon_{\mu\nu\rho\sigma} \bZ_\sigma (\bW_\rho^-
\bW_{\mu\nu}^+ + \bW_\rho^+ \bW_{\mu\nu}^-) \, , \nonumber \\
\ourop_{7} & = & i \bZ_{\mu\nu} (\bW_\mu^+ \bW_\nu^- - \bW_\nu^+
\bW_\mu^-) \, , \nonumber \\
\ourop_{8} & = & i \bB_{\mu\nu} (\bW_\mu^+ \bW_\nu^- - \bW_\nu^+
\bW_\mu^-) \, , \nonumber \\
\ourop_{9} & = & i \bZ_\mu ( \bd_\mu \bW_\nu^+ \bW_\nu^- - \bd_\mu
\bW_\nu^- \bW_\nu^+) \, , \nonumber \\
\ourop_{10} & = & i \bZ_\nu ( \bd_\mu \bW_\mu^+ \bW_\nu^- - \bd_\mu
\bW_\mu^- \bW_\nu^+) \, , \nonumber \\
\ourop_{11} & = & \bZ_{\mu\nu} \bZ_{\mu\nu} \, , \nonumber \\
\ourop_{12} & = & \bB_{\mu\nu} \bZ_{\mu\nu} \, , \nonumber \\
\ourop_{13} & = & (\bd_\mu \bW_\mu^+) (\bd_\nu \bW_\nu^-) \, ,
\nonumber \\ 
\ourop_{14} & = & (\p_\mu \bZ_\mu) (\p_\nu \bZ_\nu) \, , \nonumber \\
\ourop_{15} & = & M_W^2 \left( \bW_\mu^+ \bW_\mu^- +
{1\over4} \bZ_\mu \bZ_\mu \right) \, , \nonumber \\
\ourop_{16} & = & M_Z^2 \bZ_\mu \bZ_\mu \, , \nonumber \\
\ourop_{17} & = & \bW_{\mu\nu}^a \bW_{\mu\nu}^a \, , \nonumber \\
\ourop_{18} & = & \bB_{\mu\nu} \bB_{\mu\nu} \, .  \label{gen_lageff_4_0}
\eea
We recall that we count the gauge fields $\bW_\mu^\pm, \bZ_\mu$ and the
masses $M_W, M_Z$ as order $p$ in the low-energy expansion, therefore
the custodial symmetry breaking term $\ourop_{16}$ is of the order
$p^4$. The second term in Eq.~(\ref{gen_lageff_4}) contains all
contributions involving external sources:
\be \label{lag_4_s}
\lag_4^s = \sum_{i=1}^{76} \ourlow_i^s \ourop_i^s \, .
\ee
The operators~$\ourop_i^s$ are listed in
Appendix~\ref{app:sources_p4}. Note, that we consider CP-even terms
only. The low-energy constants~$\ourlow_i$ and~$\ourlow_i^s$ are
quantities of order $p^0$.

It is important to note, that the most general effective Lagrangian at
this order is given as a linear combination of a maximal set of
gauge-invariant terms of order~$p^4$. One can then eliminate redundant
terms by using algebraic relations of the form
\begin{equation}
	\int d^dx (\bd_\mu \bW_\nu^+) (\bd_\nu \bW_\mu^-) =
	\int d^dx \left( {1\over2} \ourop_8 + \ourop_{13} \right) \, , 
\end{equation}
which are readily verified by partial integration. On the other hand,
the Lagrangian~$\lag_4$ contributes only at the classical level.
Hence, the equations of motion~(\ref{eomWpm_eff})--(\ref{eomB_eff_2})
as well as the constraints~(\ref{eomU_pm_eff}) and~(\ref{eomU_Z_eff})
can also be used to eliminate further redundant
terms~\cite{GL84_85,Bij_Col_Eck}. Equivalently, one can also remove
terms in the effective Lagrangian by performing appropriate
reparametrizations of the fields and external sources in the path
integral~\cite{Fearing_Scherer}.  Note that we have already eliminated
all algebraically dependent terms from the lists given in
Eq.~(\ref{gen_lageff_4_0}) and in Appendix~\ref{app:sources_p4}. Thus,
we only need to employ the equations of motion and the constraints to
eliminate further redundant terms. Note that in our gauge-invariant
approach no gauge artifacts can enter through this procedure.

The constraints (\ref{eomU_pm_eff}) and (\ref{eomU_Z_eff}) yield the
following relations between the operators in the Lagrangian $\lag_4$:
\bea
\ourop_{10} & = & - 2 (1-\trho) \ourop_4 + 4 \ourop_{4}^s - 4
\ourop_{6}^s - 4 \ourop_{46}^s  \, , \label{remove_O9} \\
\ourop_{13} & = & (1-\trho)^2 \ourop_4 - 4 (1-\trho) \ourop_{4}^s + 4
(1-\trho) \ourop_{6}^s + 16 \ourop_{14}^s - 16 \ourop_{17}^s
\nonumber \\
&&\mbox{} + 16 \ourop_{19}^s + 4 (1-\trho) \ourop_{46}^s - 16
\ourop_{51}^s + 16 \ourop_{53}^s + 16 \ourop_{74}^s \, ,
\label{remove_O13} \\ 
\ourop_{14} & = & {64\over \trho^2} (2 \ourop_{10}^s - \ourop_{12}^s)
+ {64 \over \trho^2} (\ourop_{49}^s - \ourop_{52}^s) + {16\over \trho^2}
\ourop_{76}^s \, , \label{remove_O14} \\
\ourop_{41}^s & = & - (1-\trho) \ourop_{4}^s + 8 \ourop_{14}^s - 4
\ourop_{17}^s - 4 \ourop_{51}^s \, , \label{remove_O41s} \\
\ourop_{43}^s & = & - (1-\trho) \ourop_{6}^s  + 4 \ourop_{17}^s - 8
\ourop_{19}^s - 4 \ourop_{53}^s \, , \label{remove_O43s} \\
\ourop_{47}^s & = & {8\over \trho} (2 \ourop_{10}^s - \ourop_{12}^s) +
{4\over \trho} (\ourop_{49}^s - \ourop_{52}^s) \, ,
\label{remove_O47s} \\ 
\ourop_{48}^s & = & - (1-\trho) \ourop_{16}^s + 4 \ourop_{25}^s - 4
\ourop_{27}^s - 4 \ourop_{55}^s \, , \label{remove_O48s} \\
\ourop_{73}^s & = & - (1-\trho) \ourop_{46}^s + 4 \ourop_{51}^s - 4
\ourop_{53}^s - 8 \ourop_{74}^s \, , \label{remove_O73s} \\
\ourop_{75}^s & = & - {8\over \trho} (\ourop_{49}^s - \ourop_{52}^s) -
{4\over \trho} \ourop_{76}^s \, . \label{remove_O75s}
\eea
The equations of motion for $\bW_\mu^\pm$, Eq.~(\ref{eomWpm_eff}), and
$\bW_\mu^3$, Eq.~(\ref{equmow3_eff}), yield
\bea
\ourop_{11} & = & - 8 \ourop_1 + 8 \ourop_2 - 16 \ourop_3 + 16 \trho
\ourop_4 + 8 \ourop_7 - 8 \ourop_9 - 8 \ourop_{15} 
+ 2 \bc^2 \left({1\over \trho} - 2\right) \ourop_{16} - \ourop_{17} +
\ourop_{18} \nonumber \\
&&+ 32 \ourop_{4}^s - 32 \ourop_{6}^s - 32 \ourop_{46}^s - 16
\ourop_{64}^s - 16 {\bc^2 \over \trho} \ourop_{66}^s \, ,
\label{remove_O11} \\ 
\ourop_{12} & = & 8 \ourop_1 - 8 \ourop_2 + 8 \ourop_3 - 8 \trho
\ourop_4 - 4 \ourop_7 + 4 \ourop_9 + 8 \ourop_{15} 
- 2 \bc^2 \left( {1\over \trho} - 1 \right) \ourop_{16} + \ourop_{17}
- \ourop_{18} \nonumber \\
&&- 16 \ourop_{4}^s + 16 \ourop_{6}^s + 16 \ourop_{46}^s + 16
\ourop_{64}^s + 8 {\bc^2 \over \trho} \ourop_{66}^s \, ,
\label{remove_O12} \\ 
\ourop_{68}^s & = & - 4 \ourop_{1}^s + 4 \ourop_{2}^s  -
2 \ourop_{5}^s + 2 \trho \ourop_{6}^s - {32\over \trho} \ourop_{10}^s
+ {16\over \trho} \ourop_{12}^s + 8 \ourop_{17}^s 
- 16 \ourop_{19}^s + 2 \ourop_{35}^s + \ourop_{36}^s \nonumber \\
&&- 4 \ourop_{44}^s + 2 \ourop_{46}^s - {8\over \trho} (\ourop_{49}^s -
\ourop_{52}^s ) - 8 \ourop_{53}^s - 2 \ourop_{64}^s - 16 \ourop_{65}^s
\, ,  \label{remove_O68s} \\
\ourop_{70}^s & = & - 8 \ourop_{3}^s + 4 \trho \ourop_{4}^s + 32
\ourop_{14}^s - 16 \ourop_{17}^s + 4 \ourop_{34}^s  - 4 \ourop_{42}^s
- 16 \ourop_{51}^s - 2 \bc^2 \ourop_{66}^s - 8 {\bc^2 \over \trho}
\ourop_{67}^s  - \ourop_{71}^s \, . \label{remove_O70s}
\eea
Note that we have frequently employed partial integrations to derive
the Eqs.~(\ref{remove_O9})--(\ref{remove_O70s}).  Furthermore, we
have already replaced all dependent terms on the right-hand side of
Eqs.~(\ref{remove_O11})--(\ref{remove_O70s}).
Equation~(\ref{remove_O12}) can be derived by observing the identities
\be
\bW_{\mu\nu}^+ \bW_{\mu\nu}^- = - 2 \ourop_1 + 2 \ourop_2 - 2
\ourop_3 + 2 \ourop_4 + 2 \ourop_7 + \ourop_8 - 2 \ourop_9 + 2
\ourop_{10} - {1\over 4} \ourop_{11} - {1\over 2} \ourop_{12} +
{1\over 4} \ourop_{17} - {1\over 4} \ourop_{18} \, , 
\ee
and
\be \label{WmunuWmunu}
\bW_{\mu\nu}^+ \bW_{\mu\nu}^- = - \bW_\nu^+ \bd_\mu \bW_{\mu\nu}^- -
\bW_\nu^- \bd_\mu \bW_{\mu\nu}^+ \, , 
\ee
which are valid up to partial integrations.  Afterwards one can employ
the equation of motion~(\ref{eomWpm_eff}) to substitute the expression
for $\bd_\mu \bW_{\mu\nu}^\pm$ in Eq.~(\ref{WmunuWmunu}).  In the same
way one can obtain the relation~(\ref{remove_O68s}) for
$\ourop_{68}^s$.  Similarly, performing partial integrations in
$(\ourop_{11} + \ourop_{12})$ and $(\ourop_{70}^s + \ourop_{71}^s)$
lead to $\p_\mu (\bB_{\mu\nu} + \bZ_{\mu\nu})$ where the equation of
motion~(\ref{equmow3_eff}) can be applied in order to obtain
Eqs.~(\ref{remove_O11}) and (\ref{remove_O70s}). Using the
relations~(\ref{remove_O9})--(\ref{remove_O70s}) one can eliminate
the terms on the left-hand side of the corresponding equations from
the set of terms in the Lagrangian $\lag_4$. This reduces the number
of low-energy constants by 13.  Note that one has to adjust the values
of the low-energy constants of the remaining terms accordingly. We
will denote the modified low-energy constants by $\ourlow_i^\prime$
and $\ourlow_i^{s\prime}$ in order to distinguish them from the old
ones.

Finally, there are terms in the Lagrangian $\lag_4$ which are
proportional to corresponding terms in the lowest order Lagrangian
$\lag_2$. These are the operators $\ourop_{15}$, $\ourop_{16}$,
$\ourop_{17}$, $\ourop_{18}$, $\ourop_{64}^s$, $\ourop_{65}^s$,
$\ourop_{66}^s$, $\ourop_{67}^s$, and $\ourop_{71}^s$. Following the
interpretation given in Refs.~\cite{Longhitano,Feruglio} these terms
lead to a renormalization of the low-energy constants and sources at
order~$p^2$ according to
\begin{eqnarray}
 \bv^2 &\rightarrow& \bv_{eff}^2 = \bv^2
	\left( 1 + 2 \ourlow_{15} {M_W^2\over\bv^2}  \right) \, , 
        \label{v_eff_general} \\
 \brho &\rightarrow& \brho_{eff} = \brho - 2 (\brho -1)  \ourlow_{15}
 {M_W^2 \over \bv^2} + 8 \ourlow_{16} {M_Z^2 \over \bv^2} \, , 
 \label{rho_eff_general} \\
%
%
 {\bg}{}^2 &\rightarrow& {\bg}{}_{eff}^2 =  {\bg}{}^2
	\left( 1 - 4 \ourlow_{17} {\bg}{}^2  \right) \, , \\
 {\bg^\prime}{}^2 &\rightarrow& {\bg^\prime}{}_{eff}^2 = {\bg^\prime}{}^2
	\left( 1 - 4 \ourlow_{18} {\bg^\prime}{}^2  \right) \, ,
\label{gprime_eff_general} \\
 \bK_{\mu\nu} &\rightarrow& \bK_{\mu\nu; eff} = \bK_{\mu\nu} - 2
 \ourlow_{71}^s \bJ_{\mu\nu}^\Z  \, , \label{renorm_K} \\
 \bJ_\mu^{\pm} &\rightarrow& \bJ_{\mu; eff}^\pm = \bJ_\mu^\pm \left( 1 +
 \left( {1\over 2} \ourlow_{64}^s - 2
 \ourlow_{15} \right) {M_W^2 \over \bv^2} \right) \, , \\
 \bJ_\mu^{\Z} &\rightarrow& \bJ_{\mu; eff}^\Z = \bJ_\mu^\Z \left( 1 +
 \ourlow_{66}^s {M_Z^2 \over \bv^2}  - 2  \ourlow_{15} {M_W^2 \over
   \bv^2} \right) \, ,  \label{renorm_JZ} \\
 \bc_\W         &\rightarrow& \bc_{\W; eff} = \bc_\W \left( 1 +
  (2 \ourlow_{15} - \ourlow_{64}^s) {M_W^2 \over \bv^2} \right) +
  {1\over 4} \ourlow_{65}^s {M_W^2 \over \bv^2} \, ,   \\
 \bc_\Z         &\rightarrow& \bc_{\Z; eff} = \bc_\Z \left( 1 + 2
 \ourlow_{15} {M_W^2 \over \bv^2}  -  2 \ourlow_{66}^s {M_Z^2 \over \bv^2}
   \right) + \ourlow_{67}^s {M_Z^2 \over \bv^2}  \, .
\end{eqnarray}

Hence, we end up with the following set of independent operators at
order~$p^4$: 
\bea
\ourop_{1} & = & (\bW_\mu^+ \bW_\mu^-) (\bW_\nu^+ \bW_\nu^-) \, ,
\nonumber \\ 
\ourop_{2} & = & (\bW_\mu^+ \bW_\nu^-) (\bW_\mu^+ \bW_\nu^-) \, ,
\nonumber \\ 
\ourop_{3} & = & (\bZ_\mu \bZ_\mu) (\bW_\nu^+ \bW_\nu^-) \, ,
\nonumber \\ 
\ourop_{4} & = & (\bZ_\mu \bZ_\nu) (\bW_\mu^+ \bW_\nu^-) \, ,
\nonumber \\ 
\ourop_{5} & = & (\bZ_\mu \bZ_\mu)(\bZ_\nu \bZ_\nu) \, , \nonumber \\ 
\ourop_{6} & = & \epsilon_{\mu\nu\rho\sigma} \bZ_\sigma (\bW_\rho^-
\bW_{\mu\nu}^+ + \bW_\rho^+ \bW_{\mu\nu}^-) \, , \nonumber \\
\ourop_{7} & = & i \bZ_{\mu\nu} (\bW_\mu^+ \bW_\nu^- - \bW_\nu^+
\bW_\mu^-) \, , \nonumber \\
\ourop_{8} & = & i \bB_{\mu\nu} (\bW_\mu^+ \bW_\nu^- - \bW_\nu^+
\bW_\mu^-) \, , \nonumber \\
\ourop_{9} & = & i \bZ_\mu ( \bd_\mu \bW_\nu^+ \bW_\nu^- - \bd_\mu
\bW_\nu^- \bW_\nu^+) \, ,    \label{gen_lageff_4_0_reduced}
\eea
and 
\be
\ourop_{1}^s,
\ldots, \ourop_{40}^s, \ourop_{42}^s, \ourop_{44}^s,
\ourop_{45}^s, \ourop_{46}^s, \ourop_{49}^s, \ldots,
\ourop_{63}^s, \ourop_{69}^s, \ourop_{72}^s, \ourop_{74}^s,
\ourop_{76}^s \, .  
\ee
Thus, we obtain $9 + 63 = 72$ independent low-energy constants which 
we denote by $\ourlow_i^\prime$ and $\ourlow_i^{s\prime}$. 

As discussed above, since $\trho-1$ is tiny, some people set $\trho =
1$ and instead add the operator $M_Z^2 \bZ_\mu \bZ_\mu$ to the basis
at order $p^4$. In order to facilitate the comparison with the
literature, we cover this case by including the term
\be
\ourop_0 \doteq M_Z^2 \bZ_\mu \bZ_\mu \equiv \ourop_{16} \, , 
\ee
with the corresponding low-energy constant $\ourlow_0^\prime$ into the
basis from Eq.~(\ref{gen_lageff_4_0_reduced}). The total number of
independent low-energy constants in $\lag_2 + \lag_4$ remains the
same, if we trade $\trho-1$ for $\ourlow_0^\prime$. The momentum
counting, however, is different, see the discussion after
Eq.~(\ref{bare_quantities_eff}). 

Note that one cannot obtain additional relations between the operators
in $\lag_4$ from the equation of motion for $\bB_\mu$,
Eq.~(\ref{eomB_eff_2}), since it contains non-local terms involving
the projection operator $\PT_{\mu\nu}$, cf.\ Eq.~(\ref{def_PT}). Let
us consider this equation in greater detail.

The presence of non-local terms in Eq.~(\ref{eomB_eff_2}) results from
our coupling sources to the non-local charged gauge-boson fields in
Eq.~(\ref{gen_lageff_2_s}). Indeed, switching off the
sources~$\bJ_\mu^\pm$ yields
\begin{eqnarray}
\bW_\mu^\pm & = & 0 \, , \\
\partial_\mu \bY_\mu^\Z & = & 0 \ . 
\end{eqnarray}
Hence, Eq.~(\ref{eomB_eff_2}) simplifies to
\be
- \p_\mu \bB_{\mu\nu} =  \bs^2 M_Z^2 \bY^\Z_\nu
- {\bes\over \bc^2} \p_\mu \bK_{\mu\nu} \, . \label{eomB_eff_2_s}
\ee
Multiplying this equation by~$\bZ_\nu$ one obtains by partial
integration
\begin{equation} \label{add_relation}
\ourop_{12} =  2 \bs^2\ourop_{16} + 8 {\bs^2\over\brho} \ourop_{66}^s +
{{\bar e}^2\over\bc^2} \bZ_{\mu\nu}\bK_{\mu\nu} \, .
\end{equation}
This relation involves the new operator
\begin{equation}
\bZ_{\mu\nu}\bK_{\mu\nu} \, , \label{add_source_term}
\end{equation}
which we did not consider because it is physically irrelevant. In the
case of the standard model the source~$K_{\mu\nu}$ enters the
Lagrangian as in Eq.~(\ref{basic_sources}) below. As will be shown in
Sec.~\ref{sec:matching}, this in turn implies that the
corresponding effective field theory involves the
source~$\bK_{\mu\nu}$ only through the single source term introduced
in Eq.~(\ref{gen_lageff_2_s}). As long as the field~$\B_\mu$ describes
a weakly interacting~$U(1)_Y$ gauge field, this is in fact true for
any underlying theory. Hence, operators as the one shown in
Eq.~(\ref{add_source_term}) need not be considered and
Eq.~(\ref{add_relation}) cannot be used to eliminate further redundant
terms.

If the source~$\bK_{\mu\nu}$ is switched off as well,
Eq.~(\ref{add_relation}) simplifies to
\begin{equation} \label{add_relation_s}
\ourop_{12} =  2 \bs^2\ourop_{16} + 8 {\bs^2\over\brho} \ourop_{66}^s
\, . 
\end{equation}
This relation can also be derived from Eqs.~(\ref{remove_O11})
and~(\ref{remove_O12}) since the equations of motion now have the solutions 
\begin{eqnarray}
 \bW_\mu^\pm & = & 0 \\
 \bA_\mu & = & 0 \label{add_relation_equiv} \, ,
\end{eqnarray}
implying~$\bB_\mu = - \bs^2 \bZ_\mu$ and $\bW^3_\mu = \bc^2
\bZ_\mu$. Equations~(\ref{remove_O11}), (\ref{remove_O12})
and~(\ref{add_relation_s}) do, in fact, require
Eq.~(\ref{add_relation_equiv}) to be satisfied. This result shows
clearly, that one should be careful in using equations of motion to
eliminate operators in the effective Lagrangian, if (some of) their
solutions vanish. In doing so, one may accidentally remove terms that
are not redundant at all.

In the remainder of this section we will compare our results with
those obtained in the literature~\cite{Appelquist_Wu,Feruglio}. Since
no source terms have been considered in these references we will
switch off all the sources for the moment.  Furthermore, we have to
take into account that in Ref.~\cite{Feruglio} the low-energy constant
$\brho-1$ is treated as a quantity of order~$p^2$. Thus, we will
compare our 10 low-energy constants
\begin{equation}
  \ourlow_1^\prime, \ldots, \ourlow_9^\prime \mbox{  and  } \brho - 1
  \ \mbox{ (or equivalently~$\ourlow_0^\prime$) } \, ,  
\end{equation}
with those obtained in the literature. The expression for the
effective Lagrangian $\lag_2^0$ in the notation which is usually used
in the literature and the relation between our set of operators in
$\lag_4^0$ and the usual basis can be found in
Appendix~\ref{app:EW_chiral_Lag}. In
Refs.~\cite{Appelquist_Wu,Feruglio} all operators in $\lag_4^0$ that
are proportional to terms in the lowest order Lagrangian~$\lag_2^0$
have been discarded right at the beginning. Hence, the authors start
with 15 CP-even terms corresponding to the terms $\ourop_1, \ldots,
\ourop_{14}$ and $\ourop_{16}$ in Eq.~(\ref{gen_lageff_4_0}), see also
Eq.~(\ref{basis_appelquist_wu}).

By making use of the equations of motion, $\tr (\covDmat_\mu \Vmat_\mu)
= 0$ (for notations see Appendix~\ref{app:EW_chiral_Lag}),
corresponding to our constraints~(\ref{eomU_pm_eff}) and
(\ref{eomU_Z_eff}), the number of terms was reduced from 15 to 12 in
these references. In fact, the three relations
\bea
\op_{11} & = & 0 \, , \\
\op_{12} & = & 0 \, , \\
\op_{13} & = & {1\over 4} \bB_{\mu\nu} \bB_{\mu\nu} + \op_1 + \op_4 -
\op_5 - \op_6 + \op_7 + \op_8 \, ,  \label{remove_op13}
\eea
given in Ref.~\cite{Feruglio}\footnote{We obtain a different sign of
the terms $\op_4$ and $\op_5$ in Eq.~(\ref{remove_op13}) compared to
Ref.~\cite{Feruglio}.} correspond to
Eqs.~(\ref{remove_O9})--(\ref{remove_O14}), if we set all sources to
zero and assume $\trho = 1$ at leading order, i.e.\ to
\be
\ourop_{10} = 0 \, , \quad \ourop_{13} = 0 \, , \quad
\ourop_{14} = 0 \, .
\ee
Note especially that Eq.~(\ref{remove_op13}) corresponds to
$\ourop_{14} = 0$ in our basis, cf.\ the relation between the two sets
of operators which is given in Eq.~(\ref{L_i_vs_O_i}).

In addition to the constraints we furthermore use the equations of
motion for the gauge fields~(\ref{eomWpm_eff}) and (\ref{equmow3_eff})
to reduce the number of low-energy constants from 12 to 10. Since this
step was not taken in Refs.~\cite{Appelquist_Wu,Feruglio} the set of
low-energy constants used in these references is redundant. 

This is an important result and we would like to add some comments.
First of all, we stress again that we are studying for the moment a
purely bosonic effective field theory which describes any underlying
theory with the same symmetry breaking pattern as the standard model,
i.e.\ no fermions have been included in the effective Lagrangian. In
order to really compare our findings with
Refs.~\cite{Appelquist_Wu,Feruglio} one has to consider the fermions
in the analysis, which was implicitly done in these references, see
also Ref.~\cite{D_GK}. We will come back to this point below.

Using Eqs.~(\ref{remove_O11}) and (\ref{remove_O12}) we have {\it
  chosen} to remove the operators $\ourop_{11}$ and $\ourop_{12}$ from
the effective Lagrangian in Eq.~(\ref{gen_lageff_4_0}). These
operators contribute to the self-energies of the gauge bosons which
are not observable anyway. In the basis which is usually used in the
literature this corresponds to removing the operators $\op_1$ and
$\op_8$ from the basis, see Appendix~\ref{app:EW_chiral_Lag}.
Sometimes the corresponding low-energy constants $a_1$ and $a_8$ are
identified with the oblique correction parameters $S$ and
$U$~\cite{STU}. Furthermore, the parameter $T$ is identified with the
low-energy constant $a_0$ which corresponds to $\brho-1$, or,
depending on the momentum counting, to the low-energy constant
$l_{0}^\prime$ in our basis.  Before any conclusions about the oblique
parameters can be drawn, however, one has to study the inclusion of
fermions in the effective field theory. This will be done below where
we will compare our results with the experimentally determined values
for the oblique parameters $S, T,$ and $U$.

Of course, within our functional approach the source terms have to be
considered as well. Even in this case, however, only the 10 low-energy
constants $\ourlow_{1}^\prime, \ldots, \ourlow_{9}^\prime$ and $\brho
- 1$ (or equivalently $\ourlow_0^\prime$) will contribute to physical
quantities, like $S$-matrix elements, masses and decay constants of
gauge bosons.  The first group of source terms which will obviously
not contribute to physical quantities are the contact terms
$\ourop_{65}^s$, $\ourop_{67}^s$, $\ourop_{69}^s$, $\ourop_{72}^s$,
$\ourop_{74}^s$, and $\ourop_{76}^s$ with two powers of the external
sources, cf.\ Eq.~(\ref{two_factors}), and all terms in $\lag_4^s$
with three or four powers of the fields and sources which contain at
least one factor with an external source, i.e.\ the operators
$\ourop_{1}^s, \ldots, \ourop_{63}^s$ in Eqs.~(\ref{four_factors}) and
(\ref{three_factors}).  This is due to the fact that in physical
$S$-matrix elements all external lines are amputated from the Green's
functions.  The corresponding low-energy constants are thus similar to
the constants $h_i$ in the ordinary chiral Lagrangian~\cite{GL84_85}.
Furthermore, with the help of Eqs.~(\ref{remove_O68s}),
(\ref{remove_O70s}), (\ref{remove_O73s}), and (\ref{remove_O75s}), one
can remove the operators $\ourop_{68}^s, \ourop_{70}^s,
\ourop_{73}^s,$ and $\ourop_{75}^s$ from the basis.  Finally, the
operators $\ourop_{64}^s, \ourop_{66}^s$ and $\ourop_{71}^s$ lead only
to a renormalization of the sources $\bJ_\mu^\pm, \bJ_\mu^\Z,$ and
$\bK_{\mu\nu}$ in the lowest order effective Lagrangian in
Eq.~(\ref{gen_lageff_2_s}), cf.\
Eqs.~(\ref{renorm_K})--(\ref{renorm_JZ}). 

In summary, in a purely bosonic effective field theory with the same
symmetry breaking pattern as the standard model, there are only 10
instead of 12 physically relevant low-energy constants at order $p^4$
in the electroweak chiral Lagrangian. In particular, one can choose to
remove two low-energy constants $\ourlow_{11}$ and $\ourlow_{12}$
which contribute to the self-energies of the gauge bosons. An
additional number of 63 low-energy constants contributes to the
off-shell behavior of our gauge-invariant Green's functions. The
latter low-energy constants, however, do not enter physical
quantities. 

The situation is more involved, however, if fermions are included in
the analysis, since in that case the sources $\bJ_\mu^\pm$ and
$\bJ_\mu^\Z$ also contain fermionic currents. We will now comment on
this point.

\subsection{On the inclusion of fermions}
\label{sec:EFT_fermions}

The fermionic part of the effective Lagrangian is of the form
\begin{equation} \label{efffermionlag1}
  \lag^f_{eff} = \lag^f_{eff}\left(\bPsi_L^k, \bu_R^k, \bdf_R^k, \bU,
  D_\mu \Psi_L^k, D_\mu \bu_R^k, D_\mu \bdf_R^k, \bD_\mu \bU, \ldots;
  M_L^k, N_L^k , M_R^k, N_R^k \right) \, ,
\end{equation}
where~$\Psi_L^k$ denotes the left-handed iso-doublet fields
while~$d_R^k$ and~$u_R^k$ represent right-handed up- and down-type
fermion fields comprising leptons and quarks. Note that all our
fermion fields are weak eigenstates. The quantities~$M_{L,R}^k$
and~$N_{L,R}^k$ denote external sources coupling to these fermion
fields. As discussed for the bosonic part, the effective Lagrangian is
a sum of terms with an increasing number of derivatives and powers of
fields and sources corresponding to an expansion of the generating
functional in powers of the momenta and the masses.  In addition to
the counting rules discussed above we require that fermion fields are
treated as quantities of order~$\sqrt{p}$ and fermion masses, denoted
by $m_f^k,$ as of order $p$. This ensures that the low-energy
expansion is carried out at a fixed ratio~$m_f^k / p$.

The left-handed iso-doublet fields transform under~$SU(2)_L$ gauge
transformations in the following way:
\be \label{SUtwo_trafo_f_eff}
\Psi_L^k \to \V \Psi_L^k \, , \quad \V \in SU(2) \, ,  \\
\ee
and under~$U(1)_Y$ gauge transformations as follows:
\be \label{Uone_trafo_f_Psi_eff}
\Psi_L^k   \to e^{- i Y(\Psi_L^k) \UoneParam / 2} \ \Psi_L^k .
\ee
The iso-singlets transform under $U(1)_Y$ gauge transformations in the
following way:
\bea 
u_R^k   & \to & e^{- i Y(u_R^k) \UoneParam / 2} \ u_R^k \, ,
\nonumber \\ 
d_R^k   & \to & e^{- i Y(d_R^k) \UoneParam / 2} \ d_R^k \, . 
\label{Uone_trafo_f_singlets_eff}
\eea
The hypercharges for lepton fields are~$Y(\Psi_L^k) = -1$, $Y(u^k_R) =
0$ and~$Y(d^k_R) = -2$ while those for quark fields are~$Y(\Psi_L^k) =
{1\over 3}$, $Y(u^k_R) = {4\over3}$ and~$Y(d^k_R) = -{2\over3}$.  The
covariant derivatives for the fermion fields in
Eq.~(\ref{efffermionlag1}) are given by
\begin{eqnarray}
 D_\mu \Psi_L^k & = & \left(\partial_\mu - i{\tau^a\over2} \bar
W_\mu^a - i {Y(\Psi_L^k)\over2} \bB_\mu\right) \Psi_L^k \, ,   \\
D_\mu f_R^k & = & \left(\partial_\mu - i {Y(f_R^k)\over2} \bB_\mu\right)
f_R^k \ ,\qquad f = u, d \ .
\end{eqnarray}

Following our approach to the bosonic sector, we can rewrite the
effective Lagrangian~(\ref{efffermionlag1}) in terms of~$SU(2)_L$
invariant fields, which are defined as~\cite{QED_gaugeinv}
\begin{eqnarray}
	\bu_L^k & = & \tbU^\dagger \bPsi_L^k \, , \\
	\bdf_L^k & = & \bU^\dagger \bPsi_L^k \,  .
\end{eqnarray}
They transform under~$U(1)_Y$ gauge transformations as
\bea 
u_L^k   & \to & e^{- i Y(u_L^k) \UoneParam / 2} \ u_L^k \, ,
\nonumber \\
d_L^k   & \to & e^{- i Y(d_L^k) \UoneParam / 2} \ d_L^k \, , 
\label{Uone_trafo_f_gaugeinv_eff}
\eea
where~$Y(u_L^k) = Y(u_R^k)$ and~$Y(d_L^k) = Y(d_R^k)$. 

At order~$p^2$ the fermionic part of the effective Lagrangian contains
several terms 
\begin{equation}
\lag_2^f = \lag_2^{f, kin} + \lag_2^{f, Y} + \lag_2^{f, CC} +
	\lag_2^{f, NC} + \lag_2^{f, 4F} + \lag_2^{f, s} \, .
\end{equation}
They denote the kinetic part of the Lagrangian, the Yukawa couplings,
the coupling to charged and neutral currents, four-fermion
interactions and source terms. The first four terms can readily be 
inferred from the corresponding terms in the fermionic sector of the
standard model~\cite{QED_gaugeinv}  
\begin{eqnarray}
  \lag_{2}^{f, kin} & = & \sum_k \left( \bar\bdf_L^k i \Dslash \bdf_L^k +
			 \bar\bu_L^k i \Dslash \bu_L^k
	             + \bar\bdf_R^k i \Dslash \bdf_R^k
	             + \bar\bu_R^k i \Dslash \bu_R^k \right) \, , \\
 \lag_2^{f, Y} & = &   \bar v
	\sum_{ij} \left( \bg_{ij} \bar\bdf_L^i  \bdf_R^j +
		\bg_{ji}^{*}\bar\bdf_R^i \bdf_L^j +
	\bh_{ij} \bar\bu_L^i \bu_R^j +
		\bh_{ji}^{*} \bar\bu_R^i \bu_L^j \right) \, , \\
 \lag_2^{f, CC} & = &
	\sum_{ij} c_{CC}^{ij,L} \left( \bW_\mu^+ j_\mu^{L, ij -} + \bW_\mu^-
			j_\mu^{L, ij +} \right) +
	c_{CC}^{ij,R} \left( \bW_\mu^+ j_\mu^{R, ij -} + \bW_\mu^-
			j_\mu^{R, ij +} \right) \, ,  \label{CC} \\ 
 \lag_2^{f, NC} & = &  \sum_{ij} c_{NC}^{ij,L} \bZ_\mu J_\mu^{L, ij 3}
		+ \sum_{ij} c_{NC}^{ij,R} \bZ_\mu J_\mu^{R, ij 3} 
                - \bs^2 J_\mu^{f,Q} \bZ_\mu \, , 
\label{NC} 
\end{eqnarray}
where 
\begin{eqnarray}
	D_\mu f^k_{L,R} & = & \left( \partial_\mu - i Q_{f^k}
		\bA_\mu\right) f^k_{L,R} \, , \\
	j_\mu^{L/R, ij +} & = & \bar \bdf_{L/R}^i \gamma_\mu
        \bu_{L/R}^j  \, , \\ 
	j_\mu^{L/R, ij -} & = & \bar \bu_{L/R}^i \gamma_\mu
        \bdf_{L/R}^j \, , \\
        J_\mu^{L/R, ij 3} &  = & {1\over2} \left( \bar \bu_{L/R}^i
        \gamma_\mu \bu_{L/R}^j
      - \bar \bdf_{L/R}^i \gamma_\mu \bdf_{L/R}^j \right) \, , \\ 
        J_\mu^{f, Q} &  = & \sum_k \left(
        Q_{u^k} \bar \bu_L^k \gamma_\mu \bu_L^k
      + Q_{d^k} \bar \bdf_L^k \gamma_\mu \bdf_L^k
      + Q_{u^k} \bar \bu_R^k \gamma_\mu \bu_R^k
      + Q_{d^k} \bar \bdf_R^k \gamma_\mu \bdf_R^k
        \right) \, .  
\end{eqnarray}
The electromagnetic charges are given by the quantities~$Q_{f^k} =
{1\over 2} Y(f^k)$. The Yukawa coupling constants~$\bg_{ij}$
and~$\bh_{ij}$ count as quantities of order~$p$ in the low-energy
expansion. This ensures that fermion masses $m_f^k$ are treated as of
order $p$ as well. The constants~$c_{CC}^{ij}$ and~$c_{NC}^{ij}$ are
of order $p^0$. Gauge-invariant sources for fermions are readily
constructed. We do not need to discuss this point here and refer the
interested reader to Ref.~\cite{QED_gaugeinv}.

A general effective Lagrangian analysis involves, {\it a priori}, all
possible couplings between the fermions and the gauge bosons.
Invariance under~$U(1)$ gauge transformations completely determines
only the coupling between fermions and the photon. The coupling
between fermions and the massive gauge bosons, on the other hand, is
only restricted such that the constants~$c_{CC}^{ij}$
and~$c_{NC}^{ij}$ vanish if the electromagnetic charge is not
conserved at the vertex. However, from experiment one knows that many
of these low-energy constants are very small, e.g.\ the couplings of
the massive gauge bosons to right-handed fermions or those couplings
which induce flavor-changing neutral currents or lepton-number
violation. Therefore, in analogy to the low-energy constant $\trho -
1$ in the bosonic sector, one might set these low-energy constants in
$\lag_2^f$ equal to zero and consider them only at order $p^4$ in the
effective Lagrangian. In general, however, these coupling constants
are already present at order $p^2$.

It is interesting to note that the coupling to charged and neutral
currents can readily be derived from Eq.~(\ref{gen_lageff_2_s}) by
substituting 
\begin{eqnarray}
 \bv^2  \bj_\mu^+  & \rightarrow &  \bv^2  \bj_\mu^+  +
	\sum_{ij} c_{CC}^{ij,L} j_\mu^{L, ij +} +  
        \sum_{ij} c_{CC}^{ij,R} j_\mu^{R, ij +} \, , \\
 \bv^2  \bJ_\mu^\Z & \rightarrow &  \bv^2  \bJ_\mu^\Z +
        \sum_{ij} c_{NC}^{ij,L} J_\mu^{L, ij 3} +  
        \sum_{ij} c_{NC}^{ij,R} J_\mu^{R, ij 3} 
	- \bs^2 J_\mu^{f,Q}  \, .
\end{eqnarray}
For the case of four-fermion interactions this is also true. In 
substituting 
\begin{eqnarray}
 \bv^2  \bj_\mu^+  & \rightarrow &  \bv^2  \bj_\mu^+  +
	\sum_{ij} d_{CC}^{ij,L} j_\mu^{L, ij +} +  
        \sum_{ij} d_{CC}^{ij,R} j_\mu^{R, ij +} \, , \\
 \bv^2  \bJ_\mu^\Z & \rightarrow &  \bv^2  \bJ_\mu^\Z +
        \sum_{ij} d_{NC}^{ij,L} J_\mu^{L, ij 3} +  
        \sum_{ij} d_{NC}^{ij,R} J_\mu^{R, ij 3} 
	- \bs^2 J_\mu^{f,Q} \, , 
\end{eqnarray}
all four-fermion interactions of the current-current type can be
generated from the last two terms in Eq.~(\ref{gen_lageff_2_s}). One
should note, however, that there are other four-fermion interactions,
which are not of this type and which cannot be generated in this way.
The same procedure works at order~$p^4$. Using our source terms given
in Appendix~\ref{app:sources_p4} one can generate a host of terms
involving the interaction of fermionic currents.  Again, a
considerable number of the corresponding low-energy constants is,
however, either irrelevant to the current experimental situation or is
very small. All terms involving four powers of currents and / or gauge
fields, for example, contribute to eight-fermion processes only.

One should also note, that terms of order~$p^4$ are already of
next-to-next-to-leading order if fermions are present. This is due to
the fact that fermionic fields count as order~$\sqrt{p}$. Hence, the
effective Lagrangian also contains terms of order~$p^3$, for example
\begin{equation}
	\bar \bdf_L^i i \Dslash \bdf_L^i \bar \bu_L^j \bu_L^j \ ,
\ldots \, . 
\end{equation}
This is well known from the effective Lagrangian analysis of
pion-nucleon physics~\cite{PionNucleon}.

Now we are in the position to resume the comparison of our findings
for the number of independent low-energy constants in the electroweak
chiral Lagrangian with the results found in
Refs.~\cite{Appelquist_Wu,Feruglio,D_GK}. Furthermore, we want to
clarify the role of the oblique correction parameters $S, T,$ and
$U$~\cite{STU} within our effective field theory analysis.

Obviously the analysis presented in the preceding subsection is not
affected by the presence of the fermions. One can use the equations of
motion to eliminate the same operators. The only difference is that
these equations now depend on a linear combination of external and
fermionic currents. In particular, one can again remove the low-energy
constants~$\ourlow_{11}$ and~$\ourlow_{12}$. This will renormalize the
external currents~$\bJ_\mu^\Z$ and~$\bJ_\mu^\pm$ as well as the
coupling constants~$c_{CC}^{ij}$ and~$c_{NC}^{ij}$ in Eqs.~(\ref{CC})
and (\ref{NC}) among other quantities. Hence, the complete low-energy
analysis of a strongly interacting electroweak symmetry breaking
sector does not involve the low-energy constants $\ourlow_{11}$ and
$\ourlow_{12}$, or equivalently, the low-energy constants $a_1$ and
$a_8$ in the usual basis. These constants contribute to the
self-energies of the gauge bosons which are not observable anyway.
Note, that the situation here is similar to the one described in the
purely bosonic effective field theory. The low-energy constants
$\bv^2$ and $\trho -1$ in $\lag_2$, Eq.~(\ref{gen_lageff_2_0}), are of
order $p^0$, however, there are terms in $\lag_4$ which renormalize
these low-energy constants as described in Eqs.~(\ref{v_eff_general})
and (\ref{rho_eff_general}). In the same way, removing $\ourlow_{11}$
and $\ourlow_{12}$ modifies two of the coupling
constants~$c_{CC}^{ij}$ and~$c_{NC}^{ij}$ at order $p^2$. Therefore,
it is not possible to remove two of the parameters $c_{CC}^{ij}$
and~$c_{NC}^{ij}$ instead of $\ourlow_{11}$ and $\ourlow_{12}$.  It
should be noted, however, that the reduction of the number of
operators does not affect the result for any physical quantity
evaluated by employing the effective Lagrangian.

As already mentioned in the previous subsection, the step to remove
the two low-energy constants $a_1$ and $a_8$ from the basis was not
taken in Refs.~\cite{Appelquist_Wu,Feruglio}. These authors were
interested to parametrize the electroweak symmetry breaking sector by
means of an effective chiral Lagrangian involving only the bosonic
degrees of freedom (without the usual Higgs boson). The couplings of
the fermions to the gauge bosons were assumed to have their standard
model values.  In this respect, no complete effective Lagrangian
analysis was attempted in these references. The constraint equations
then relate $\tr (\covDmat_\mu \Vmat_\mu)$ to a four-fermion term
which can be transformed further by employing the equations of motion
for the fermions.  The quantity $\tr (\covDmat_\mu \Vmat_\mu)$ is then
proportional to the square of the fermion masses which are small for
external light fermions. Only in this approximate sense the terms
$\op_{11}, \op_{12},$ and $\op_{13}$ have been removed from the basis
in Refs.~\cite{Appelquist_Wu,Feruglio}.  The application of the
equations of motion for the gauge fields, on the other hand, leads to
fermionic operators which would modify the usual couplings of the
fermions to the gauge bosons.  Therefore, no reduction of the number
of independent terms can be achieved in this framework. This interplay
of bosonic and fermionic operators when employing the equations of
motion was also noted in Ref.~\cite{D_GK}. In that paper a heavy Higgs
boson is integrated out of the standard model including the fermions.
However, no complete effective field theory analysis including the
most general couplings of the fermions to the gauge bosons was given
in that reference. Furthermore, only the constraint equations, not the
equations of motion for the gauge fields, have been used to reduce the
number of operators in the basis.

The low-energy constants $a_1$ and $a_8$ are sometimes identified with
the oblique correction parameters $S$ and $U$~\cite{STU}. What is the
relation of the above findings to the experimentally determined values
for the oblique parameters\footnote{The oblique parameter $T$ is often
identified with the low-energy constant $a_0$ which corresponds to
$\brho-1$, or, depending on the momentum counting, to the low-energy
constant $l_{0}^\prime \equiv l_{16}$ in our basis.} $S, T,$ and $U$
quoted by the particle data group~\cite{PDG_98} ?

From our point of view it is not possible to directly identify the
low-energy constants $\ourlow_{11}, \ourlow_{12},$ and $\ourlow_{16}$,
or equivalently, $a_0, a_1,$ and $a_8$ with the oblique correction
parameters $S, T,$ and $U$.  The reason is the following: the
definition of the oblique parameters by Peskin and Takeuchi~\cite{STU}
is intended to parametrize the effects of heavy new physics {\it
beyond} the standard model on the self-energies of the gauge bosons.
In particular, it is assumed that there exists an elementary Higgs
boson and that the full Lagrangian can be decomposed in the form
$\lag_{full} = \lag_{SM} + \lag_{new}$. This is also reflected by the
fact that one has always to specify a reference value for the Higgs
boson mass when quoting results for $S, T,$ and $U$. In contrast to
that, the parametrization of new physics by means of the electroweak
chiral Lagrangian assumes that the electroweak symmetry breaking is
mediated by a strongly interacting theory. This might either be the
standard model with a heavy Higgs boson or another, genuinely strongly
interacting model like technicolor where no Higgs particle exists at
all. In order to make contact between the two descriptions one could
try to mimic any strongly interacting symmetry breaking sector by
studying the large Higgs boson mass limit. Note, however, that one
cannot completely remove the Higgs particle from the theory in this
way, since for $M_H \to \infty$, the Higgs sector becomes strongly
interacting and non-perturbatively. The decoupling
theorem~\cite{decoupling_theorem} does not apply in this case.

Let us go back to Eqs.~(\ref{Lag_eff_series}) and~(\ref{lag_k_exp})
and assume that fermions are included and that redundant terms have
{\it not} yet been removed. The low-energy constants have the
following form:
\be
        l_i^{(k)} = \delta_i^{(k)} \Lambda_\epsilon + l_i^{(k),
r}(\mu) \, .
\ee
They contain a pole term~$\delta_i^{(k)} \Lambda_\epsilon$, with
$\pole \doteq (\mu^{d-4} / 16\pi^2) \left( 1 / (d-4) -
{1\over2} [ \ln(4\pi) + \Gamma'(1) + 1] \right),$ and a renormalized
low-energy constant~$l_i^{(k), r}(\mu)$. Apart from redundancy the
constants~$\delta_i^{(k)}$ are universal, i.e.\ independent of the
underlying theory. We now assume that the finite, renormalized
low-energy constants can be decomposed as follows:
\be
        l_i^{(k), r}(\mu) = l_i^{(k), SM}(\mu) + l_i^{(k), new}(\mu)
\, , 
\ee
where the first terms describe the contributions for the standard
model with a heavy Higgs boson, i.e.\ the results given below for the
bosonic sector up to order~$p^4$, and the second terms describe new
physics effects. In general, for $k \geq 4$ the contributions
$l_i^{(k), SM}(\mu)$ diverge for $M_H \to \infty$, indicating that one
enters the strongly interacting regime where the perturbative analysis
breaks down.

The definition of~$S, T,$ and $U$ given by Peskin and
Takeuchi~\cite{STU} now amounts to setting~$l_i^{(k), new}(\mu) = 0$
for all~$i$ and~$k$ except for~$k=4$ and~$i=11, 12$ and~$16$. This
introduces three finite parameters independent of each other to
describe new physics effects.  At this point the effective Lagrangian
still involves a redundant set of operators~$\ourop_i^{(k)}$ which can
be reduced by employing the equations of motion. Hence, one can again
remove the operators~$\ourop_{11}$ and~$\ourop_{12}$. In the present
situation, however, this does not reduce the number of independent
parameters. It merely moves them to some other operators. 

To close this section we note that appropriate source terms for the
fermions are given in Ref.~\cite{QED_gaugeinv}. They are
gauge-invariant and yield local equations of motion for the fermion
fields. These equations can then be used to eliminate additional terms
in the effective Lagrangian at order~$p^3$ and at order~$p^4$.  A
complete analysis including the fermions and the corresponding source
terms is, however, beyond the scope of the present work.


\section{A manifestly gauge-invariant approach to the standard model}
\label{sec:SM_gauge_inv}

\subsection{The Lagrangian and the gauge-invariant gen\-er\-at\-ing
functional} 
\label{sec:SM_lagrangian}

The standard model with a heavy Higgs boson can be described by an
effective Lagrangian as introduced in the previous section. For this
specific case, the corresponding low-energy constants can be
calculated explicitly in perturbation theory if the coupling constant
of the Higgs boson is not too large. The effective Lagrangian can be
evaluated by matching the standard model and the effective theory at
low energies. In this section we will briefly introduce our
gauge-invariant approach to the bosonic sector of the standard model,
following the discussion in Ref.~\cite{SM_gaugeinv} to which we refer
for more details. The matching calculation will be presented in
Sec.~\ref{sec:matching}.

The Lagrangian of the standard model without fermions is of the form
\be
\label{Lag_SM_before}
\lag  =
{1\over2} D_\mu\Bphi^\dagger D_\mu\Bphi -{1\over 2} m^2
\Bphi^\dagger\Bphi +{\lambda\over4} (\Bphi^\dagger\Bphi)^2 + {1\over
4\gs} W_{\mu\nu}^a W_{\mu\nu}^a + {1\over 4\gps} \B_{\mu\nu}
\B_{\mu\nu} \, ,
\ee
where $\Bphi = \left( \begin{array}{c} \phi^1 \\ \phi^2 \end{array}
\right)$ denotes the Higgs boson doublet which is coupled to the
$SU(2)_L$ gauge fields $W_\mu^a \, (a=1,2,3)$ and the $U(1)_Y$ gauge
field $B_\mu$ through the covariant derivative
\be
\label{cov_deriv}
D_\mu \Bphi = \left( \p_\mu - i {\tau^a\over 2} W_\mu^a
                          - i {1\over 2} B_\mu \right) \Bphi \, .
\ee
We have again absorbed the coupling constants $g$ and $\gp$ into the
gauge fields $W_\mu^a$ and $B_\mu$, respectively. The field strengths
are defined analogously to Eqs.~(\ref{Wmunu}) and (\ref{Bmunu}).  The
Higgs field $\Bphi$ transforms under $SU(2)_L$ gauge transformations in
the following way:
\be \label{SUtwo_trafo}
\Bphi \to \V \Bphi \, , \quad \V \in SU(2) \, , 
\ee
and under $U(1)_Y$ gauge transformations as follows:
\be \label{Uone_trafo}
\Bphi   \to  e^{- i \UoneParam / 2} \ \Bphi \, . 
\ee

For $m^2 > 0$ the classical potential has its minimum at a nonzero
value $\Bphi^\dagger \Bphi = m^2 / \lambda$ and the $SU(2)_L \times
U(1)_Y$ symmetry is spontaneously broken down to $\Uoneem$.
Accordingly, the field $\Bphi$ describes one massive mode, the Higgs
particle, and three Goldstone bosons which render the gauge fields $W$
and $Z$ massive. Finally, the spectrum contains the massless photon.
At tree level, the masses and the electric coupling constant $e$ are
given by the relations
\be \label{bare_masses_couplings}
M_H^2 = 2 m^2 \, , \, M_W^2 = {m^2 g^2 \over 4 \lambda} \, , \, M_Z^2
= {m^2 (g^2 + \gp^2) \over 4 \lambda} \, , e^2 = {g^2 \gp^2 \over g^2
+ \gp^2} \ .
\ee
We will use the same definition of the weak mixing angle as in the
effective field theory, cf.\ Eq.~(\ref{cos_theta_EFT}).

In order to have nontrivial solutions of the equations of motion, we
furthermore couple external sources to the gauge fields and the Higgs
boson. As in the preceding section we will couple sources only to
gauge-invariant operators. Again we introduce another set of fields
for the dynamical degrees of freedom which are already invariant under
the non-Abelian group $SU(2)_L$ and, in parts, under the Abelian group
$U(1)_Y$ as well. It is convenient to use a polar representation for
the Higgs doublet field
\be \label{polar}
\Bphi = {m\over \sqrt{\lambda}} R U \, , 
\ee
where the unitary field~$U$, satisfying~$U^\dagger U = 1$, describes
the three Goldstone bosons, while the radial component~$R$ represents
the Higgs boson. Furthermore, we define the $Y$-charge conjugate doublet
\be
\tBphi = i \tau_2 \Bphi^* \, . 
\ee

\noindent
We introduce the following operators:
\bea
V_\mu^1 & = & i \tBphi^\dagger D_\mu\Bphi + i \Bphi^\dagger
D_\mu\tBphi \, = {m^2\over\lambda} R^2 {\cal W}_\mu^1 \, , \nonumber \\
V_\mu^2 & = & - \tBphi^\dagger D_\mu\Bphi +  \Bphi^\dagger 
D_\mu\tBphi = {m^2\over\lambda} R^2 {\cal W}_\mu^2 \, , \nonumber \\
V_\mu^3 & = & i \tBphi^\dagger D_\mu\tBphi - i \Bphi^\dagger
D_\mu\Bphi \, = {m^2\over\lambda} R^2 {\cal Z}_\mu \, ,
\label{bcomposite}  
\eea
and
\be
V_\mu^\pm = {1\over 2} (V_\mu^1 \mp i V_\mu^2) \, ,
\ee
where the $SU(2)_L$ gauge-invariant fields~$\W_\mu^a$ and $\Z_\mu$ are
defined analogously to Eqs.~(\ref{bfieldsf_EFT})--(\ref{defZ_EFT}). Up
to a constant factor the operators~$V_\mu^i$ in Eq.~(\ref{bcomposite})
correspond to the currents of the global symmetry~$SU(2)_R$.

In terms of these composite fields the
Lagrangian from Eq.~(\ref{Lag_SM_before}) reads
\be
\label{Lag_SM_inv}
\lag_\SM^0 = {1\over 2} {m^2\over \lambda}
\left[ \p_\mu R \p_\mu R - m^2 R^2 + {m^2\over 2} R^4 + R^2
\left(\W_\mu^+ \W_\mu^- + {1\over 4} \Z_\mu \Z_\mu \right) \right]
+ {1\over 4\gs} \W_{\mu\nu}^a \W_{\mu\nu}^a
+ {1\over 4\gps} \B_{\mu\nu} \B_{\mu\nu} \, , 
\ee
where $\W_{\mu\nu}^a$ is defined similarly to Eq.~(\ref{defcalWmunu}).  

In order to calculate Green's functions from which we then can extract
physical masses, coupling constants and $S$-matrix elements, we have
to introduce external sources which emit one-particle states of the
Higgs field and the gauge bosons. In analogy to the Abelian
case~\cite{LSM} we couple sources to the $SU(2)_L \times U(1)_Y$
gauge-invariant operator $\Bphi^\dagger \Bphi$ and the field strength
$B_{\mu\nu}$. As discussed in the previous section, for the massive
gauge bosons the situation is more involved. Compensating the residual
gauge dependence of the currents $V_\mu^\pm$ under the $U(1)_Y$ gauge
transformations from Eqs.~(\ref{Uone_trafo_EFT}) and
(\ref{Uone_trafo})
\be
V_\mu^{\pm} \to e^{\mp i \UoneParam} V_\mu^{\pm} \, , 
\label{calV_trafo}
\ee
by a phase
factor~\cite{Steinmann,Horan_Lavelle_McMullan,QED_gaugeinv}, we can
write appropriate $SU(2)_L \times U(1)_Y$ gauge-invariant source terms
for all the fields as follows:
\be
\label{basic_sources}
\lag_{source}^1 =
- \half h \Bphi^\dagger \Bphi
- \half K_{\mu\nu} B_{\mu\nu}
+ J_\mu^a \phase^{ab} V_\mu^b \, , 
\ee
with external sources $h, K_{\mu\nu}$, and $J_\mu^a (a=1,2,3)$. The
phase factor in Eq.~(\ref{basic_sources}) is defined by
\be \label{phase_matrix}
\phase(x)     = \exp\left({T \int d^dy \, \G_0(x-y) \,
\partial_\mu \B_\mu(y)}\right) \, , 
\ee
with
\be \label{matrix_T}
T  =  \left(
\begin{array}{ccc}
0  & 1 & 0 \\
-1 & 0 & 0 \\
0  & 0 & 0
\end{array}
\right) \, , 
\ee
and $\G_0(x-y)$ is given by Eq.~(\ref{G_zero}). Since the vacuum in
the spontaneously broken phase corresponds to the value~$R=1$, Green's
functions of the field $\Bphi^\dagger \Bphi$ contain one-particle
poles of the Higgs boson, whereas those of $\phase^{ab} V_\mu^b$ have
one-particle poles of the gauge bosons $W$ and $Z$.

In Ref.~\cite{QED_gaugeinv} it was shown to all orders in perturbation
theory that a phase factor $\phase$ which is defined analogously to
Eq.~(\ref{phase_matrix}) does not spoil the renormalizability of QED.
Since the proof did not rely on any particular feature of QED, the
same should be true for the present case as well. This is due to the
fact that the phase factor only contains the Abelian gauge degree of
freedom which does not affect the dynamics of the theory. Since the
operator $\Bphi^\dagger \Bphi$ and the currents $V_\mu^a$ from
Eq.~(\ref{bcomposite}) have dimension less than four, source terms
involving these operators do not spoil the renormalizability
either. The reader should note, however, that we do not have a formal
proof of renormalizability to all orders in perturbation theory for
the present case. As was shown in Ref.~\cite{SM_gaugeinv}, at the
one-loop level everything works fine and on physical grounds we expect
this to happen at all orders. 

Green's functions of the operators in Eq.~(\ref{basic_sources}) are,
however, more singular at short distances than (gauge-dependent)
Green's functions of the fields $\Bphi, W_\mu^a,$ and $B_\mu$
themselves.  Time ordering of these operators gives rise to
ambiguities, and the corresponding Green's functions are only unique
up to contact terms. In order to make the theory finite, these contact
terms of dimension four need to be added to the Lagrangian which is
then given by
\be \label{lagSMfull}
\lag_\SM = \lag_\SM^0 + \widehat\lag_{source}^1 + \lag_{source}^2 \, .
\ee
The first term in Eq.~(\ref{lagSMfull}) is defined in
Eq.~(\ref{Lag_SM_inv}). The second term is given by
\be
\label{lag_sources_1_hat}
\widehat\lag_{source}^1 =
- \half \widehat h \Bphi^\dagger \Bphi
- \half \widehat K_{\mu\nu} B_{\mu\nu}
+ J_\mu^a \phase^{ab} V_\mu^b \, , 
\ee
where
\bea
\widehat h  & = & h + 4 v_{jj} J_{\mu}^+ J_{\mu}^-
        + c_{jj} J_{\mu}^\Z J_{\mu}^\Z + 4 J_\mu^a J_\mu^a \, , \\
\widehat K_{\mu\nu} & = & K_{\mu\nu} + c_{Bj} \jmunut
        - 2 i c_{Bjj} (J_\mu^+ J_\nu^- - J_\mu^- J_\nu^+) \, .
\eea
The last term in Eq.~(\ref{lagSMfull}) is defined by
\bea
\lag_{source}^2 & = & 
\mbox{}- v_{djj} J_{\nu}^\Z [ i (d_\mu j_{\nu}^+ - d_\nu j_{\mu}^+)
j_\mu^- - i (d_\mu j_{\nu}^- - d_\nu j_{\mu}^-) j_\mu^+]
+v_{dj} (d_\mu j_{\nu}^+ - d_\nu j_{\mu}^+)  (d_\mu
j_{\nu}^- - d_\nu j_{\mu}^-) \nonumber \\
&&\mbox{}- {i\over 2} c_{djj} \jmunut (J_\mu^+ J_\nu^- - J_\mu^-
J_\nu^+)  
+{1\over4} c_{dj} \jmunut \jmunut \nonumber\\
&&\mbox{}+ 16 v_{JJ2} (J_{\mu}^+ J_{\mu}^-)^2
+ 4 v_{JJJJ} (J_{\mu}^+ J_{\nu}^- + J_\mu^- J_\nu^+)^2
+ c_{JJ2} (J_{\mu}^\Z J_{\mu}^\Z )^2 \nonumber\\
&&\mbox{}+ 4 v_{J2ZZ} J_{\mu}^+ J_{\mu}^- J_{\nu}^\Z J_{\nu}^\Z
+ 2 v_{JJZZ} (J_{\mu}^+ J_{\nu}^- + J_\mu^- J_\nu^+) J_{\mu}^\Z
J_{\nu}^\Z  \nonumber\\
&&\mbox{}+ c_{hh} h^2 + c_{mh} m^2 h 
+ 4 c_{hJJ} h  J_{\mu}^+ J_{\mu}^-
+ 4 c_{mJJ} m^2  J_{\mu}^+ J_{\mu}^- 
+ c_{hZZ} h J_{\mu}^\Z J_{\mu}^\Z
+ c_{mZZ} m^2 J_{\mu}^\Z J_{\mu}^\Z \, , \label{lagsourcetwo}
\eea
where we introduced the quantities
\be
J_\mu^\pm  = {1\over 2} (J_\mu^1 \mp i J_\mu^2) \ , \quad
J_\mu^\Z        \equiv  J_\mu^3 \, . 
\ee
The quantities $d_\mu j_\nu^\pm$ and $j_\mu^\pm$ are defined
analogously to Eqs.~(\ref{d_Wpm_EFT}) and (\ref{def_jpm}). 
The contact terms in $\lag_{source}^2$ will not contribute to any
physical $S$-matrix elements.

For later use we introduce the quantities
\bea
\u V_\mu^a  & = & \phase^{ab} V_\mu^b \, ,  \label{def_uV} \\
\Y_\mu^\pm  & = & \W_\mu^\pm + 4 j_\mu^\pm \ , \
\Y_\mu^\Z     =  \Z_\mu + 4 J_\mu^\Z  \ . \label{def_Y}
\eea

The generating functional $W_\SM[h,K_{\mu\nu},J_\mu^a]$ for the
gauge-invariant Green's functions is defined by the path integral 
\begin{equation} \label{genfunc_pathint}
   e^{-W_\SM [h,K_{\mu\nu},J_\mu^a]} = \int \d\mu[\Bphi,W_\mu^a,B_\mu]
        e^{-\intdx \lag_\SM} \ .
\end{equation}
Note that we still integrate over the original fields $\Bphi,
W_\mu^a,$ and $B_\mu$ in Eq. (\ref{genfunc_pathint}). Furthermore, we
have absorbed an appropriate normalization factor into the measure
$\d\mu[\Bphi,W_\mu^a,B_\mu]$. Derivatives of this functional with
respect to the field $h$ generate Green's functions of the scalar
density $\Bphi^\dagger\Bphi$, derivatives with respect to the source
$K_{\mu\nu}$ generate Green's functions of the field strength
$B_{\mu\nu}$, while derivatives with respect to $J_\mu^a$ generate
Green's functions for the currents $\u V_\mu^a$.

In the spontaneously broken phase, these Green's functions have
one-particle poles from the Higgs boson as well as the gauge bosons.
Thus, one can extract $S$-matrix elements for the physical degrees of
freedom from the generating functional in Eq.~(\ref{genfunc_pathint}).
Due to the equivalence theorem~\cite{equivalence_theorem} these
$S$-matrix elements will be identical to the ones obtained from those
Green's functions which are used in the usually employed formalism.
The presence of the contact terms in $\lag_{source}^2$ in
Eq.~(\ref{lagsourcetwo}) reflects the fact that the off-shell
continuation of the $S$-matrix is not unambiguously defined.  Note
that this is a general feature of any field theory and not particular
to those involving a gauged symmetry.  The continuation we choose has
the virtue of being gauge-invariant.

As was pointed out in
Refs.~\cite{Abelian_Higgs,QED_gaugeinv,SM_gaugeinv} it is possible to
evaluate the path integral in Eq.~(\ref{genfunc_pathint}) without the
need to fix a gauge as will be shown below.

\subsection{Tree level}
\label{sec:tree_level}

At tree level, the generating functional for the bosonic sector of the
standard model is given by
\begin{equation} \label{genfunc_tree}
        W_\SM[h,K_{\mu\nu},J_\mu^a]
                = \intdx \, \lag_\SM(R^{cl}, \W_\mu^{cl,\pm},
                \Z_\mu^{cl}, \A_\mu^{cl}) \ ,
\end{equation}
where $R^{cl}, \W_\mu^{cl,\pm}, \Z_\mu^{cl},$ and $\A_\mu^{cl}$
are determined by the equations of motion
\bea
- \Box R & = & - \left[ m^2 (R^2 - 1) + \Y_\mu^+ \Y_\mu^- +
{1\over4} \Y_\mu^\Z \Y_\mu^\Z - \widehat h  \right] R \, ,
\label{eomRcomp} \\ 
- d_\mu \W^\pm_{\mu\nu}
& = & - M_W^2 R^2 \Y^\pm_\nu
\pm i ( \Z_{\mu\nu} + \B_{\mu\nu}) \W^\pm_\mu \mp i
\W^\pm_{\mu\nu}\Z_\mu 
\mp i (\p_\mu \Z_\mu) \W^\pm_\nu \pm i (\p_\mu \Z_\nu) 
\W^\pm_\mu \nonumber \\ 
&&\pm i \Z_\nu d_\mu \W^\pm_\mu \mp i \Z_\mu d_\mu \W^\pm_\nu
- (\Z_\mu \Z_\mu) \W^\pm_\nu + (\Z_\mu \Z_\nu) \W^\pm_\mu
\pm 2 \W^\pm_\mu (\W_\mu^+ \W_\nu^- - \W_\nu^+ \W_\mu^- ) \, , 
\label{eomWcomppm} \\
-\p_\mu\Z_{\mu\nu} & = & \PT_{\nu\mu}\left(-M_Z^2 R^2 \Y_\mu^\Z +
\DeltaEoM_\mu\right) + {e^2\over c^2} \p_\mu \widehat K_{\mu\nu}
+ {e^2\over c^2} \PT_{\nu\mu} S_\mu \, , \label{eomZcomp} \\
-\p_\mu\A_{\mu\nu} &=& s^2 \PT_{\nu\mu}\DeltaEoM_\mu - e^2 \p_\mu
\widehat K_{\mu\nu} - e^2 \PT_{\nu\mu} S_\mu \label{eomAcomp} \ .
\eea
Furthermore, the equations for the Goldstone boson field $U$
correspond to
\bea
d_\mu\yv^\pm_\mu &=& - 2 {\p_\mu R\over R} \yv^\pm_\mu \pm i \Z_\mu
\yv^\pm_\mu \mp i \y_\mu\W^\pm_\mu \, , \label{eomUcomppm} \\
\p_\mu\y_\mu &=& - 2 {\p_\mu R\over R} \y_\mu - 8 i (
j^+_\mu\W^-_\mu - j_\mu^-\W^+_\mu) \ .
\label{eomUcompZ}
\eea
In order to simplify the notation we have omitted the prescription
``cl'' in the equations above. In
Eqs.~(\ref{eomRcomp})--(\ref{eomUcompZ}) we have introduced the
quantities 
\bea
\A_{\mu\nu} & = & \p_\mu \A_\nu - \p_\nu \A_\mu \, , \\
\DeltaEoM_\mu &=&  2 \Z_\rho ( \W_\rho^+ \W_\mu^- + \W_\mu^+
\W_\rho^- ) - 4 \Z_\mu \W_\rho^+ \W_\rho^-  
+ 2 i ( \W^+_{\rho\mu} \W^-_\rho  -
\W^-_{\rho\mu} \W^+_\rho ) \nonumber \\
& &\mbox{} - 2 i ( d_\rho\W_\rho^+\W_\mu^-  -
d_\rho\W_\rho^-\W_\mu^+
- d_\rho\W_\mu^+\W_\rho^- + d_\rho\W_\mu^-\W_\rho^+)
\, , \label{DeltaEoM} \\
S_\mu   & = & - v_{djj} J^\Z_{\rho} (J_{\rho}^+ J_{\mu}^- + J_\rho^-
        J_\mu^+) + 2 v_{djj} J^\Z_{\mu} J_{\rho}^+ J_{\rho}^-
        \nonumber \\
        &&\mbox{}- 2 v_{dj} [i (d_\rho j_{\mu}^+ - d_\mu j_{\rho}^+)
        j_\rho^- - i (d_\rho j_{\mu}^- - d_\mu j_{\rho}^-) j_\rho^+]
        \, . 
\eea
The projector $\PT_{\mu\nu}$ has been defined in Eq.~(\ref{def_PT}).
The quantities $\W_{\mu\nu}^\pm$ and $\Z_{\mu\nu}$ are defined
analogously to Eqs.~(\ref{def_caldWmunu}) and (\ref{def_Zmunu}). The
covariant derivatives in $d_\mu \W_\mu^\pm, d_\mu j_\nu^\pm, d_\mu
\Y_\nu^\pm,$ and $d_\mu \W_{\mu\nu}^\pm$ are defined in the same way
as in Eq.~(\ref{d_Wpm_EFT}).

The equations of motion (\ref{eomRcomp})--(\ref{eomUcompZ}) have
similar properties as those in the effective field theory, see the
discussion after Eq.~(\ref{def_PT}) above. We only note here that the
radial variable $R$ which is related to the massive Higgs boson is
determined by Eq.~(\ref{eomRcomp}). Solutions for the massive gauge
boson fields $\phase^\mp \W_\mu^\pm$ and $\Z_\mu$ follow from
Eqs.~(\ref{eomWcomppm}) and (\ref{eomZcomp}).  Finally,
Eq.~(\ref{eomAcomp}) determines the transverse component of the
massless photon field $\A_\mu^T = \PT_{\mu\nu} \A_\nu$.  The solutions
of the equations of motion for the physical degrees of freedom in
powers of the external sources can be found in
Ref.~\cite{SM_gaugeinv}.

\subsection{One-loop level}
\label{sec:one_loop}

The one-loop contribution to the generating functional can be
evaluated with the saddle-point method.  If we write the fluctuations
$\fluct$ around the classical fields $\Allfields^{cl}$ as $\Allfields =
\cAllfields + \fluct$, we obtain the following representation for the
one-loop approximation to the generating functional:
\begin{equation} \label{genfunc_saddlepoint}
  e^{-W_\SM[h,K_{\mu\nu},J_\mu^a]} = e^{-\intdx\lag_\SM^{cl}}
     \int\d\mu[\fluct] e^{- (1/2) \intdx \fluct^T \tildeD \fluct} \ .
\end{equation}
Gauge invariance implies that the operator $\tildeD$ has zero
eigenvalues corresponding to fluctuations $\fluct$ which are
equivalent to infinitesimal gauge transformations. Treating these
zero modes appropriately~\cite{Abelian_Higgs,SM_gaugeinv}, see also
Sec.~\ref{subsec:EFT_W_4} above, one can evaluate the path-integral
representation for the generating functional at the one-loop level
without the need to fix a gauge and without introducing ghost
fields. Up to an irrelevant infinite constant one obtains the
following result for the one-loop generating functional from
Eq.~(\ref{genfunc_saddlepoint}):
\begin{equation} \label{1loopgf}
  W_\SM[h,K_{\mu\nu},J_\mu^a] = \intdx \lag_\SM
  + \half {\ln\det}^\prime\tildeD
  -\half \ln\det P^T P \ .
\end{equation}
The first term on the right-hand side represents the classical action
which describes the tree-level contributions to the generating
functional. In the second term, the determinant ${\det}^\prime\tildeD$
is defined as the product of all non-zero eigenvalues of the operator
$\tildeD$. The last term originates from the path integral measure.
The sum of the last two terms in Eq.~(\ref{1loopgf}) corresponds to
the one-loop contributions to the generating functional. The operator
$P$ satisfies the relation $ P^T \tildeD  =  \tildeD P  = 0$.

For the explicit evaluation of the one-loop contributions to the
generating functional in Eq.~(\ref{genfunc_saddlepoint}) it is very
important to choose an appropriate parametrization of the physical
modes and their quantum fluctuations. Otherwise the expression for the
differential operator becomes too complicated. We introduce
fluctuations $f,\eta^a,\w_\mu^a,$ and $\b_\mu$ around the Higgs field
$R$, the Goldstone boson field $U$, the three $SU(2)_L$ gauge fields
$W_\mu^a$ and the $U(1)_Y$ gauge field $B_\mu$, respectively.
Furthermore, we collect the fluctuations of the gauge fields in a
vector $\tq_\mu^A \doteq \left( \w_\mu^a, \b_\mu \right)$.  Following
the steps described in Ref~\cite{SM_gaugeinv}, the generating functional
at the one-loop level can then be written in the form
\begin{equation} \label{1loopgf_1}
        W_\SM[h,K_{\mu\nu},J_\mu^a] = \intdx \lag_\SM
           + \half {\ln\det}\left(\Dfull \right) - \ln\det P^T P \, , 
\end{equation}
where the solutions of the equations of motion
(\ref{eomRcomp})--(\ref{eomAcomp}) have to be
inserted. Eq.~(\ref{1loopgf_1}) represents all tree-level and one-loop
contributions of the bosonic sector of the standard model.  Note that
in order to obtain Eq.~(\ref{1loopgf_1}) we have used the identity
\begin{equation} \label{detprimeD}
  \ln{\det}^\prime \tildeD = \ln\det\left(\tildeD + P P^T +
  \delta_P\right) - \ln\det(P^T P) \ ,
\end{equation}
to rewrite the determinant ${\det}^\prime\tildeD$, i.e.\ the product
of all non-zero eigenvalues of the differential operator $\tildeD$,
which appears in Eq.~(\ref{1loopgf}). Equation~(\ref{detprimeD}),
which is valid up to an irrelevant infinite constant, follows from the
fact that zero and non-zero eigenvectors are orthogonal to each other.

The explicit expressions for the components of the differential
operator $\Dfull$, which we parametrize by
\be
\Dfull \doteq  \left( \begin{array}{ccc}
                        d       & \delta        & \delta_\nu  \\
                    \delta^T    & D             & \Delta_\nu  \\
                   \delta_\mu^T & \Delta_\mu^T  & D_{\mu\nu}
                \end{array} \right) , \label{defD} \\
\ee
can be found in Eqs.~(\ref{firstcomp_defD})--(\ref{lastcomp_defD})
in Appendix~\ref{app:diffop_HH}. The operators $P P^T$, $P^T P$, and
$\delta_P$ are listed in Eqs.~(\ref{def_PPT})--(\ref{def_deltaP}).
The $3\times 3$-matrix of the differential operator $\Dfull$ from
Eq.~(\ref{defD}) is acting on the 3-dimensional space of fluctuations
$y = \left( f, \eta^a, q_\mu^A \right)$.

We would like to stress an important point here. At the classical
level only physical modes propagate. The classical Goldstone boson
field $U^{cl}$ represents the $SU(2)_L$ gauge degrees of freedom. At
the quantum level, however, the situation is different. Quantum
fluctuations around the classical field $U^{cl}$, denoted by $\eta^a$,
imply virtual Goldstone boson modes propagating within loops. Note
that these modes are absent in any gauge-dependent approach based on
the unitary gauge.  They are, however, necessary in order to ensure a
decent high-energy behavior of the theory.

In order to separate the heavy Higgs boson mode from the light modes
of the Goldstone and the gauge bosons it is useful to diagonalize the
differential operator $\Dfull$. First, we introduce some additional
quantities
\bea
\varD_{\mu\nu} & = & D_{\mu\nu} - \delta_\mu^T d^{-1} \delta_\nu
               - \vartheta_\mu^T \Theta^{-1} \vartheta_\nu \, , 
\label{varD} \\
\Theta         & = & D - \delta^T d^{-1} \delta \, , \\
\vartheta_\nu  & = & \Delta_\nu - \delta^T d^{-1} \delta_\nu \ .
\label{vartheta}
\eea
Using the identity
\be \label{trafowithT}
\Trafo^T \left( \Dfull \right) \Trafo =
\mathrm{diag} \left( d, \, \Theta, \, \varD_{\mu\nu} \right) \, , 
\ee
where
\be
\Trafo         =  \left( \begin{array}{ccc}
  1 & - d^{-1} \delta  & - d^{-1} \delta_\nu + d^{-1} \delta
                          \Theta^{-1} \vartheta_\nu   \\
  0 & 1  & - \Theta^{-1} \vartheta_\nu \\
  0 & 0  & \delta_{\mu\nu}
\end{array} \right) , 
\ee
and the fact that the transformation matrix $\Trafo$ has unit
determinant, one obtains the following result for the
generating functional:
\be \label{1loopgf_2}
      W_\SM[h,K_{\mu\nu},J_\mu^a] = \intdx \lag_\SM
           + \half \ln\det d + \half \ln\det \Theta + \half \ln\det
\varD - \ln\det P^T P \ . 
\ee

Equation~(\ref{1loopgf_1}) and the equivalent form in Eq.~(\ref{1loopgf_2})
represent our result for the generating
functional $W_{\SM}[h,K_{\mu\nu}, J_\mu^a]$ for the gauge-invariant
Green's functions for the bosonic sector of the standard model. These
formulae encode the full tree-level and one-loop effects of the
theory. If one expands the generating functional up to a given order
in powers of the external sources one can extract any $n$-point
Green's functions for the gauge-invariant operators $\Bphi^\dagger
\Bphi, B_{\mu\nu},$ and $\u V_\mu^a$.

As noted before, the generating functional $W_{\SM}[h,K_{\mu\nu},
J_\mu^a]$ from Eq.~(\ref{1loopgf_1}) or Eq.~(\ref{1loopgf_2}) can be
renormalized by an appropriate choice of renormalization prescriptions
for the fields, the mass parameter $m^2$, the coupling constants, and
the sources. The full list can be found in Appendix~B of
Ref.~\cite{SM_gaugeinv}. The relations between bare and renormalized
fields, masses and coupling constants which will be needed in
Sec.~\ref{sec:renorm} are given by
\bea
W_\mu^a & = & W_\mu^{a,r} \, , \label{W_mu_ren} \\
B_\mu   & = & B_\mu^r \, , \label{B_mu_ren} \\
\phi    & = & Z_\phi^{1/2} \phi_r \, , \\
Z_\phi  & = & 1 - (6 \grs + 2 \gprs) [\polemr + \delta z] \, , 
\label{Zphi_ren} \\
m^2     & = & \mrs \left[ 1 - {1\over2} (24\lr + 3\grs + \gprs) 
        [\polemr + \delta m^2] - (Z_\phi - 1) \right] , \label{m_ren}
\\
\lambda & = & \lr \Bigg[ 1 - \left( 24 \lr + 3 \grs + \gprs +
        {3\over8} {(\grs + \gprs)^2 + 2 \gr^4 \over\lr} \right)
        [\polemr + \delta \lambda] - 2 (Z_\phi - 1) \Bigg] , \\
g^2     & = & \grs \left[ 1 + {43\over3} \grs
        [\polemr + \delta g^2] \right] , \\
\gp^2   & = & \gprs \left[ 1 - {1\over3} \gprs [\polemr + \delta \gps]
        \right] , \label{gprime_ren}
\eea
where we denoted the pole term by
\be
\polemr   \doteq {\mu^{d-4} \over 16\pi^2} \left( {1 \over d-4} -
{1\over2} [ \ln(4\pi) + \Gamma'(1) + 1] \right)
+ {1\over 32\pi^2} \ln \left( {2\mrs \over \mu^2} \right) \ .
\ee
The finite renormalization constants $\delta m^2, \ldots , \delta
\gps$ which appear in the Eqs.~(\ref{m_ren})--(\ref{gprime_ren}) are
determined by the renormalization scheme, cf.\ 
Ref.~\cite{SM_gaugeinv}.

With the renormalization conditions from
Eqs.~(\ref{W_mu_ren})--(\ref{gprime_ren}) and the corresponding
relations for the sources~\cite{SM_gaugeinv}, the generating
functional for the standard model, $W_{\SM}[h,K_{\mu\nu},J_\mu^a]$,
can be renormalized at the one-loop level. In this way we have
completely defined our theory at the one-loop level. The
expression~(\ref{1loopgf_2}) for the generating functional will be
used as the starting point of the matching calculation for the case of
the standard model with a heavy Higgs boson, which will be discussed
in the next section.


\section{Matching}
\label{sec:matching}

\subsection{Evaluating the matching relation for the case of a heavy
  Higgs boson}

The effective Lagrangian for the case of a heavy Higgs boson is
determined by requiring that both the full and the effective theory
yield the same Green's functions in the low-energy region, i.e.\ by
the matching relation:
\begin{equation}  \label{match1}
   W_{eff}[\bh, \bK_{\mu\nu}, \bJ_\mu^a] = W_{\SM}[h, K_{\mu\nu},
   J_\mu^a] \, .
\end{equation}
Note that Eq.~(\ref{match1}) should not be understood as an identity
but rather as an asymptotic equality in the low-energy region. See
Refs.~\cite{LSM,Abelian_Higgs} for a more detailed discussion of this
point. Furthermore, we note that in the standard model we have
introduced a source $h$ coupled to the scalar density
$\Bphi^\dagger\Bphi$, cf.\ Eq.~(\ref{basic_sources}). Therefore, in
this specific case the effective Lagrangian will also contain terms
involving a source $\bh$, cf.\ Ref.~\cite{LSM}. As mentioned before,
we will consider only Green's functions of gauge-invariant operators in
the matching relation~(\ref{match1}). At low energies, these Green's
functions have non-local contributions involving only the vector
bosons, which are the light particles in the theory. These
contributions drop out of the matching relation. The remaining
contributions involve the propagator of the heavy Higgs boson and
allow a systematic low-energy expansion.  In order to evaluate this
expansion one has to understand the counting of loops in the full
theory and of the low-energy expansion in the effective theory, cf.\
Ref.~\cite{Abelian_Higgs}.

The loop expansion in the full theory generates a power series in the
coupling constants~$\lambda, \gs,$ and $\gps$, while the low-energy
expansion produces powers of the momenta and the gauge boson
masses~$M_W$ and $M_Z$. It is, however, not possible to treat these
six quantities as independent of each other, since the gauge boson
masses depend on the coupling constants through the
relations~(\ref{bare_masses_couplings}). These expressions also
indicate that it will not be very transparent to count mass factors in
terms of the quantities~$\lambda, \gs,$ and~$\gps$.  The loop
expansion in the full theory generates positive powers of the
coupling~$\lambda$, while the low-energy expansion produces negative
powers thereof. It is possible, however, to discard the coupling
constants~$\g$ and $\gp$ from the counting scheme.  This is a
consequence of the definition of the vector fields~$W_\mu^a$ and
$B_\mu$ in Eq.~(\ref{cov_deriv}), which are scaled such that the
coupling constants do not explicitly occur in the covariant
derivative. As a result, these coupling constants naturally enter all
loop corrections only through the gauge boson masses~$M_W$ and $M_Z$
as well as through the weak mixing angle $\sin\theta_W$.  Regarding
the one-loop contributions to the generating functional, this can
readily be inferred from the results for the differential operators
listed in Appendix~\ref{app:diffop_HH}.  With this bookkeeping powers
of~$\llambda$ count the number of loops in the full theory.

In order to evaluate the low-energy expansion at a given loop-level,
we treat the covariant derivative~$D_\mu$, the gauge boson masses
$M_W$ and $M_Z$, the momenta and the external source $J_\mu^a$ as in
the effective theory, i.e.~as quantities of order~$p$. The external
source~$h$ is of order~$p^2$, while the scalar field $\Bphi$, the mass
parameter $m$, the coupling constant $\lambda,$ and the external
source~$K_{\mu\nu}$ are quantities of order $p^0$.

If the coupling constant $\lambda$ of the Higgs field is not too strong,
the low-energy constants $\low_i$ from Eq.~(\ref{lag_k_exp}) admit an
expansion in powers of the parameter~$\lambda$,
\begin{equation}  \label{lowconstexp}
  \low_i = {1 \over \lambda} \low_i^{tree} + \low_i^{1-loop} + \lambda
  \low_i^{2-loop} + \cdots \, ,
\end{equation}
corresponding to the loop expansion in the full theory. In this case
the accuracy of the effective field theory description is controlled
by the order of both the momentum and the coupling constant~$\lambda$.
For values of~$\lambda$ close to the strong coupling region, one may
consider higher orders in the expansion~(\ref{lowconstexp}). Large
values of the momentum or the gauge boson masses may require including
higher orders in Eq.~(\ref{Lag_eff_series}).  In the following, we will
determine the effective Lagrangian up to order~$p^4$, and the
low-energy constants up to order~$\llambda^0$, i.e.  at the one-loop
level.

In order to evaluate the low-energy constants, one can calculate the
generating functional in both the full and the effective theory, and
solve the matching relation~(\ref{match1}).  It turns out, however,
that the evaluation of the one-loop contributions to the generating
functional in the effective theory for the case of a general
coefficient $\trho \neq 1$ in $\lag_2^0$ in Eq.~(\ref{gen_lageff_2_0})
is quite involved.  Therefore, we proceed in a similar way as in the
Abelian case~\cite{Abelian_Higgs} and make use of the fact that powers
of the constant~$\lambda$ count the number of loops in the full
theory.  At leading order in $\lambda$, i.e.\ $\lambda^{-1}$, we get
contributions to the parameters~$\low_i^{tree}$ in the terms $\lag_2$
and $\lag_4$. Only the parameters $\low_i^{tree}$ in $\lag_2$ will,
however, be relevant to evaluate the one-loop contribution to the
generating functional of the effective theory up to order~$\lambda^0$.

The leading contributions in $\lambda$ to the effective Lagrangian can
be read off from the low-energy expansion of the classical action of
the full theory, i.e., from
\be \label{clactR}
 \intdx \lag_\SM = \intdx \Bigg(-{m^4\over 4\lambda} R^4
                + {1\over4\gs} \W_{\mu\nu}^a \W_{\mu\nu}^a
                + {1\over4\gps} B_{\mu\nu} B_{\mu\nu}  
                - \half \widehat K_{\mu\nu} B_{\mu\nu} +
\lag_{source}^2 \Bigg) \, . 
\ee
The Lagrangian $\lag_{source}^2$ was defined in
Eq.~(\ref{lagsourcetwo}).  For slowly varying external fields, the
behavior of the massive mode~$R$ is under control and the equation of
motion~(\ref{eomRcomp}) can be solved algebraically. The result is a
series of local terms with increasing order in~$p^2$:
\bea
R   & = & 1 + r_2 + r_4 + \cdots \, , \quad r_n = \order(p^n) \, , \\
r_2 & = & {1\over 2m^2} \left(-{1\over 4} \Y_\mu^a \Y_\mu^a + \widehat h
\right) \, , \\
r_4 & = & - {1\over 8m^4} \left(-{1\over 4} \Y_\mu^a \Y_\mu^a + \widehat h
\right)^2 + {1\over 2m^2} \Box r_2  \, .
\eea
Inserting the solution for $R$ into the classical action
Eq.~(\ref{clactR}) we obtain the following tree level contributions to
the effective Lagrangian up to order $p^4$:
\bea
\lag_2^{tree} & = & - {m^2\over 2 \lambda}  \left(-{1\over 4} \Y_\mu^a
\Y_\mu^a + \widehat h \right)
+ {1\over4\gs} \W_{\mu\nu}^a \W_{\mu\nu}^a
+ {1\over4\gps} B_{\mu\nu} B_{\mu\nu} 
- \half K_{\mu\nu} B_{\mu\nu} \nonumber \\
&&\mbox{}+ c_{mh} m^2 h
+ 4 c_{mJJ} m^2  J_{\mu}^+ J_{\mu}^-
+ c_{mZZ} m^2 J_{\mu}^\Z J_{\mu}^\Z \, , \label{lag2_tree} \\
\lag_4^{tree} & = & - {1 \over 4 \lambda} \left(-{1\over 4} \Y_\mu^a
\Y_\mu^a + \widehat h \right)^2 
- \half B_{\mu\nu} \left[ c_{Bj} \jmunut - 2 i c_{Bjj} (J_\mu^+
J_\nu^- - J_\mu^- J_\nu^+) \right] \nonumber \\
& &\mbox{}- v_{djj} J_{\nu}^\Z [ i (d_\mu j_{\nu}^+ - d_\nu j_{\mu}^+)
j_\mu^- - i (d_\mu j_{\nu}^- - d_\nu j_{\mu}^-) j_\mu^+]  
+v_{dj} (d_\mu j_{\nu}^+ - d_\nu j_{\mu}^+)  (d_\mu
j_{\nu}^- - d_\nu j_{\mu}^-) \nonumber \\
&&\mbox{}- {i\over 2} c_{djj} \jmunut (J_\mu^+ J_\nu^- - J_\mu^-
J_\nu^+) +{1\over4} c_{dj} \jmunut \jmunut \nonumber\\
&&\mbox{}+ 16 v_{JJ2} (J_{\mu}^+ J_{\mu}^-)^2
+ 4 v_{JJJJ} (J_{\mu}^+ J_{\nu}^- + J_\mu^- J_\nu^+)^2
+ c_{JJ2} (J_{\mu}^\Z J_{\mu}^\Z )^2 
+ 4 v_{J2ZZ} J_{\mu}^+ J_{\mu}^- J_{\nu}^\Z J_{\nu}^\Z
\nonumber \\ 
&&\mbox{}+ 2 v_{JJZZ} (J_{\mu}^+ J_{\nu}^- + J_\mu^- J_\nu^+) J_{\mu}^\Z
J_{\nu}^\Z  
+ c_{hh} h^2 + 4 c_{hJJ} h  J_{\mu}^+ J_{\mu}^-
+ c_{hZZ} h J_{\mu}^\Z J_{\mu}^\Z \, . \label{lag4_tree}
\eea
Hence, at leading order in $\lambda$ the parameters and low-energy
constants in $\lag_2$ are given by
\be \label{vsquare_rho_tree}
\bv^2 = {m^2 \over \lambda} \ , \ \trho = 1 
\ , \ \bg = g \ , \ \bg^\prime = g^\prime \ , 
\ee
and
\bea
\bJ_\mu^\pm & = & J_\mu^\pm \ , \quad \bJ_\mu^\Z = J_\mu^\Z \
, \quad \bK_{\mu\nu} = K_{\mu\nu} \ , \quad \bh = h \, , \nonumber \\
\bar c_h & = & - {1\over 2} + c_{mh} \lambda  \ , \quad
\bar c_\W = - {1\over 2} v_{jj} + c_{mJJ} \lambda  \ , \quad 
\bar c_\Z = - {1\over 2} c_{jj} + c_{mZZ} \lambda  \, , 
\eea 
where $\bar c_h$ denotes the coefficient of $\bv^2 \bh$ in $\lag_2$. 
Since there are no custodial symmetry breaking effects in the standard
model at tree level we get $\trho = 1$.  Note that the matching
condition~(\ref{match1}) determines the low-energy constants and
the sources in the effective theory.

From $\lag_4^{tree}$ in Eq.~(\ref{lag4_tree}) we obtain the following
tree-level contributions to the low-energy constants $\ourlow_i$ in
$\lag_4^0$ in Eq.~(\ref{lag_4_0}):
\be
\ourlow_{1}^{tree} = - {1\over 4\lambda}  \, , \quad
\ourlow_{3}^{tree} = - {1\over 8\lambda}  \, , \quad
\ourlow_{5}^{tree} = - {1\over 64\lambda} \, .
\ee
All other low-energy constants $\ourlow_i$ in $\lag_4^0$ vanish at
tree level. From Eq.~(\ref{lag4_tree}) we can also read off the
tree-level contributions to the low-energy constants of the source
terms at order $p^4$. Only some of the 76~terms which appear in the
general expression $\lag_4^s$ in Eq.~(\ref{lag_4_s}) are non-zero
at tree level for the present case. It will not be necessary later on
to list these contributions here explicitly.

Now one can evaluate the one-loop contribution to the generating
functional in the effective theory using the technique described in
Sec.~\ref{subsec:EFT_W_4}. At order~$\llambda^0$, the matching
relation~(\ref{match1}) is of the form [cf.\ Eq.~(\ref{1loopgf_2})]:
\begin{eqnarray} 
  \lefteqn{ \intdx (\lag_2 + \lag_4) +  \half \ln\det \bD
+  \half \ln\det \bvarD - \ln\det \bP^T \bP }
\nonumber \\
& = &   \intdx \lag_\SM + \half \ln\det d
    + \half \ln\det \Theta +  \half \ln\det \varD - \ln\det P^T P \,
. \label{match2} 
\end{eqnarray}
The first terms on both sides of Eq.~(\ref{match2}) represent the tree
level contributions in the effective and full theory, respectively.
The differential operators on the right-hand side, describing the
one-loop contributions in the full theory, are defined in
Eqs.~(\ref{firstcomp_defD})--(\ref{lastcomp_defD}),
(\ref{varD})--(\ref{vartheta}), and (\ref{def_PTP}).  The differential
operators on the left-hand side, indicated with a bar, represent the
one-loop contributions in the effective theory. Using the iterative
matching procedure described above, these differential operators can
be inferred from the corresponding operators in the full theory by
taking the limit $R \to 1$ and by disregarding all operators which
involve the fluctuations $f$ for the radial component~$R$ of the Higgs
field.  Furthermore, we make the identifications $\bv^2 = m^2 /
\lambda$ and $\trho = 1$, cf.\ Eq.~(\ref{vsquare_rho_tree}).

Note that the quantities on the left-hand side of the matching
relation~(\ref{match2}) involve the solutions of the equations of
motion in the effective theory, while those on the right-hand side
depend on the solutions of the equations of motion in the full theory.
At the stationary point, however, the corresponding corrections are of
second order in the shift of the fields and beyond the present
accuracy.  Thus, our notation will not distinguish between the two
solutions from now on.

The last three terms on the right-hand side of Eq.~(\ref{match2})
contain non-local contributions from loops which involve only the
light degrees of freedom. They are, however, canceled by the
corresponding contributions in the effective theory on the left-hand
side of the matching condition.

The fact that all the infrared effects of the massless and light
particles cancel out of the matching relation~(\ref{match2}) is a
considerable advantage of the matching of Green's functions. In
contrast to that, matching $S$-matrix elements in the full and the
effective theory involves the evaluation of all infrared effects.

For completeness sake, we list below all one-loop corrections to the
generating functional of the full theory which will contribute to the
effective Lagrangian up to the order $p^4$.

One obtains the following terms from the first determinant on the
right-hand side of Eq.~(\ref{match2}) which involve only the
propagator of the massive Higgs mode:
\be \label{dloops}
	\half \ln\det d  = \half \ln\det d_m
		+ \half \Tr \left( d_m^{-1} \sigma_m \right)
		- \fourth \Tr \left( (d_m^{-1} \sigma_m)^2 \right)
			\, .
\ee
Here we used the decomposition $d = d_m + \sigma_m, d_m = - \Box +
2m^2$.  The explicit form of $\sigma_m$ can be inferred from
Eq.~(\ref{firstcomp_defD}). The second term in
Eq.~(\ref{dloops}), a tadpole graph, is of order $p^2$, whereas the
third term is of order $p^4$.

Mixed loops, which contain Higgs and Goldstone boson propagators, are
given by
\bea \label{Ddloops}
\half \ln\det \Theta
& \equiv & \half \ln\det \left( D - \delta^T d^{-1} \delta \right)
\nonumber \\ 
& = & \half  \ln\det D
- \half \Tr \left( \delta D^{-1} \delta^T d_m^{-1} \right) 
+ \half \Tr \left( \delta D^{-1} \delta^T d_m^{-1} \sigma_m
d_m^{-1} \right) 
- \fourth \Tr \left( (\delta D^{-1} \delta^T d_m^{-1})^2 \right)
 \, .
\eea 
As noted above, the term $\half \ln\det D$ on the right-hand side
cancels against the corresponding contribution in the effective
theory. The next term is of order $p^2$, whereas the last two terms
lead to contributions of order $p^4$.

Finally, the following terms involve the gauge boson propagators:
\be \label{gaugeloops}
\half \ln\det \varD  = \half \ln\det \bvarD
          + \half \Tr \left( \bvarD^{-1} \delta \varD \right) \, , 
\ee
where we used the decomposition $\varD_{\mu\nu} = \bvarD_{\mu\nu} +
\delta\varD_{\mu\nu}, \delta\varD_{\mu\nu} = \order(p^4)$.
Again the first term on the right-hand side of
Eq.~(\ref{gaugeloops}) cancels against the corresponding contribution
in the effective theory. The second term is of order~$p^4$.

Finally we note that the difference between the contribution from the
path integral measure in the full theory, $\ln\det P^T P$, and in the
effective theory, $\ln\det \bP^T \bP$, in the matching
relation~(\ref{match2}) is of order $p^6$.

Techniques to evaluate the low-energy expansion of the traces in
Eqs. (\ref{dloops}), (\ref{Ddloops}), and~(\ref{gaugeloops}) are
discussed in detail in Ref.~\cite{LSM}. The results for the
terms~(\ref{dloops}) and (\ref{Ddloops}) can be inferred
from the expressions given there. The evaluation of the second term in
Eq.~(\ref{gaugeloops}), involving the gauge bosons, proceeds in the
same way with the result
\be
\half \Tr \left( \bvarD^{-1} \delta \varD \right) = \intdx
  \Bigg( \polem M_W^2 \Y_\mu^a \Y_\mu^a 
+ \left( {3\over4} \polem + {1\over 16} {1\over 16 \pi^2} \right)
(M_Z^2 - M_W^2) \Y_\mu^3 \Y_\mu^3 \Bigg)
+ \order(p^6) \, , \label{result_gaugeloops} 
\ee
with
\be \label{polembare}
\polem   \doteq {\mu^{d-4} \over 16\pi^2} \left( {1 \over d-4} -
{1\over2} (\ln(4\pi) +    \Gamma'(1) + 1) \right)
+ {1\over 32\pi^2} \ln \left( {2 m^2 \over \mu^2} \right) \, .
\ee

\subsection{The bare effective Lagrangian}

Collecting all contributions we obtain the following result for the
bare effective Lagrangian for the standard model with a heavy Higgs
boson, up to order~$p^4$ and up to $\lambda^0$, i.e.\ at the one
loop level:
\bea
\lag_2 &  = & \left({1\over 4\lambda} - 3 \polem +
{1\over 4}
  {1\over 16 \pi^2}\right) (2m^2) \left( \W_\mu^+ \W_\mu^- +
  {1\over4} \Z_\mu \Z_\mu \right) 
+ {1\over 4\gs}  \W_{\mu\nu}^a \W_{\mu\nu}^a
+ {1\over 4\gps} \B_{\mu\nu} \B_{\mu\nu} + \lag_2^s \, , 
\label{lag_eff_2_bare_ourop} \\
\lag_4 & = & \sum_{i=1}^{18} \ourlow_i^b \ourop_i + \lag_4^s \, ,
\label{lag_eff_4_bare_ourop}
\eea
with the following results for the bare low-energy constants $\ourlow_i^b$:
\bea
\ourlow_1^b & = & - {1\over 4\lambda} + 5 \polem + {19\over 12}
  {1\over 16 \pi^2} \, , \nonumber \\
\ourlow_2^b & = & 0 \, , \nonumber \\
\ourlow_3^b & = & - {1\over 8\lambda} + {5 \over 2} \polem + {19\over 24}
  {1\over 16 \pi^2} \, , \nonumber \\
\ourlow_4^b & = & 0 \, , \nonumber \\
\ourlow_5^b & = & {1\over 16} \left(- {1\over 4\lambda} + 5 \polem +
{19\over 12} {1\over 16 \pi^2}\right) \, , \nonumber \\
\ourlow_{6}^b & = & 0 \, , \nonumber \\
\ourlow_7^b & = & {1\over 6} \polem - {11\over 72} {1\over 16 \pi^2}
\, , \nonumber \\
\ourlow_8^b & = & - {1\over 6} {1\over 16 \pi^2} \, , 
\nonumber \\
\ourlow_9^b & = & - {1\over 6} \polem + {11\over 72} {1\over 16 \pi^2} 
\, , \nonumber \\
\ourlow_{10}^b & = & {1\over 6} \polem - {11\over 72} {1\over 16
\pi^2} \, , \nonumber \\
\ourlow_{11}^b & = & 0 \, , \nonumber \\
\ourlow_{12}^b & = & {1\over 12} \polem + {1\over 144} {1\over 16
\pi^2} \, , \nonumber \\
\ourlow_{13}^b & = & - {1\over 12} {1\over 16 \pi^2} \, , 
\nonumber \\
\ourlow_{14}^b & = & - {1\over 48} {1\over 16 \pi^2} \, , \nonumber \\
\ourlow_{15}^b & = & 3 \polem + {1\over 4} {1\over 16 \pi^2} \, ,
\nonumber \\
\ourlow_{16}^b & = & s^2 \left( {3\over 4} \polem + {1\over16} {1\over
16 \pi^2} \right) \, , \nonumber \\
\ourlow_{17}^b & = & - {1\over 24} \polem - {1\over 288} {1\over 16
  \pi^2} \, , \nonumber \\
\ourlow_{18}^b & = & {1\over 24} \polem + {1\over 288} {1\over 16
  \pi^2} \, . \label{ourlowbare} 
\eea
Note that only bare quantities (coupling constants, masses, fields) appear
in the result for the effective Lagrangian in
Eqs.~(\ref{lag_eff_2_bare_ourop})--(\ref{ourlowbare}).

In order to simplify the expressions for the effective Lagrangian and
to compare our results with other calculations in the literature we
have not explicitly written down the contributions from the source
terms $\lag_2^s$ and $\lag_4^s$ in Eqs.~(\ref{lag_eff_2_bare_ourop})
and (\ref{lag_eff_4_bare_ourop}) respectively. The contributions
including the sources at tree-level are given in
Eqs.~(\ref{lag2_tree}) and (\ref{lag4_tree}).  All contributions from
the source terms at the one-loop level can be calculated from
Eqs.~(\ref{dloops}), (\ref{Ddloops}), and (\ref{result_gaugeloops}),
if one inserts the explicit expressions for the differential operators
given in Appendix~\ref{app:diffop_HH}.
Note that we have not yet used the equations of motion to reduce the
number of terms in the basis of $\lag_4$.

The result for the bare electroweak chiral Lagrangian in the usually
employed notation and the corresponding bare low-energy constants
$a_i^b$ in the usual basis at order $p^4$ can be found in
Appendix~\ref{app:EW_chiral_Lag}, Eqs.~(\ref{hv_bare}) and
(\ref{a_i_bare}). Following the conventions used in chiral
perturbation theory~\cite{GL84_85} we have included some additional,
finite terms in our definition of the pole term $\polem$,
Eq.~(\ref{polembare}), compared to the conventions used in
Refs.~\cite{H_M,E_M,D_GK}.  Taking this into account the results for
the bare low-energy constants $a_0^b, \ldots, a_{14}^b$ agree with
those obtained in Ref.~\cite{H_M}.

The results for the bare low-energy constants $\ourlow^b_{15},
\ourlow^b_{17}$, and $\ourlow^b_{18}$, or equivalently, the low-energy
constants $a^b_{15}, a^b_{16}$, and $a^b_{17}$ in
Eq.~(\ref{a_i_bare}), which correspond to operators in $\lag_4^0$ that
are proportional to terms in $\lag_2^0$, agree with the results
obtained in Ref.~\cite{D_GK}.

In the following section we are going to express the bare effective
Lagrangian from Eqs.~(\ref{lag_eff_2_bare_ourop})--(\ref{ourlowbare})
in terms of physical quantities. 


\section{Renormalization}
\label{sec:renorm}

\subsection{Physical input parameters from gauge-invariant
  Green's functions }
\label{sec:inputparameter}

In this section we want to express the bare parameters which appear in
the effective
Lagrangian~(\ref{lag_eff_2_bare_ourop})--(\ref{ourlowbare}) through
physical quantities.  As physical input parameters we choose the
masses of the Higgs and the $W$- and $Z$-bosons, and the electric
charge (on-shell scheme).  The physical mass of the Higgs boson, which
we denote by $\MHps$, is determined by the pole position of the
two-point function
\be \label{twopoint_Higgs}
\langle 0 | T (\Bphi^\dagger \Bphi)(x) (\Bphi^\dagger \Bphi)(y)| 0
\rangle \, .
\ee
The physical masses of the $W$-boson, $\MWps$, and the $Z$-boson,
$\MZps$, are defined by the pole positions of the two-point function
\be \label{twopoint_V}
\langle 0 | T (\u V_\mu^a)(x) (\u V_\nu^b)(y)| 0 \rangle \, .
\ee
As discussed in Ref.~\cite{SM_gaugeinv} one can define a renormalized
electric charge as the residue at the photon pole of the two-point function
\be \label{twopoint_Bmunu}
\langle 0 | T B_{\mu\nu}(x) B_{\rho\sigma}(y)| 0 \rangle \, .
\ee
We will denote the corresponding coupling constant by $\ers$. As was
shown in Ref.~\cite{SM_gaugeinv} by an explicit one-loop calculation,
the coupling constant $\ers$ agrees with the usual result for the
electric charge in the Thompson limit. We note that the residue of the
two-point function of the field strength $B_{\mu\nu}$ in
Eq.~(\ref{twopoint_Bmunu}) differs from unity and that it is uniquely
determined.  This can be traced back to our normalization of the gauge
field $B_\mu$ in the covariant derivative in Eq.~(\ref{cov_deriv}).
Gauge invariance requires that this field is not renormalized, cf.\
Eq.~(\ref{B_mu_ren}). The same statement holds for the gauge field
$W_\mu^a$, cf.\ Eq.~(\ref{W_mu_ren}).

For the determination of the two-point functions in
Eqs.~(\ref{twopoint_Higgs})--(\ref{twopoint_Bmunu}) we need the
generating functional $W_\SM[h,K_{\mu\nu},J_\mu^a]$ up to second order
in the external sources.  The calculation of the physical masses and
the coupling constant $\ers$ was performed in Ref.~\cite{SM_gaugeinv}
at the one-loop level. Below we will use the relations between the
bare and physical masses and electric charge which were obtained in
that reference. Because we are interested here in expressing the bare
effective Lagrangian from
Eqs.~(\ref{lag_eff_2_bare_ourop})--(\ref{ourlowbare}) in terms of
physical quantities we will only write down the low-energy expansion
of the physical quantities.

In order to determine the effective Lagrangian up to order $p^4$ we
need the physical Higgs boson mass $\MHps$ up to order $p^0$
\bea
\MHps & = & M_H^2 \left(1 + \lambda \delta M_{H,0}^2 +
\order(p^2)  \right) \, , \label{MH_physical} \\
\delta M_{H,0}^2 & = & 12 \poleMHbare - {1\over 16 \pi^2} (12 - 3
\sqrt{3} \pi) \ .
\eea
On the right-hand side of the equations only bare quantities appear.
Furthermore, we have introduced the abbreviations
\be
M_H^2 \equiv 2m^2 \ , \
\lambda \equiv {1\over 8} {e^2 \over s^2}
{M_H^2 \over M_W^2} \ , \
c^2 \equiv {M_W^2 \over M_Z^2} \ .
\ee

For the physical masses of the gauge bosons, $\MWps$ and
$\MZps$, we need the low-energy expansion up to order $p^4$. For the
$W$-boson mass we get
\bea
\MWps & = & M_W^2 \left( 1 + \lambda \delta M_{W,0}^2 +
\lambda {M_W^2 \over M_H^2} \delta M_{W,2}^2 + \order(p^4) \right) \,
, \label{MW_physical} \\
\delta M_{W,0}^2 & = & - 12 \poleMHbare + {1\over 16 \pi^2} \, , \\
\delta M_{W,2}^2 & = & c_1^W \poleMHbare + c_2^W \ln({M_W^2\over M_H^2}) +
c_3^W \ln({M_Z^2\over M_H^2}) + c_4^W + c_5^W \left[ \root \ln\left({1
+ \root \over 1 - \root}\right) \right] \, , \label{delta_MW_4}
\eea
where
\be
\root =  \sqrt{1 - 4 M_W^2 / M_Z^2} \ , 
\ee
and
\bea
c_1^W & = & {1\over c^2} \left(- {272\over 3} c^2 + 12 \right) \, , 
\nonumber \\
c_2^W & = & {1\over \pi^2 c^6} \left( - {13\over 4} c^6 + {17\over 8}
c^4 - {7\over 24} c^2 - {1\over 48} \right) \, , \nonumber \\
c_3^W & = & {1\over \pi^2 c^6} \left( - {7\over 4} c^4 + {7\over 24}
c^2 + {1\over 48} \right) \, , \nonumber \\
c_4^W & = & {1\over \pi^2 c^6} \left({461\over 72} c^6 - {7\over 12}
c^4 - {1\over 24} c^2 \right) \, , \nonumber \\
c_5^W & = & {1\over \pi^2 c^6} \left( c^6 + {17\over 12}
c^4 - {1\over 3} c^2 - {1\over 48} \right) \, . 
\eea
For the $Z$-boson mass we obtain the expression
\bea
\MZps & = & M_Z^2 \left( 1 + \lambda \delta M_{Z,0}^2 +
\lambda {M_Z^2 \over M_H^2} \delta M_{Z,2}^2 + \order(p^4) \right) \,
, \label{MZ_physical} \\
\delta M_{Z,0}^2 & = & - 12 \poleMHbare + {1\over 16 \pi^2} \, , \\
\delta M_{Z,2}^2 & = & c_1^Z \poleMHbare + c_2^Z \ln \left({M_W^2\over
  M_H^2} \right) + c_3^Z + c_4^Z \left[\root
\ln\left({\root - 1 \over \root + 1}\right) \right] \, ,
\label{delta_MZ_4} 
\eea
with
\bea
c_1^Z & = & -112 c^4 + {56\over 3} c^2 + {44\over 3} \, , \nonumber \\
c_2^Z & = & {1\over \pi^2} \left( - {7\over 2} c^4 + {7\over 12} c^2 +
{1\over 24} \right) \, , \nonumber \\
c_3^Z & = & {1\over \pi^2} \left(4 c^6 + {13\over 6} c^4 - {7\over 18}
c^2 \right) \, , \nonumber \\
c_4^Z & = & {1\over \pi^2} \left(2 c^6 + {17\over 6} c^4 - {2\over 3}
c^2 - {1\over 24} \right) \, . 
\eea
Note that the low-energy expansion for the physical gauge boson masses
starts at order $p^2$ since $M_W^2, M_Z^2 = O(p^2)$.  Furthermore, the
factors $\delta M^2$ in Eqs.~(\ref{MH_physical})--(\ref{delta_MZ_4})
count as quantities of order $p^0$ in the low-energy expansion. The
$p^2$-weighted prefactors have been extracted explicitly.

Finally, we get the following relation between the physical coupling
constant $\ers$ and the bare coupling constant $e^2$:
\bea
\ers & = & e^2 \left(1 + e^2 \delta e^2_2 +
\order(p^4)  \right) \, ,  \label{e_res_physical} \\
\delta e^2_2 & = & -14  \left[ \poleMHbare + {1\over 32 \pi^2} \ln \left(
{M_W^2\over M_H^2} \right) \right] - {19\over 3} {1\over 16 \pi^2} \, .
\label{delta_e_4}
\eea
We recall that the coupling constant $e^2$ is a quantity of order $p^2$
according to our momentum counting rules. The factor $\delta
e^2_2$ counts as order $p^0$ in the low-energy expansion.  As noted
above the result for $\ers$ agrees with the usual definition of the
electric charge in the Thompson limit \cite{delta_e_Thompson} in the
absence of fermion contributions.

The expressions for the physical masses, Eqs.~(\ref{MH_physical}),
(\ref{MW_physical}), (\ref{MZ_physical}) and the coupling constant
$\ers$, Eq.~(\ref{e_res_physical}), are finite if we insert the
renormalization prescriptions (\ref{W_mu_ren})--(\ref{gprime_ren})
for the bare quantities on the right-hand side. Of course, this is
true for the complete results for the masses, not only for the
expressions after the low-energy expansion has been carried out.
Furthermore, in the limit $\gp \to 0$, which implies $c^2 \to 1$, we
get $\MWps \equiv \MZps$ as expected.

\subsection{The effective Lagrangian}

We are now in the position to express the bare parameters which appear
in the effective Lagrangian in
Eqs.~(\ref{lag_eff_2_bare_ourop})--(\ref{ourlowbare}) in terms of
physical quantities using the relations from
Eqs.~(\ref{MH_physical})--(\ref{delta_e_4}). Note that the gauge
fields $\W_\mu^\pm, \Z_\mu$ and $B_\mu$ are not renormalized due to
gauge invariance, cf.\ Eqs.~(\ref{W_mu_ren}) and (\ref{B_mu_ren}).  At
the one-loop level and up to order $p^4$ in the low-energy expansion
we obtain the following expression for the effective Lagrangian for
the standard model with a heavy Higgs boson:
\bea
\lag_2 &  = & \left(2 \MWps {s_p^2 \over \ers} \right)
\left( \W_\mu^+ \W_\mu^- + {1\over4} \Z_\mu \Z_\mu \right) +
{s_p^2\over 4 \ers}  \W_{\mu\nu}^a \W_{\mu\nu}^a 
+ {c_p^2 \over 4 \ers} \B_{\mu\nu} \B_{\mu\nu} + \lag_2^s \, , 
\label{lag_eff_2_ren_ourop} \\
\lag_4 & = & \sum_{i=1}^{18} \ourlow_i \ourop_i + \lag_4^s \, , 
\label{lag_eff_4_ren_ourop}
\eea
with
\bea
\ourlow_1 & = & - \pole - {2 s_p^2 \MWps \over \ers
  \MHps} - {1\over 2} {1\over 16 \pi^2} \ln
\left( {\MHps \over \mu^2} \right)  + {1\over 16 \pi^2} {58 - 9
  \sqrt{3} \pi \over 12}  \, , \nonumber   \\
\ourlow_2 & = & 0 \, , \nonumber \\
\ourlow_3 & = & - {1\over 2} \pole - {s_p^2 \MWps \over \ers
  \MHps} - {1\over 4} {1\over 16 \pi^2} \ln
\left( {\MHps \over \mu^2} \right) + {1\over 16 \pi^2} {58 - 9
  \sqrt{3} \pi \over 24} \, , \nonumber \\
\ourlow_4 & = & 0 \, , \nonumber \\
\ourlow_5 & = & - {1\over 16} \pole - {s_p^2 \MWps \over 8 \ers
  \MHps} - {1\over 32} {1\over 16 \pi^2} \ln
\left( {\MHps \over \mu^2} \right) + {1\over 16 \pi^2} {58 - 9
  \sqrt{3} \pi \over 192} \, , \nonumber \\
\ourlow_{6} & = & 0 \, , \nonumber \\
\ourlow_7 & = & {1\over 6} \pole + {1\over 12} {1\over 16 \pi^2} \ln
\left( {\MHps \over \mu^2} \right) - {11\over 72} {1\over 16 \pi^2} \, 
, \nonumber \\
\ourlow_8 & = & - {1\over 6} {1\over 16 \pi^2} \, , 
\nonumber \\
\ourlow_9 & = & - {1\over 6} \pole - {1\over 12} {1\over 16 \pi^2} \ln
\left( {\MHps \over \mu^2} \right) + {11\over 72} {1\over 16 \pi^2} \, 
, \nonumber \\
\ourlow_{10} & = & {1\over 6} \pole + {1\over 12} {1\over 16 \pi^2} \ln
\left( {\MHps \over \mu^2} \right) - {11\over 72} {1\over 16 \pi^2} \, 
, \nonumber \\
\ourlow_{11} & = & 0 \, , \nonumber \\
\ourlow_{12} & = & {1\over 12} \pole + {1\over 24} {1\over 16 \pi^2} \ln
\left( {\MHps \over \mu^2} \right) + {1\over 144} {1\over 16 \pi^2} \, 
, \nonumber \\
\ourlow_{13} & = & - {1\over 12} {1\over 16 \pi^2} \, , \nonumber \\
\ourlow_{14} & = & - {1\over 48} {1\over 16 \pi^2} \, , \nonumber \\
\ourlow_{15} & = & 3 \pole + {3\over 2} {1\over 16 \pi^2} \ln
\left( {\MHps \over \mu^2} \right)  + {1\over4} {1\over 16 \pi^2} - {1\over
  4} (1 - {c_p^2 \over s_p^2}) \delta M_{W,2}^2 - {1 \over 4 s_p^2}
\delta M_{Z,2}^2 + 2 s_p^2 \delta e^2_2 \, , \nonumber \\
\ourlow_{16} & = & s_p^2 \left( {3\over 4} \pole + {3\over 8} {1\over 16
  \pi^2} \ln \left( {\MHps \over \mu^2} \right) + {1\over16} {1\over
16 \pi^2} \right) \, , \nonumber \\
\ourlow_{17} & = & - {1\over 24} \pole - {1\over 48} {1\over 16 \pi^2} \ln
\left( {\MHps \over \mu^2} \right) - {1\over 288} {1\over
  16\pi^2} + {c_p^2 \over 32 s_p^2} \delta M_{W,2}^2 - {1\over 32
s_p^2} \delta M_{Z,2}^2 + {s_p^2 \over 4} \delta e^2_2 \, , \nonumber \\
\ourlow_{18} & = & {1\over 24} \pole + {1\over 48} {1\over 16 \pi^2} \ln
\left( {\MHps \over \mu^2} \right) + {1\over 288} {1\over
  16\pi^2} - {c_p^2 \over 32 s_p^2} \delta M_{W,2}^2 + {1 \over 32
s_p^2} \delta M_{Z,2}^2 + {c_p^2 \over 4} \delta e^2_2 \, . \label{ourlow_ren}
\eea
The results for the low-energy constants $\ourlow_1, \ourlow_3$, and
$\ourlow_5$ are obtained by expressing the bare coupling constant
$\lambda$ which appears in Eq.~(\ref{ourlowbare}) through physical
quantities.  In order to obtain $\ourlow_{15}, \ourlow_{17}$, and
$\ourlow_{18}$ one has to express the bare quantities $m^2 / \lambda,
g^2,$ and $\gp^2$ in Eq.~(\ref{lag_eff_2_bare_ourop}) through physical
quantities. The quantities $\delta M_{W,2}^2, \delta M_{Z,2}^2$, and
$\delta e^2_2$ are defined in Eqs.~(\ref{delta_MW_4}),
(\ref{delta_MZ_4}), and (\ref{delta_e_4}), respectively.  We use the
on-shell definition for the weak mixing angle
\be \label{def_cp_sp}
c_p^2 \doteq {\MWps  \over \MZps} \, , \quad
s_p^2 \doteq 1 - c_p^2 \, .
\ee
The pole term in $d=4$ dimensions is given by
\be \label{pole}
\pole   \doteq {\mu^{d-4} \over 16\pi^2} \left( {1 \over d-4} -
{1\over2} [\ln(4\pi) + \Gamma'(1) + 1] \right) \, .
\ee
In order to simplify the expressions we have not explicitly written
down the results for the source terms $\lag_2^s$ and $\lag_4^s$ in
Eqs.~(\ref{lag_eff_2_ren_ourop}) and (\ref{lag_eff_4_ren_ourop}),
respectively.

As discussed in Sec.~\ref{subsec:EFT_W_4} we can reduce the number of
terms in the effective Lagrangian $\lag_4$ by making use of the
equations of motion in the effective field theory and by renormalizing
the parameters and low-energy constants in the lowest order Lagrangian
$\lag_2$.  The source terms in $\lag_4^s$ will thereby not affect the
terms $\lag_4^0$ without sources. Switching off the sources
altogether, we then obtain the following result for the effective
Lagrangian:
\be \label{lag_eff_2_ren_redefined}
\lag_2 = {\bv_{eff}^2 \over 2}
\left( \W_\mu^+ \W_\mu^- + {1\over4} \Z_\mu \Z_\mu \right) 
+ {1\over 4 \bg_{eff}^2}  \W_{\mu\nu}^a \W_{\mu\nu}^a
+ {1\over 4 \bgp_{eff}^2} \B_{\mu\nu} \B_{\mu\nu} \, , 
\ee
with
\bea
\bv_{eff}^2 & = & 4 \MWps {s_p^2 \over \ers} \Bigg( 1 + {\ers \over
  s_p^2} \bigg[ {11\over 6} \pole
+ {11\over 12} {1\over 16
  \pi^2} \ln \left( {\MHps \over \mu^2} \right) + {11\over 72} {1\over
16 \pi^2}  \nonumber \\
& &\mbox{}- {1\over 8} \left(1 - {c_p^2
  \over s_p^2}\right) \delta M_{W,2}^2 
- {1 \over 8 s_p^2} \delta
M_{Z,2}^2 + s_p^2 \delta e^2_2 \bigg] \Bigg) \, , \label{v_eff} \\
\bg_{eff}^2 & = & {\ers \over s_p^2} \Bigg( 1 + {\ers \over s_p^2}
\bigg[ - {1\over 6} \pole - {1\over 12} {1\over
  16 \pi^2} \ln \left( {\MHps \over \mu^2} \right) - {1\over 72}
{1\over 16 \pi^2}  
- {c_p^2 \over 8 s_p^2}
\delta M_{W,2}^2 + {1\over 8 s_p^2} \delta M_{Z,2}^2
- s_p^2 \delta e^2_2 \bigg] \Bigg) \, , \label{g_eff} \\
\bgp_{eff}^2 & = & {\ers \over c_p^2} \Bigg(1 + {\ers \over c_p^2}
\bigg[ {1\over 6} \pole + {1\over 12} {1\over 16
  \pi^2} \ln \left( {\MHps \over \mu^2} \right) + {1\over 72} {1\over
16 \pi^2}  
+ {c_p^2 \over 8 s_p^2}
\delta M_{W,2}^2 - {1 \over 8 s_p^2} \delta M_{Z,2}^2
- c_p^2 \delta e^2_2 \bigg] \Bigg) \, . \label{gprime_eff}
\eea

At order $p^4$ we obtain the result
\be
\lag_4 = \sum_{i=0}^{9} \ourlow_i^\prime \ourop_i \, , 
\label{lag_eff_4_ren_redefined}
\ee
where the low-energy constants $\ourlow_i^\prime$ corresponding to the
independent terms in the Lagrangian $\lag_4$ are given by
\bea
\ourlow_{0}^\prime  & = & s_p^2 \left( {3\over 4} \pole + {3\over 8}
{1\over 16 \pi^2} \ln \left( {\MHps \over \mu^2} \right) + {1\over16} 
{1\over 16 \pi^2} \right) \, , \nonumber \\ 
\ourlow_1^\prime & = & - {1\over 3} \pole - {2 s_p^2 \MWps \over \ers
  \MHps} - {1\over 6} {1\over 16 \pi^2} \ln
\left( {\MHps \over \mu^2} \right)  + {1\over 16 \pi^2} {176 - 27
  \sqrt{3} \pi \over 36} \, , \nonumber   \\
\ourlow_2^\prime & = & - {2\over 3} \pole - {1\over 3} {1\over 16 \pi^2} 
\ln \left( {\MHps \over \mu^2} \right) - {1\over 18} {1\over 16 \pi^2} 
\, , \nonumber \\
\ourlow_3^\prime & = & {1\over 6} \pole - {s_p^2 \MWps \over \ers
  \MHps} + {1\over 12} {1\over 16 \pi^2} \ln
\left( {\MHps \over \mu^2} \right) + {1\over 16 \pi^2} {178 - 27
  \sqrt{3} \pi \over 72} \, , \nonumber \\
\ourlow_4^\prime & = & - {2\over 3} \pole - {1\over 3} {1\over 16 \pi^2} 
\ln \left( {\MHps \over \mu^2} \right) - {1\over 18} {1\over 16 \pi^2} 
\, , \nonumber \\
\ourlow_5^\prime & = & - {1\over 16} \pole - {s_p^2 \MWps \over 8 \ers
  \MHps} - {1\over 32} {1\over 16 \pi^2} \ln
\left( {\MHps \over \mu^2} \right) + {1\over 16 \pi^2} {58 - 9
  \sqrt{3} \pi \over 192} \, , \nonumber \\
\ourlow_{6}^\prime & = & 0 \, , \nonumber \\ 
\ourlow_7^\prime & = & - {1\over 6} \pole - {1\over 12} {1\over 16 \pi^2} 
\ln \left( {\MHps \over \mu^2} \right) - {13\over 72} {1\over 16
  \pi^2} \, , \nonumber \\
\ourlow_8^\prime & = & - {1\over 6} {1\over 16 \pi^2} \, , 
\nonumber \\
\ourlow_9^\prime & = & {1\over 6} \pole + {1\over 12} {1\over 16 \pi^2} \ln
\left( {\MHps \over \mu^2} \right) + {13\over 72} {1\over 16 \pi^2}
\, . \label{ourlow_eom}
\eea

\subsection{Discussion}
\label{sec:discussion_Leff_ren}

Equations~(\ref{lag_eff_2_ren_redefined})--(\ref{ourlow_eom})
represent our final result for the effective Lagrangian for the
standard model with a heavy Higgs boson, expressed through the
physical masses of the Higgs boson, the $W$- and the $Z$-boson, as
well as the electric charge. The effective Lagrangian includes all
contributions at one-loop in the standard model and up to order $p^4$
in the low-energy expansion.

Let us first discuss the lowest order Lagrangian $\lag_2$ in
Eq.~(\ref{lag_eff_2_ren_redefined}) and the corresponding low-energy
constants in Eqs.~(\ref{v_eff})--(\ref{gprime_eff}). Comparing with
the general effective Lagrangian in Eq.~(\ref{gen_lageff_2_0}) we note
that $\trho = 1$. This is due to the fact that in the standard model
with a heavy Higgs boson the custodial symmetry violating effects in
$\Delta \rho$ are proportional to $\gps$, i.e.\ they are of higher
order in the momentum expansion. We recall that $\gp = \order(p)$
according to the counting rules discussed in Sec.~\ref{sec:def_EFT}.

The additional terms proportional to $\ers / s_p^2$ in $\bv_{eff}^2,
\bg_{eff}^2$, Eqs.~(\ref{v_eff}), (\ref{g_eff}), and the additional terms
proportional to $\ers / c_p^2$ in $\bgp_{eff}^2$,
Eq.~(\ref{gprime_eff}), deserve some comments. Employing our counting
rules these terms will contribute only at order $p^4$. They originate
from the low-energy constants $\ourlow_{15}, \ourlow_{17},$ and
$\ourlow_{18}$ in Eq.~(\ref{ourlow_ren}) before removing redundant
terms from the effective Lagrangian. These low-energy constants are
not independently observable and only renormalize the low-energy
constants $\bv_{eff}^2, \bg_{eff}^2,$ and $\bgp_{eff}^2$ in the lowest
order Lagrangian $\lag_2$, nevertheless their contributions have to be
kept in order to fully describe all effects for the standard model
with a heavy Higgs boson up to order $p^4$.

For convenience, we have included these contributions into the
low-energy constants $\bv_{eff}^2, \bg_{eff}^2,$ and
$\bgp_{eff}^2$. Thus, $\bv_{eff}^2$ contains terms of order $p^0$ and
$p^2$, while $\bg_{eff}^2$ and $\bgp_{eff}^2$ contain terms of order
$p^2$ and $p^4$. Since we have chosen the on-shell scheme,
low-energy physics enters the effective Lagrangian after the
renormalization through the input parameters $\MWp, \MZp$, and
$\eres$, leading to these nonanalytic terms.  We note that the same
happens in ordinary chiral perturbation theory for low-energy QCD. The
relations between the parameters $F$ and $M$ in the effective
Lagrangian and the physical pion decay constant $F_\pi$ and the
physical pion mass $M_\pi$ both contain a nonanalytic chiral logarithm
$\ln(M_\pi^2)$, see Ref.~\cite{GL84_85}.

Thus, it is important to distinguish between the general, local
effective Lagrangian with arbitrary bare low-energy constants that
have to be determined from experiment from the explicit result for the
effective Lagrangian for a given underlying theory, here the standard
model with a heavy Higgs boson, evaluated in a given regularization
and renormalization scheme.

Next we turn to the result for the effective Lagrangian $\lag_4$ in
Eq.~(\ref{lag_eff_4_ren_redefined}) and the corresponding low-energy
constants~$\ourlow_i^\prime$ in Eq.~(\ref{ourlow_eom}). Since $\trho =
1$ we have 10 independent low-energy constants in $\lag_4$.  Only the
low-energy constants $\ourlow_0^\prime \equiv \ourlow_{16},
\ourlow_5^\prime, \ourlow_6^\prime,$ and $ \ourlow_{8}^\prime$ in 
Eq.~(\ref{ourlow_eom}) are equal to their counterparts $\ourlow_i$ in
Eq.~(\ref{ourlow_ren}) before the elimination of redundant terms.

The low-energy constants in Eqs.~(\ref{ourlow_ren}) and
(\ref{ourlow_eom}) have the following general form:
\bea
\ourlow_i & = & \delta_i \pole + \ourlow_i^r(\mu) \, , \nonumber \\
\ourlow_i^\prime & = & \delta_i^\prime \pole + \ourlow_i^{\prime \,
  r}(\mu) \, , 
\eea
i.e.\ they contain a pole term proportional to $\pole$, cf.\
Eq.~(\ref{pole}), and a scale dependent part. We denote the coupling
constants $\ourlow_i^r(\mu)$ and $\ourlow_i^{\prime \, r}(\mu)$ as
renormalized low-energy constants. The renormalization group running
of the coupling constants $\ourlow_i^r(\mu)$ and $\ourlow_i^{\prime \,
r}(\mu)$ is determined by the coefficient $\delta_i$ and
$\delta_i^\prime$ of the respective pole term.  These coefficients are
determined by the one-loop divergences of the gauged nonlinear sigma
model described by $\lag_2^0$ and have been calculated long time
ago~\cite{heavy_Higgs,Longhitano}. They are universal, i.e.\
independent of any underlying strongly interacting model with the same
symmetry breaking pattern as the standard model. Note that we obtain
these universal pole terms only after the renormalization has been
carried out in the standard model. The pole terms of the low-energy
constants $\ourlow_1, \ourlow_3,$ and $\ourlow_5$, which receive a
tree-level contribution in the standard model with a heavy Higgs
boson, have changed compared to the results for the bare low-energy
constants in Eq.~(\ref{ourlowbare}) which contain a term $1/\lambda$,
where $\lambda$ is the bare, divergent scalar coupling constant.

The effective Lagrangian given in
Eqs.~(\ref{lag_eff_2_ren_redefined})--(\ref{ourlow_eom}) can now be
used to calculate physical quantities like scattering amplitudes up to
order $p^4$, by adding tree-level diagrams from $\lag_2 + \lag_4$ and
contributions from one-loop graphs with the Lagrangian $\lag_2$. Note
that the contributions from the source terms and from the path
integral measure have to be taken into account as well. As discussed
above, the renormalization has, however, been carried out already. In
particular, there is no need to calculate once more the masses of the
light particles, like the $W$ or the $Z$-boson, in the effective field
theory. Note that the effective Lagrangian~$\lag_4$ in
Eqs.~(\ref{lag_eff_4_ren_redefined}) and (\ref{ourlow_eom}) contains
pole terms $\pole$ even after the renormalization. This fact is well
known from chiral perturbation theory~\cite{Weinberg,GL84_85}.
One-loop graphs with vertices from $\lag_2$ generate divergences which
are canceled by the corresponding pole terms in the low-energy
constants from $\lag_4$. In this way, physical quantities will be
finite.

We would like to add a few comments about the size of the renormalized
low-energy constants $\ourlow_i^{\prime \, r}(\mu)$ in
Eq.~(\ref{ourlow_eom}). First of all we note that due to the Veltman
screening theorem~\cite{Veltman_screening}, there are only logarithmic
non-decoupling terms of the form $\ln(\MHps)$ in the low-energy
constants $\ourlow_i^{\prime \, r}(\mu)$ in Eq.~(\ref{ourlow_eom}) at
the one-loop level. In addition, the low-energy constants
$\ourlow_1^{\prime \, r}(\mu), \ourlow_3^{\prime \, r}(\mu),$ and
$\ourlow_5^{\prime \, r}(\mu)$ contain a tree-level contribution
proportional to $1/\MHps$. Even though we assume that the Higgs boson
is heavy, we cannot simply take $\MHp \to \infty$ and drop these
terms. This would be equivalent to the assumption that the one-loop
terms dominate over the tree-level contributions. Since our matching
calculation was done by using perturbation theory this is certainly
not permissible. The renormalized low-energy constants
$\ourlow_i^{\prime \, r}(\mu)$ depend on a reference scale $\mu$. We
will vary this scale between the mass of the $Z$-boson, $M_Z$, and a
value of $2$~TeV, which lies in the resonance region of a truly
strongly interacting symmetry breaking sector, e.g.\ this scale
corresponds roughly to the mass of a technirho in technicolor
models~\cite{Technicolor}. We thus follow the conventions usually
adopted in chiral perturbation theory~\cite{GL84_85} for QCD where the
setting $\mu = M_\rho$ is used to quote values for the renormalized
low-energy constants.  The Higgs boson mass is varied between $M_Z$ and
$2$~TeV as well, although for Higgs boson masses above $1$~TeV the
applicability of perturbation theory is certainly questionable. We
then find that the values of those renormalized low-energy constants
$\ourlow_i^{\prime r}(\mu)$ which receive only contributions from
loops are of the size which one would expect from using naive
dimensional analysis~\cite{dim_analysis}, i.e.\ they are of the order
of $1 / (16 \pi^2)$.  On the other hand, as mentioned above, the
low-energy constants $\ourlow_1^{\prime \, r}(\mu), \ourlow_3^{\prime
\, r}(\mu),$ and $\ourlow_5^{\prime \, r}(\mu)$ contain a tree-level
contribution proportional to $1/\MHps$.  For all values of $\mu$ in
the range between $M_Z$ and $2$~TeV this term dominates for Higgs
boson masses below $1$~TeV. In fact, in the low-energy constant
$\ourlow_1^{\prime \, r}(\mu)$ the tree level and the one-loop term
are of the same order of magnitude only for Higgs boson masses of the
order of $2.5$~TeV, due to an accidental cancellation in the one-loop
contribution.

Some phenomenological consequences of the analysis
presented here for models of a strongly interacting electroweak
symmetry breaking sector can be found in Ref.~\cite{LC_Note}. In
particular, we compare in that paper the results for the reduced set
of independent low-energy constants $\ourlow_i^{\prime \, r}(\mu)$ for
the standard model with a heavy Higgs boson with those for a simple
technicolor model. 

Finally, we would like to compare our result for the effective
Lagrangian from
Eqs.~(\ref{lag_eff_2_ren_redefined})--(\ref{ourlow_eom}) for the
standard model with a heavy Higgs boson after the renormalization with
those obtained in the literature~\cite{H_M,E_M,D_GK}. As noted above,
the result for the bare effective Lagrangian in
Eqs.~(\ref{lag_eff_2_bare_ourop})--(\ref{ourlowbare}) agreed with the
literature.  In order to facilitate the comparison we will use the
usual notation for the electroweak chiral Lagrangian and discuss the
low-energy constants $a_i$ expressed through physical quantities as
given in Appendix~\ref{app:EW_chiral_Lag}, in Eq.~(\ref{a_i_ren}), and
the low-energy constants $a_i^\prime$ after the elimination of
redundant terms as given in Eq.~(\ref{a_i_EoM_ren}). First of all, the
expression of the lowest order effective Lagrangian $\lag_2$,
Eq.~(\ref{lag_eff_2_ren_op}), agrees with Refs.~\cite{H_M,E_M,D_GK},
i.e.\ we have $\trho=1$. At order $p^4$ our result for the pole terms
and the finite parts of the low-energy constants $a_i, i = 0, \ldots,
14,$ given in Eq.~(\ref{a_i_ren}), agrees with the results obtained in
Refs.~\cite{H_M}.  Note that we have included some finite parts in the
definition of the pole term $\pole$, cf.\ Eq.~(\ref{pole}), compared
to the conventions used in that reference. 

Reducing the number of terms as outlined in Sec.~\ref{subsec:EFT_W_4}
leads to the results for the low-energy constants $a_i^\prime$ as
given in Eq.~(\ref{a_i_EoM_ren}). Only the value of the low-energy
constant $a_3^\prime$ has changed compared to $a_3$ in
Eq.~(\ref{a_i_ren}).  Note, however, that $a_1$ and $a_8$ have
disappeared from the list of independent low-energy constants. In this
respect our result differs from the literature since this further
elimination of redundant terms was not carried out in
Refs.~\cite{H_M,E_M,D_GK}. 

Furthermore, the expressions for $\bv_{eff}^2, \bg_{eff}^2$, and
$\bgp_{eff}^2$ in $\lag_2$ as given in Eqs.~(\ref{v_eff}),
(\ref{g_eff}), and (\ref{gprime_eff}), respectively, differ from the
results obtained in Refs.~\cite{H_M,E_M,D_GK}. This is due to the fact
that we went one step further in the low-energy expansion of the mass
for the Higgs boson in Eq.~(\ref{MH_physical}), of the masses for the
gauge bosons in Eqs.~(\ref{MW_physical}), (\ref{MZ_physical}), and of
the electric charge in Eq.~(\ref{e_res_physical}). As mentioned above
this is necessary in order to obtain all contributions in the
effective field theory up to order $p^4$, if the low-energy constants
in the effective Lagrangian are expressed through these physical input
parameters.

As was noted already in Ref.~\cite{H_M} the results for the
low-energy constants agree with those obtained in the ungauged
$O(4)$-linear sigma model~\cite{GL84_85,LSM}, in all cases where such
a comparison is possible. Note that there are more low-energy
constants in the present case, since the symmetry is $SU(2)_L \times
U(1)_Y$ instead of $SU(2)_L \times SU(2)_R$ for the case of the sigma
model.  Employing a functional approach this agreement can easily be
inferred from the matching relation~(\ref{match2}). After the
diagonalization of the differential operator in the full theory, those
loops which contain gauge bosons are separated from the loops
involving the Higgs and the Goldstone bosons.  A similar observation
was made in Ref.~\cite{D_GK}.  Since we count powers of $g^2$ and
$\gps$ as quantities of order $p^2$, any correction from gauge-boson
loops to the low-energy constants in $\lag_4$ must be of order $p^6$
in the effective field theory.  Therefore within the standard model
with a heavy Higgs boson, the effects from gauge-boson loops are
suppressed compared to the contributions from the Higgs and the
Goldstone bosons.


\section{Summary and Discussion}
\label{sec:summary}

In this article we have reanalyzed the electroweak chiral Lagrangian
which describes the low-energy structure of a strongly interacting
electroweak symmetry breaking sector. We have employed a manifestly
gauge-invariant functional approach that was introduced
recently~\cite{SM_gaugeinv}. It is well suited to analyze two issues
related to gauge invariance where there are some subtleties involved,
because one has to deal with off-shell quantities. First, we
determined the number of independent low-energy constants in the
electroweak chiral Lagrangian. By employing the equations of motion we
found that the set of parameters currently used in the
literature~\cite{Appelquist_Wu,Feruglio} is redundant.  The second
topic of this paper was the evaluation of the low-energy constants in
the effective Lagrangian by matching the full and effective theory at
low energies. As an example we studied the standard model with a heavy
Higgs boson\footnote{Since all recent fits to electroweak precision
data prefer a light Higgs boson~\cite{Higgs_fit}, we regard the
standard model with a heavy Higgs boson only as a testing ground for
our method of matching.} where the calculation can be performed by
using perturbative methods.

We first introduced the effective field theory for the bosonic part of
a strongly interacting electroweak symmetry breaking sector under the
assumption that $p^2, M_W^2, M_Z^2 \ll M^2$, where $p$ is a typical
momentum and $M$ is the mass scale for heavy particles in the
underlying theory, e.g.~a heavy Higgs boson in the standard model or a
technirho in some technicolor model~\cite{Technicolor}. In order to
preserve the gauge symmetry we employed the gauge-invariant functional
approach presented in Ref.~\cite{SM_gaugeinv}.  Its essential feature
is to consider Green's functions of gauge-invariant operators which
excite one-particle states of the photon, the $W$-, and the $Z$-boson,
respectively. The effective field theory is then described by an
effective Lagrangian which is gauge-invariant and depends on the
Goldstone boson field~$\bU$, the vector fields~$\bar W_\mu^a,
\bB_\mu$, and external sources.

We have constructed the effective Lagrangian including appropriate
source terms up to order $p^4$ in the low-energy expansion. The lowest
order effective Lagrangian $\lag_2$ involves the four physical
parameters $\bar e, M_W, M_Z,$ and $\brho$, corresponding to the
electric charge, the masses of the gauge bosons and the
$\brho$-parameter in the effective field theory,
respectively. Furthermore, there are two additional low-energy
constants from the source terms. At order $p^4$ the effective
Lagrangian is given as a linear combination of a maximal set of
gauge-invariant terms.  One can then eliminate redundant terms by
using algebraic relations which follow by partial integration. Since
the Lagrangian~$\lag_4$ contributes only at the classical level one
can also use the equations of motion to eliminate further redundant
terms~\cite{GL84_85,Bij_Col_Eck}.  We note that in our
gauge-invariant approach no gauge artifacts can enter through this
procedure, because there is no gauge-fixing term and the sources
respect the gauge symmetry.  Finally, there are terms in the
Lagrangian $\lag_4$ which are proportional to corresponding terms in
the lowest order Lagrangian $\lag_2$. These terms lead to a
renormalization of the low-energy constants and sources at order~$p^2$
and therefore have no observable effect.

In this way we find that if one considers a purely bosonic effective
field theory with the same symmetry breaking pattern as the standard
model there are 10 physically relevant low-energy constants at order
$p^4$ in the electroweak chiral Lagrangian. In particular, by
employing the equations of motion of the gauge fields, one can choose
to remove two low-energy constants, usually denoted by $a_1$ and
$a_8$~\cite{Appelquist_Wu}, which contribute to the self-energies of
the gauge bosons. This is in contrast to the number of 12 low-energy
constants which is quoted in the
literature~\cite{Appelquist_Wu,Feruglio}. An additional number of 63
low-energy constants contributes to the off-shell behavior of our
gauge-invariant Green's functions.  The latter low-energy constants,
however, do not enter physical quantities.

If fermions are included the situation changes as follows.  There are
many more terms present in the effective Lagrangian, including sources
coupled to the fermions. Therefore, a host of additional low-energy
constants enters the effective Lagrangian. Many of them are, however,
strongly bounded by experiments or irrelevant to the current
experimental situation.  A complete effective field theory analysis
including the fermions was beyond the scope of the present work.
Nevertheless, even when fermions are included, it is possible to
eliminate the same two terms in the effective Lagrangian at order
$p^4$ which contribute to the self-energies of the gauge bosons. This
will only lead to a renormalization of the external sources as well as
the couplings of the gauge fields to the fermions. Hence, even in the
presence of fermions, the complete low-energy analysis of a strongly
interacting symmetry breaking sector does not involve the low-energy
constants $a_1$ and $a_8$.

These two low-energy constants are often identified with the oblique
parameters $S$ and $U$~\cite{STU}. As discussed in
Sec.~\ref{sec:EFT_fermions} this identification is not
possible. The oblique parameters $S, T,$ and $U$ describe new physics
{\it beyond} the standard model with an elementary Higgs boson,
whereas the low-energy constants in the electroweak chiral Lagrangian
describe any strongly interacting symmetry breaking sector, even if
there is no Higgs boson at all.  From the point of view of an
effective Lagrangian analysis the parametrization of new physics
effects by Peskin and Takeuchi amounts to setting all low-energy
constants to their standard model values (assuming a heavy Higgs
boson), except for three parameters contributing to gauge-boson
self-energies. Employing the equations of motion one can still remove
the terms corresponding to $a_1$ and $a_8$, however, two other
low-energy constants will then differ from their values in the
standard model and the total number of parameters to describe new
physics remains three.

In the second part of the paper we have investigated the issue of
evaluating the effective Lagrangian for a given underlying theory. The
effective field theory can be defined by requiring, for instance, that
corresponding Green's functions in the full and in the effective
theory have the same low-energy structure.  In order to make sure that
no gauge artifacts can enter in this matching procedure, we propose to
match gauge-invariant Green's functions. As an example we have
considered the standard model with a heavy Higgs boson where the
low-energy constants can explicitly be calculated using perturbative
methods, if the scalar coupling constant is not too large. We briefly
recapitulated the main results from our manifestly gauge-invariant
approach~\cite{SM_gaugeinv} to the bosonic sector of the standard
model. We then evaluated the matching condition at the one-loop level
and at order $p^4$ in the low-energy expansion, employing functional
techniques that have been discussed in detail in Ref.~\cite{LSM}. In
this way we obtained the effective Lagrangian expressed through bare
quantities. The results agree with the literature~\cite{H_M,E_M,D_GK}.

We then expressed the low-energy constants in the effective Lagrangian
through physical quantities. As physical input parameters we chose the
mass of the Higgs boson, the masses of the $W$- and $Z$-boson, and the
electric charge (on-shell scheme) which have been extracted from
two-point functions of appropriately chosen gauge-invariant operators
in Ref.~\cite{SM_gaugeinv}. We went one step further in the low-energy
expansion of the physical masses for the Higgs boson and the gauge
bosons and the electric charge compared to
Refs.~\cite{H_M,E_M,D_GK}. In this way we obtained explicit
expressions for the effective low-energy constants $\bv_{eff}^2,
\bg_{eff}^2$, and $\bgp_{eff}^2$ which appear in $\lag_2$. As
discussed in Sec.~\ref{sec:discussion_Leff_ren} this is necessary in
order to obtain all contributions in the effective field theory up to
order $p^4$, if the low-energy constants in the effective Lagrangian
are expressed through these physical input parameters. Furthermore, we
removed the redundant terms in the effective Lagrangian by the
procedure outlined in Sec.~\ref{subsec:EFT_W_4}.

The effective Lagrangian given in
Eqs.~(\ref{lag_eff_2_ren_redefined})--(\ref{ourlow_eom}) can now be
used to calculate physical quantities like scattering amplitudes up to
order $p^4$, by adding tree-level diagrams from $\lag_2 + \lag_4$ and
contributions from one-loop graphs with the Lagrangian $\lag_2$. Note
that the contributions from the source terms and from the path
integral measure have to be taken into account as well. The
renormalization has, however, been carried out already. In particular,
there is no need to calculate once more the masses of the light
particles, like the $W$ or the $Z$-boson, in the effective field
theory.

As was noted in Ref.~\cite{H_M} the results for the low-energy
constants at order $p^4$ agree with those obtained in the ungauged
$O(4)$-linear sigma model~\cite{GL84_85,LSM}, in all cases where such
a comparison is possible. This can easily be understood within our
functional framework from the matching relation and the counting of
powers of $g^2$ and $\gps$ as quantities of order $p^2$. We note that
this counting rule is needed for the consistency of the effective
field theory. Therefore within the standard model with a heavy Higgs
boson, the effects from gauge-boson loops are suppressed compared to
the contributions from the Higgs and the Goldstone bosons. The
situation is, however, different, if higher orders in the momentum
expansion or in the loop expansion are evaluated or if other theories
are considered.  A well defined matching procedure which deals only
with gauge-invariant quantities as proposed in this paper is mandatory
in such cases.

Some phenomenological consequences of the analysis presented in this
article for models of a strongly interacting electroweak symmetry
breaking sector are discussed in Ref.~\cite{LC_Note}. In particular,
we compare in that paper the results for the reduced set of
independent low-energy constants $l_i^\prime$ (in the bosonic sector)
for the standard model with a heavy Higgs boson with those for a
simple technicolor model. The low-energy constants for the technicolor
model have been estimated assuming that the exchange of the lowest
lying resonances dominates the numerical values of the renormalized
low-energy constants in the resonance region.  This assumption works
reasonably well for the coefficients in the ordinary chiral Lagrangian
for QCD~\cite{GL84_85,ChPT_Resonances} and can be justified using
large-$N_c$ arguments and constraints from
sum rules~\cite{largeNc_LMD}. Since the pattern of the low-energy
constants is very different in these two models it may be misleading
to mimic any strongly interacting symmetry breaking sector by a heavy
Higgs boson as done in Ref.~\cite{PDG_98}. From our investigation we
conclude, in accordance with Ref.~\cite{BaggerFalkSwartz}, that
current electroweak precision data do not really rule out such
strongly interacting models.


\section*{Acknowledgments}

We are grateful to F.~Jegerlehner and V.~Ravindran for enlightening
discussions, a careful reading of the manuscript and suggestions for
improvements.  We are furthermore indebted to J.~Gasser, M.~Knecht,
H.~Leutwyler, E.~de Rafael, R.~Sommer, J.~Stern, O.~Veretin, and
A.~Vicini for useful discussions. A.N.\ is grateful to the members of
the Yale Physics department for their kind hospitality during the
early stages of this project. He also acknowledges financial support
by Schweizerischer Na\-tio\-nal\-fonds during that period.


\appendix 

\section{Source terms at order $\lowercase{p}^4$}
\label{app:sources_p4}

In this Appendix we list all algebraically independent CP-even source
terms which appear at order $p^4$ in the electroweak chiral Lagrangian
in Eq.~(\ref{lag_4_s}). We have not yet used the equations of motion
to reduce the number of terms. The terms are grouped according to the
total number of fields and sources.

Terms with four powers of fields and external sources:
\begin{eqnarray}
\ourop^s_{1}  &=& (\bW_\mu^+ \bW_\mu^-) (\bW_\nu^+ \bj_\nu^- +
\bW_\nu^- \bj_\nu^+) \, , \nonumber \\
\ourop^s_{2}  &=& (\bW_\mu^+ \bW_\nu^-) (\bW_\mu^+
\bj_\nu^- + \bW_\nu^- \bj_\mu^+) \, , \nonumber \\
\ourop^s_{3}  &=& (\bW_\mu^+ \bW_\mu^-) (\bZ_\nu \bJ^\Z_\nu) \, , 
\nonumber \\
\ourop^s_{4}  &=& (\bW_\mu^+ \bW_\nu^- + \bW_\mu^- \bW_\nu^+)
(\bZ_\mu \bJ^\Z_\nu) \, , \nonumber \\
\ourop^s_{5}  &=& (\bZ_\mu \bZ_\mu) (\bW_\nu^+ \bj_\nu^- + \bW_\nu^-
\bj_\nu^+) \, , \nonumber \\
\ourop^s_{6}  &=& (\bZ_\mu \bZ_\nu) (\bW_\mu^+ \bj_\nu^- + \bW_\mu^-
\bj_\nu^+) \, , \nonumber \\
\ourop^s_{7}  &=& (\bZ_\mu \bZ_\mu)(\bZ_\nu \bJ^\Z_\nu) \, ,
\nonumber \\ 
%
\ourop^s_{8}  &=&  (\bW_\mu^+ \bW_\mu^-) (\bj_\nu^+ \bj_\nu^-) \, , 
\nonumber \\
\ourop^s_{9}  &=&  (\bW_\mu^+ \bW_\nu^-) (\bj_\mu^+ \bj_\nu^-) \, , 
\nonumber \\
\ourop^s_{10}  &=&  (\bW_\mu^+ \bW_\nu^-) (\bj_\mu^- \bj_\nu^+) \, , 
\nonumber \\
\ourop^s_{11}  &=&  (\bW_\mu^+ \bj_\nu^-) (\bW_\mu^+ \bj_\nu^-)  +
        (\bW_\mu^- \bj_\nu^+) (\bW_\mu^- \bj_\nu^+) \, ,  \nonumber \\
\ourop^s_{12}  &=&  (\bW_\mu^+ \bj_\mu^-) (\bW_\nu^+ \bj_\nu^-)  +
        (\bW_\mu^- \bj_\mu^+) (\bW_\nu^- \bj_\nu^+) \, ,  \nonumber \\
\ourop^s_{13}  &=&  (\bW_\mu^+ \bW_\mu^-) (\bJ^\Z_\nu \bJ^\Z_\nu) \, , 
\nonumber \\
\ourop^s_{14} &=& (\bW_\mu^+ \bW_\nu^-) (\bJ^\Z_\mu
\bJ^\Z_\nu) \, , \nonumber \\
\ourop^s_{15}  &=& (\bZ_\mu \bJ^\Z_\mu) (\bW_\nu^+ \bj_\nu^- +
\bW_\nu^- \bj_\nu^+) \, , \nonumber \\
\ourop^s_{16}  &=& (\bZ_\mu \bJ^\Z_\nu) (\bW_\mu^+ \bj_\nu^- +
\bW_\mu^- \bj_\nu^+) \, , \nonumber \\
\ourop^s_{17}  &=& (\bZ_\mu \bJ^\Z_\nu) (\bW_\nu^+ \bj_\mu^- +
\bW_\nu^- \bj_\mu^+) \, , \nonumber \\
\ourop^s_{18} &=& (\bZ_\mu \bZ_\mu)(\bj_\nu^+ \bj_\nu^-) \, , 
\nonumber  \\
\ourop^s_{19} &=& (\bZ_\mu \bZ_\nu) (\bj_\mu^+ \bj_\nu^-) \, , 
\nonumber \\
\ourop^s_{20}  &=& (\bZ_\mu \bZ_\mu)(\bJ^\Z_\nu \bJ^\Z_\nu) \, , 
\nonumber  \\
\ourop^s_{21}  &=& (\bZ_\mu \bZ_\nu)(\bJ^\Z_\mu \bJ^\Z_\nu) \, , 
\nonumber  \\
%
\ourop^s_{22}  &=& (\bj_\mu^+ \bj_\mu^-) (\bj_\nu^+ \bW_\nu^- +
\bj_\nu^- \bW_\nu^+) \, , \nonumber \\
\ourop^s_{23}  &=& (\bj_\mu^+ \bj_\nu^-) (\bj_\mu^+
\bW_\nu^- + \bj_\nu^- \bW_\mu^+) \, , \nonumber \\
\ourop^s_{24}  &=& (\bJ^\Z_\mu \bJ^\Z_\mu) (\bj_\nu^+ \bW_\nu^- +
\bj_\nu^- \bW_\nu^+) \, , \nonumber \\
\ourop^s_{25}  &=& (\bJ^\Z_\mu \bJ^\Z_\nu) (\bj_\mu^+
\bW_\nu^- + \bj_\mu^- \bW_\nu^+) \, ,  \nonumber \\
\ourop^s_{26}  &=& (\bZ_\mu \bJ^\Z_\mu) (\bj_\nu^+ \bj_\nu^-) \, , 
\nonumber \\
\ourop^s_{27}  &=& (\bZ_\mu \bJ^\Z_\nu) (\bj_\mu^+
\bj_\nu^- + \bj_\mu^- \bj_\nu^+) \, , \nonumber \\
\ourop^s_{28}  &=& (\bZ_\mu \bJ^\Z_\mu)(\bJ^\Z_\nu \bJ^\Z_\nu) \, , 
\nonumber \\
%
\ourop^s_{29}  &=& (\bj_\mu^+ \bj_\mu^-) (\bj_\nu^+ \bj_\nu^-) \, , 
\nonumber \\
\ourop^s_{30}  &=& (\bj_\mu^+ \bj_\nu^-) (\bj_\mu^+ \bj_\nu^-) \, , 
\nonumber \\
\ourop^s_{31}  &=& (\bJ^\Z_\mu \bJ^\Z_\mu) (\bj_\nu^+ \bj_\nu^-) \, , 
\nonumber \\
\ourop^s_{32}  &=& (\bJ^\Z_\mu \bJ^\Z_\nu) (\bj_\mu^+ \bj_\nu^-) \, , 
\nonumber \\
\ourop^s_{33}  &=& (\bJ^\Z_\mu \bJ^\Z_\mu)(\bJ^\Z_\nu \bJ^\Z_\nu) \,
. \label{four_factors}
\end{eqnarray}

\noindent Terms with three powers of fields and external sources:
\begin{eqnarray}
\ourop^s_{34}  &=& i \bJ^\Z_{\mu\nu} (\bW_\mu^+ \bW_\nu^- - \bW_\nu^+
\bW_\mu^- ) \, , \nonumber   \\
\ourop^s_{35}  &=& i \bZ_{\mu\nu} (\bW_\mu^+ \bj_\nu^- - \bW_\nu^+
\bj_\mu^- - \bW_\mu^- \bj_\nu^+ + \bW_\nu^- \bj_\mu^+) \, , 
\nonumber  \\
\ourop^s_{36}  &=& i \bB_{\mu\nu} (\bW_\mu^+ \bj_\nu^- - \bW_\nu^+
\bj_\mu^- - \bW_\mu^- \bj_\nu^+ + \bW_\nu^- \bj_\mu^+) \, , 
\nonumber   \\
\ourop^s_{37}  &=& i \bJ^\Z_{\mu\nu} (\bW_\mu^+ \bj_\nu^- - \bW_\nu^+
\bj_\mu^- - \bW_\mu^- \bj_\nu^+ + \bW_\nu^- \bj_\mu^+) \, , 
\nonumber \\
\ourop^s_{38}  &=& i \bZ_{\mu\nu} (\bj_\mu^+ \bj_\nu^- - \bj_\nu^+
\bj_\mu^-) \, , \nonumber  \\
\ourop^s_{39}  &=& i \bB_{\mu\nu} (\bj_\mu^+ \bj_\nu^- - \bj_\nu^+
\bj_\mu^-) \, , \nonumber   \\
\ourop^s_{40}  &=& i \bJ^\Z_{\mu\nu} (\bj_\mu^+ \bj_\nu^- - \bj_\nu^+
\bj_\mu^-) \, , \nonumber \\
\ourop^s_{41}  &=& i \bJ^\Z_\nu ( \bd_\mu \bW_\mu^+ \bW_\nu^- - \bd_\mu
\bW_\mu^- \bW_\nu^+) \, ,  \nonumber \\
\ourop^s_{42}  &=& i \bJ^\Z_\mu ( \bd_\mu \bW_\nu^+ \bW_\nu^- - \bd_\mu
\bW_\nu^- \bW_\nu^+) \, , \nonumber \\
\ourop^s_{43}  &=& i \bZ_\nu ( \bd_\mu \bW_\mu^+ \bj_\nu^- - \bd_\mu
\bW_\mu^- \bj_\nu^+) \, , \nonumber \\
\ourop^s_{44}  &=& i \bZ_\mu ( \bd_\mu \bW_\nu^+ \bj_\nu^- - \bd_\mu
\bW_\nu^- \bj_\nu^+) \, , \nonumber \\
\ourop^s_{45}  &=& i \bZ_\mu ( \bd_\nu \bW_\mu^+ \bj_\nu^- -
\bd_\nu \bW_\mu^- \bj_\nu^+) \, , \nonumber \\
\ourop^s_{46}  &=& i \bZ_\nu ( \bd_\mu \bj_\mu^+ \bW_\nu^- - \bd_\mu
\bj_\mu^- \bW_\nu^+) \, , \nonumber \\
\ourop^s_{47} & = & i (\p_\mu \bZ_\mu) (\bW_\nu^+ \bj_\nu^- -
\bW_\nu^- \bj_\nu^+) \, , \nonumber \\
%
\ourop^s_{48}  &=& i \bJ^\Z_\nu ( \bd_\mu \bW_\mu^+ \bj_\nu^- - \bd_\mu
\bW_\mu^- \bj_\nu^+) \, , \nonumber \\
\ourop^s_{49}  &=& i \bJ^\Z_\mu ( \bd_\mu \bW_\nu^+ \bj_\nu^- - \bd_\mu
\bW_\nu^- \bj_\nu^+) \, , \nonumber \\
\ourop^s_{50}  &=& i \bJ^\Z_\mu ( \bd_\nu \bW_\mu^+ \bj_\nu^- -
\bd_\nu \bW_\mu^- \bj_\nu^+) \, , \nonumber \\
\ourop^s_{51}  &=& i \bJ^\Z_\nu ( \bd_\mu \bj_\mu^+ \bW_\nu^- - \bd_\mu
\bj_\mu^- \bW_\nu^+) \, , \nonumber \\
\ourop^s_{52}  &=& i \bJ^\Z_\mu ( \bd_\mu \bj_\nu^+ \bW_\nu^- - \bd_\mu
\bj_\nu^- \bW_\nu^+) \, , \nonumber \\
\ourop^s_{53}  &=& i \bZ_\nu ( \bd_\mu \bj_\mu^+ \bj_\nu^- - \bd_\mu
\bj_\mu^- \bj_\nu^+) \, , \nonumber \\
\ourop^s_{54}  &=& i \bZ_\mu ( \bd_\mu \bj_\nu^+ \bj_\nu^- - \bd_\mu
\bj_\nu^- \bj_\nu^+) \, , \nonumber \\
%
\ourop^s_{55}  &=& i \bJ^\Z_\nu ( \bd_\mu \bj_\mu^+ \bj_\nu^- - \bd_\mu
\bj_\mu^- \bj_\nu^+) \, , \nonumber \\
\ourop^s_{56}  &=& i \bJ^\Z_\mu ( \bd_\mu \bj_\nu^+ \bj_\nu^- - \bd_\mu
\bj_\nu^- \bj_\nu^+) \, , \nonumber\\
\ourop^s_{57} &=& \epsilon_{\mu\nu\rho\sigma} \bJ^\Z_\sigma
(\bW_\rho^- \bW_{\mu\nu}^+ + \bW_\rho^+ \bW_{\mu\nu}^-) \, ,
\nonumber \\ 
\ourop^s_{58} &=& \epsilon_{\mu\nu\rho\sigma} \bZ_\sigma (\bj_\rho^-
\bW_{\mu\nu}^+  + \bj_\rho^+ \bW_{\mu\nu}^-) \, , \nonumber \\
\ourop^s_{59} &=& \epsilon_{\mu\nu\rho\sigma} \bZ_\sigma (\bW_\rho^-
\bj_{\mu\nu}^+  + \bW_\rho^+ \bj_{\mu\nu}^-) \, , \nonumber \\
\ourop^s_{60} &=& \epsilon_{\mu\nu\rho\sigma} \bJ^\Z_\sigma (\bj_\rho^-
\bW_{\mu\nu}^+  + \bj_\rho^+ \bW_{\mu\nu}^-) \, , \nonumber \\
\ourop^s_{61} &=& \epsilon_{\mu\nu\rho\sigma} \bJ^\Z_\sigma (\bW_\rho^-
\bj_{\mu\nu}^+  + \bW_\rho^+ \bj_{\mu\nu}^-) \, , \nonumber \\
\ourop^s_{62} &=& \epsilon_{\mu\nu\rho\sigma} \bZ_\sigma (\bj_\rho^-
\bj_{\mu\nu}^+  + \bj_\rho^+ \bj_{\mu\nu}^-) \, , \nonumber \\
\ourop^s_{63} &=& \epsilon_{\mu\nu\rho\sigma} \bJ^\Z_\sigma (\bj_\rho^-
\bj_{\mu\nu}^+  + \bj_\rho^+ \bj_{\mu\nu}^-) \, . \label{three_factors}
\end{eqnarray}

\noindent Terms with two powers of fields and external sources:
\begin{eqnarray}
\ourop^s_{64} &=& M_W^2 (\bW_\mu^+ \bj_\mu^- + \bW_\mu^- \bj_\mu^+) \,
,  \nonumber \\
\ourop^s_{65} &=& M_W^2 \bj_\mu^+ \bj_\mu^- \, , \nonumber \\
\ourop^s_{66} &=& M_Z^2 \bZ_\mu \bJ^\Z_\mu \, , \nonumber \\
\ourop^s_{67} &=& M_Z^2 \bJ^\Z_\mu \bJ^\Z_\mu \, , \nonumber \\
\ourop^s_{68} &=& \bW_{\mu\nu}^+ \bj_{\mu\nu}^- + \bW_{\mu\nu}^-
\bj_{\mu\nu}^+ \, , \nonumber  \\
\ourop^s_{69} &=& \bj_{\mu\nu}^+ \bj_{\mu\nu}^- \, , \nonumber  \\
\ourop^s_{70} &=& \bZ_{\mu\nu} \bJ^\Z_{\mu\nu} \, , \nonumber \\
\ourop^s_{71} &=& \bB_{\mu\nu} \bJ^\Z_{\mu\nu} \, , \nonumber \\
\ourop^s_{72} &=& \bJ^\Z_{\mu\nu} \bJ^\Z_{\mu\nu} \, , \nonumber \\
\ourop^s_{73} &=& (\bd_\mu \bW_\mu^+) (\bd_\nu \bj_\nu^-) + (\bd_\mu
\bW_\mu^-) (\bd_\nu \bj_\nu^+) \, ,  \nonumber \\
\ourop^s_{74} &=& (\bd_\mu \bj_\mu^+) (\bd_\nu \bj_\nu^-) \, , \nonumber \\
\ourop^s_{75} &=& (\p_\mu \bZ_\mu) (\p_\nu \bJ^\Z_\nu) \, , 
\nonumber \\
\ourop^s_{76} &=& (\p_\mu \bJ^\Z_\mu) (\p_\nu \bJ^\Z_\nu) \, , 
\label{two_factors}
\end{eqnarray}
where we introduced the quantities
\bea
\bj_{\mu\nu}^\pm & = & \bd_\mu \bj_\nu^\pm - \bd_\nu \bj_\mu^\pm \, , \\
\bJ_{\mu\nu}^\Z & = & \partial_\mu \bJ_\nu^\Z - \partial_\nu
\bJ_\mu^\Z \, .
\eea


\section{The electroweak chiral Lagrangian}
\label{app:EW_chiral_Lag}

It became customary in the literature to describe the low-energy
effective field theory of the bosonic sector of strongly interacting
models of electroweak symmetry breaking in terms of the so called
electroweak chiral Lagrangian, introduced in
Refs.~\cite{heavy_Higgs,Longhitano,Appelquist_Wu,Feruglio}. Before we
write down the effective Lagrangian in the notation employed in these
references, we would like to add some comments.  Following the first
paper of Ref.~\cite{Longhitano} and Ref.~\cite{Appelquist_Wu} we
include a custodial symmetry breaking term proportional to $\trho -1$
already at order $p^2$ in the low-energy expansion, cf.\
Eq.~(\ref{gen_lageff_2_0}). This is in contrast to the recent
literature which follows mostly the conventions used in the second
paper of Ref.~\cite{Longhitano} or those of the second paper of
Ref.~\cite{H_M}. These conventions may be recovered in our approach by
setting $\trho = 1$.  Furthermore, we include in the list of operators
at order $p^4$ the four terms $\ourop_{15}, \ourop_{16}, \ourop_{17},$
and $\ourop_{18}$, cf.\ Eq.~(\ref{gen_lageff_4_0}), which are
proportional to corresponding terms in $\lag_2$. The use of the
equations of motion and the renormalization of the low-energy
constants in $\lag_2$ in order to reduce the number of terms in
$\lag_4$ will be discussed later. Finally, no external sources have
been introduced in
Refs.~\cite{heavy_Higgs,Longhitano,Appelquist_Wu,Feruglio,H_M,E_M,D_GK}.
We therefore list here only the terms $\lag_2^0$,
Eq.~(\ref{gen_lageff_2_0}), and $\lag_4^0$, Eq.~(\ref{lag_4_0}), which
do not contain external sources.

Following Refs.~\cite{heavy_Higgs,Longhitano,Appelquist_Wu,Feruglio} we
introduce a $SU(2)$ matrix notation for the Goldstone bosons and the
gauge fields:
\bea
\Umat & = & \exp\left( i {\tau^a \pi^a \over \bv}
\right) \in SU(2) \, , \nonumber \\
\Wmat & = & W_\mu^a {\tau^a \over 2} \ , \
\Bmat = B_\mu {\tau^3 \over 2} \, , \nonumber \\
\covDmat_\mu \Umat & = & \p_\mu \Umat - i \Wmat_\mu \Umat + i
\Umat \Bmat_\mu \, , \nonumber \\
\Wmat_{\mu\nu} & = & \p_\mu \Wmat_\nu - \p_\nu \Wmat_\nu
- i [\Wmat_\mu, \Wmat_\nu] \, . \label{def_matrix_notation}
\eea
The effective Lagrangian can then be written in the following way:
\bea 
\lag_2^0 & = &
{1\over 4} \bv^2 \tr ( \covDmat_\mu \Umat^\dagger \covDmat_\mu \Umat)
- (\trho - 1) {\bv^2 \over 8} \left[ \tr(\Tmat \Vmat_\mu) \right]^2
+ {1\over 2\bgs} \tr ( \Wmat_{\mu\nu} \Wmat_{\mu\nu} ) +
{1\over 2\bgps} \tr ( \Bmat_{\mu\nu} \Bmat_{\mu\nu} ) \, , 
\label{lag_eff_2_op} \\
\lag_4^0 & = & \sum_{i=0}^{17} a_i \op_i \, , \label{lag_eff_4_op}
\eea
with the basic set of operators (CP-even terms only):
\bea
\op_0 & = & {\bv^2\over 4} \bgps \left[ \tr(\Tmat \Vmat_\mu) \right]^2
\, , \nonumber \\
\op_1 & = & - {1\over2} B_{\mu\nu} \tr(\Tmat \Wmat_{\mu\nu} ) \, , 
\nonumber \\
\op_2 & = & i {1\over2} B_{\mu\nu} \tr(\Tmat [\Vmat_\mu, \Vmat_\nu])
\, , \nonumber \\
\op_3 & = & - i \tr(\Wmat_{\mu\nu} [\Vmat_\mu, \Vmat_\nu]) \, ,
\nonumber \\ 
\op_4 & = & - \left[ \tr(\Vmat_\mu \Vmat_\nu) \right]^2 \, , \nonumber
\\ 
\op_5 & = & - \left[ \tr(\Vmat_\mu \Vmat_\mu) \right]^2 \, , \nonumber
\\ 
\op_6 & = & - \tr(\Vmat_\mu \Vmat_\nu) \tr(\Tmat \Vmat_\mu)
\tr(\Tmat \Vmat_\nu) \, , \nonumber \\
\op_7 & = & - \tr(\Vmat_\mu \Vmat_\mu) \left[ \tr(\Tmat \Vmat_\nu)
\right]^2 \, , \nonumber \\
\op_8 & = & {1\over4} \left[ \tr(\Tmat \Wmat_{\mu\nu}) \right]^2 \, , 
\nonumber \\
\op_9 & = & - i \half \tr(\Tmat \Wmat_{\mu\nu}) \tr(\Tmat [\Vmat_\mu,
\Vmat_\nu]) \, , \nonumber \\
\op_{10} & = & - \left[ \tr(\Tmat \Vmat_\mu) \tr(\Tmat \Vmat_\mu)
\right]^2 \, , \nonumber \\
\op_{11} & = & - \tr((\covDmat_\mu \Vmat_\mu)^2) \, , \nonumber \\
\op_{12} & = & - \tr(\Tmat \covDmat_\mu \covDmat_\nu \Vmat_\nu)
\tr(\Tmat \Vmat_\mu) \, , \nonumber \\
\op_{13} & = & - \half \left[ \tr(\Tmat \covDmat_\mu \Vmat_\nu)
\right]^2 \, , \nonumber \\
\op_{14} & = & \epsilon_{\mu\nu\rho\sigma} \tr(\Wmat_{\mu\nu}
\Vmat_\rho) \tr(\Tmat \Vmat_\sigma) \, , \nonumber \\
\op_{15} & = & M_W^2 \tr (\covDmat_\mu \Umat^+ \covDmat_\mu \Umat)
\, , \nonumber \\
\op_{16} & = & \tr ( \Wmat_{\mu\nu} \Wmat_{\mu\nu} ) \, , \nonumber \\
\op_{17} & = & \tr ( \Bmat_{\mu\nu} \Bmat_{\mu\nu} ) \, .
\label{basis_appelquist_wu}
\eea
In Eqs.~(\ref{lag_eff_2_op})--(\ref{basis_appelquist_wu}) we used the
building blocks 
\bea
\Tmat & = & \Umat \tau^3 \Umat^\dagger \ , \ 
\Vmat_\mu = (\covDmat_\mu \Umat) \Umat^\dagger \, , \nonumber \\
\covDmat_\mu \Vmat_\nu & = & \p_\mu \Vmat_\nu - i [\Wmat_\mu,
\Vmat_\nu]  \, . \label{def_appelquist_wu}
\eea

We recall that we count the gauge coupling constants $g, \gp$ as order
$p$ in the low-energy expansion, therefore the custodial symmetry
breaking term $\op_{0}$ is of the order $p^4$. Note that we have used
in Eqs.~(\ref{lag_eff_2_op})--(\ref{def_appelquist_wu}) a different
convention for the signs of the gauge coupling constants compared to
the literature. Specifically, we have $\g \to - \bg$ and $\gp \to -
\bg^\prime$ compared to Ref.~\cite{H_M}. Furthermore, we have again
absorbed the gauge coupling constants into the gauge fields, cf.\
Eq.~(\ref{cov_deriv_EFT}). Note that $\bv$ corresponds to the pion decay
constant $F_\pi$ in chiral perturbation theory. 

The relations between the two sets of operators, $\ourop_i$ from
Eq.~(\ref{gen_lageff_4_0}) and the $\op_i$, read
\bea
\op_0 & = & - \bs^2 \ourop_{16} \, , \nonumber  \\
\op_1 & = & \ourop_{8} - {1\over 2} \ourop_{12}  - {1\over 2}
\ourop_{18} \, , \nonumber \\
\op_2 & = & - \ourop_8 \, , \nonumber \\
\op_3 & = & - 4 \ourop_1 + 4 \ourop_2 - 4 \ourop_3 + 4 \ourop_4 + 2
\ourop_7 + \ourop_8 - 2 \ourop_9 + 2 \ourop_{10} \, , \nonumber \\
\op_4 & = & - 2 \ourop_1 - 2 \ourop_2 - 2 \ourop_4 - {1\over 4}
\ourop_5 \, , \nonumber \\
\op_5 & = & - 4 \ourop_1 - 2 \ourop_3 - {1\over 4} \ourop_5 \, ,
\nonumber \\ 
\op_6 & = & - 2 \ourop_4 - {1\over 2} \ourop_5 \, , \nonumber \\
\op_7 & = & - 2 \ourop_3 - {1\over 2} \ourop_5 \, , \nonumber \\
\op_8 & = & 2 \ourop_1 - 2 \ourop_2 - \ourop_7 - \ourop_8 + {1\over 4}
\ourop_{11} + {1\over 2} \ourop_{12} + {1\over 4} \ourop_{18} \, ,
\nonumber \\ 
\op_9 & = & - 4 \ourop_1 + 4 \ourop_2 + \ourop_7 + \ourop_8 \, ,
\nonumber \\ 
\op_{10} & = & - \ourop_5 \, , \nonumber \\
\op_{11} & = & 2 \ourop_{13} + {1\over 2} \ourop_{14} \, , \nonumber \\
\op_{12} & = & 2 \ourop_{10} - \ourop_{14} \, ,  \nonumber \\
\op_{13} & = & 4 \ourop_1 - 4 \ourop_2 - \ourop_7 + {1\over 4}
\ourop_{11} + {1\over 2} \ourop_{14} \, ,  \nonumber \\
\op_{14} & = & i \ourop_{6} \, , \nonumber \\
\op_{15} & = & 2 \ourop_{15} \, , \nonumber \\
\op_{16} & = & {1\over 2} \ourop_{17} \, , \nonumber \\
\op_{17} & = & {1\over 2} \ourop_{18} \, ,  \label{L_i_vs_O_i}
\eea
which are valid up to partial integrations. 

As discussed in Sec.~\ref{subsec:EFT_W_4} the equations of motion in
the effective field theory lead to relations between the operators
$\op_i$ in $\lag_4$, cf.\ the relations in
Eqs.~(\ref{remove_O9})--(\ref{remove_O70s}) between the operators
$\ourop_i$. From the constraint equations~(\ref{eomU_pm_eff}) and
(\ref{eomU_Z_eff}), which are equivalent to $\tr (\covDmat_\mu
\Vmat_\mu) = 0$ in the usually employed notation, we obtain the
following relations:
\bea
\op_{11} & = & 0 \, , \label{remove_op11_app} \\
\op_{12} & = & 0 \, , \label{remove_op12_app} \\
\op_{13} & = & {\bc^2 \over 2 \bs^2} \op_0 + \op_3 + \op_4 - \op_5 -
\op_6 + \op_7 - \op_9 - \op_{15} - {1\over 2} \op_{16} + {1\over 2}
\op_{17} \, . \label{remove_op13_app}
\eea
The equations of motion for the gauge fields in
Eqs.~(\ref{eomWpm_eff}) and (\ref{equmow3_eff}) lead to the relations
\bea
\op_1 & = & \op_3 - 2 \op_{15} - \op_{16} \, , \label{remove_op1_app} \\ 
\op_8 & = & {\bc^2 \over 2 \bs^2} \op_0 - \op_9 + \op_{15} + {1\over
2} \op_{16} \, . \label{remove_op8_app}
\eea
For simplicity we have set $\trho = 1 $ and switched off the external
sources. The general relations can be inferred from
Eqs.~(\ref{remove_O9})--(\ref{remove_O70s}) by making use of
Eq.~(\ref{L_i_vs_O_i}) to convert the basis with the operators
$\ourop_i$ into the basis with the operators $\op_i$.  Note that
Eq.~(\ref{remove_op13_app}) has changed compared to
Eq.~(\ref{remove_op13}) because we have replaced above the operators
$\op_1$ and $\op_8$ on the right-hand side of the equation.
Furthermore, we note that we get a different sign of the terms $\op_4$
and $\op_5$ in Eq.~(\ref{remove_op13_app}) compared to
Ref.~\cite{Feruglio}. 

With the help of Eqs.~(\ref{remove_op11_app})--(\ref{remove_op8_app})
we can remove the operators $\op_1, \op_8, \op_{11}, \op_{12},$ and
$\op_{13}$ from the basis.  Furthermore, we can remove the terms
$\op_{15}, \op_{16},$ and $\op_{17}$, which are proportional to terms
in the Lagrangian $\lag_2^0$, by renormalizing the parameters and
low-energy constants in the lowest order Lagrangian, cf.\
Eqs.~(\ref{v_eff_general})--(\ref{gprime_eff_general}).

\subsection{The effective Lagrangian for the standard model with a heavy
Higgs boson}

The result for the bare effective Lagrangian from
Eqs.~(\ref{lag_eff_2_bare_ourop})--(\ref{ourlowbare}) for the
standard model with a heavy Higgs boson translates into the following
expression for the bare low-energy constant $\bv_b^2$ in $\lag_2$,
Eq.~(\ref{lag_eff_2_op}), 
\be \label{hv_bare}
\bv_b^2 = {m^2\over \lambda} \left[ 1 + \lambda \left( - 12 \polem +
{1\over 16 \pi^2} \right) \right] \, .
\ee
Furthermore we obtain $\trho = 1$ in Eq.~(\ref{lag_eff_2_op}). The
bare low-energy constants $a_i^b$ in $\lag_4$,
Eq.~(\ref{lag_eff_4_op}), are given by
\bea
a_0^b & = & - {3\over 4} \polem - {1\over 16} {1\over 16 \pi^2} \, , 
\nonumber \\
a_1^b & = & - {1\over 6} \polem - {1\over 72} {1\over 16 \pi^2} \, , 
\nonumber \\
a_2^b & = & - {1\over 12} \polem + {11\over 144} {1\over 16 \pi^2} \, , 
\nonumber \\
a_3^b & = & {1\over 12} \polem - {11\over 144} {1\over 16 \pi^2} \, , 
\nonumber \\
a_4^b & = & {1\over 6} \polem - {11\over 72} {1\over 16 \pi^2} \, , 
\nonumber \\
a_5^b & = & {1 \over 16 \lambda} - {17\over 12} \polem - {35\over 144}
{1\over 16 \pi^2} \, , \nonumber \\
a_{11}^b & = & - {1\over 24} {1\over 16 \pi^2} \, , 
\nonumber \\
a_{15}^b & = & {3\over 2} \polem + {1\over 8} {1\over 16 \pi^2} \, , 
\nonumber \\
a_{16}^b & = & - {1\over 12} \polem - {1\over 144} {1\over
  16\pi^2} \, , \nonumber \\
a_{17}^b & = & - {1\over 12} \polem - {1\over 144} {1\over
  16\pi^2} \, . \label{a_i_bare} 
\eea
All other bare low-energy constants $a_i^b$ vanish. Note that we have
included some additional, finite terms into our definition of the pole
term $\polem$, cf.\  Eq.~(\ref{polembare}), compared to the
conventions used in Refs.~\cite{H_M,E_M,D_GK}.

Inserting the physical masses and coupling constants from
Sec.~\ref{sec:renorm}, the effective Lagrangian reads
\bea
\lag_2 &  = & \left(\MWps {s_p^2 \over \ers} \right)
\tr ( \covDmat_\mu \Umat^+ \covDmat_\mu \Umat) + {s_p^2\over 2 \ers}
\tr ( \Wmat_{\mu\nu} \Wmat_{\mu\nu} ) 
+ {c_p^2 \over 2 \ers} \tr ( \Bmat_{\mu\nu} \Bmat_{\mu\nu} ) \, , 
\label{lag_eff_2_ren_op} \\
\lag_4 & = & \sum_{i=0}^{17} a_i \op_i \, , \label{lag_eff_4_ren_op}
\eea
with the non-vanishing low-energy constants
\bea
a_0  & = & - {3\over 4} \pole - {3\over 8} {1\over 16 \pi^2} 
\ln \left( {\MHps \over \mu^2} \right) - {1\over 16} {1\over 16 \pi^2}
\, , \nonumber \\
a_1  & = & - {1\over 6} \pole - {1\over 12} {1\over 16 \pi^2}
\ln \left( {\MHps \over \mu^2} \right) - {1\over 72} {1\over 16 \pi^2}
\, , \nonumber \\
a_2  & = & - {1\over 12} \pole - {1\over 24} {1\over 16 \pi^2}
\ln \left( {\MHps \over \mu^2} \right) + {11\over 144} {1\over 16 \pi^2}
\, , \nonumber \\
a_3  & = & {1\over 12} \pole + {1\over 24} {1\over 16 \pi^2}
\ln \left( {\MHps \over \mu^2} \right) - {11\over 144} {1\over 16 \pi^2}
\, , \nonumber \\
a_4  & = & {1\over 6} \pole + {1\over 12} {1\over 16 \pi^2}
\ln \left( {\MHps \over \mu^2} \right) - {11\over 72} {1\over 16 \pi^2}
\, , \nonumber \\
a_5  & = & {1\over 12} \pole + {s_p^2 \MWps \over 2 \ers \MHps} +
{1\over 24} {1\over 16 \pi^2} \ln \left( {\MHps \over \mu^2} \right)
- {1\over 16 \pi^2} {152 - 27 \sqrt{3} \pi \over 144} \, , \nonumber \\
a_{11}  & = & - {1\over 24} {1\over 16 \pi^2} \, , 
\nonumber \\
a_{15}  & = & {3\over 2} \pole + {3\over 4} {1\over 16 \pi^2} \ln
\left( {\MHps \over \mu^2} \right) + {1\over8} {1\over 16 \pi^2}
- {1\over 8} (1 - {c_p^2 \over s_p^2}) \delta M_{W,2}^2 - {1 \over 8
s_p^2} \delta M_{Z,2}^2 + s_p^2 \delta e^2_2 \, , \nonumber \\
a_{16} & = & - {1\over 12} \pole - {1\over 24} {1\over 16 \pi^2} \ln
\left( {\MHps \over \mu^2} \right) - {1\over 144} {1\over
  16\pi^2} + {c_p^2 \over 16 s_p^2} \delta M_{W,2}^2 - {1\over 16
s_p^2} \delta M_{Z,2}^2 + {1 \over 2} s_p^2 \delta e^2_2 \, , \nonumber \\
a_{17}  & = & - {1\over 12} \pole - {1\over 24} {1\over 16 \pi^2} \ln
\left( {\MHps \over \mu^2} \right) - {1\over 144} {1\over
  16\pi^2} - {c_p^2 \over 16 s_p^2} \delta M_{W,2}^2 + {1 \over 16
s_p^2} \delta M_{Z,2}^2 + {1\over 2} c_p^2 \delta e^2_2 \,
. \label{a_i_ren} 
\eea
The pole term in $d=4$ dimensions, $\pole$, is defined in
Eq.~(\ref{pole}). We denoted the pole-masses of the Higgs boson, the
$W$- and the $Z$-boson by $\MHp$, $\MWp$, and $\MZp$,
respectively. The electric charge is denoted by $\eres$. The
quantities $\delta M_{W,2}^2, \delta M_{Z,2}^2$, and $\delta e^2_2$
are defined in Eqs.~(\ref{delta_MW_4}), (\ref{delta_MZ_4}), and
(\ref{delta_e_4}), respectively. Furthermore, we use the on-shell
definition for the weak mixing angle $c_p^2, s_p^2$, cf.\
Eq.~(\ref{def_cp_sp}).

Finally, we can remove the redundant terms $\op_1, \op_8,
\op_{11}, \op_{12},$ and $\op_{13}$ from the basis by employing the 
Eqs.~(\ref{remove_op11_app})--(\ref{remove_op8_app}) and the terms
$\op_{15}, \op_{16},$ and $\op_{17}$ by renormalizing the parameters
in the lowest order Lagrangian $\lag_2^0$. In this way we obtain the
expression for the Lagrangian $\lag_2$ as given in
Eqs.~(\ref{lag_eff_2_ren_redefined})--(\ref{gprime_eff}) and the
following results for the 10 low-energy constants corresponding to
independent terms in the Lagrangian $\lag_4^0$:
\bea
a_0^\prime  & = & - {3\over 4} \pole - {3\over 8} {1\over 16 \pi^2}
\ln \left( {\MHps \over \mu^2} \right) - {1\over 16} {1\over 16 \pi^2}
\, , \nonumber \\
a_2^\prime  & = & - {1\over 12} \pole - {1\over 24} {1\over 16 \pi^2}
\ln \left( {\MHps \over \mu^2} \right) + {11\over 144} {1\over 16 \pi^2}
\, , \nonumber \\
a_3^\prime  & = & - {1\over 12} \pole - {1\over 24} {1\over 16 \pi^2}
\ln \left( {\MHps \over \mu^2} \right) - {13\over 144} {1\over 16 \pi^2}
\, , \nonumber \\
a_4^\prime  & = & {1\over 6} \pole + {1\over 12} {1\over 16 \pi^2}
\ln \left( {\MHps \over \mu^2} \right) - {11\over 72} {1\over 16 \pi^2}
\, , \nonumber \\
a_5^\prime  & = & {1\over 12} \pole + {s_p^2 \MWps \over 2 \ers \MHps} +
{1\over 24} {1\over 16 \pi^2} \ln \left( {\MHps \over \mu^2} \right)
- {1\over 16 \pi^2} {152 - 27 \sqrt{3} \pi \over 144} \, , \nonumber \\
a_6^\prime & = & 0 \, , \nonumber \\
a_7^\prime & = & 0 \, , \nonumber \\
a_9^\prime & = & 0 \, , \nonumber \\
a_{10}^\prime & = & 0 \, , \nonumber \\
a_{14}^\prime & = & 0 \, . \label{a_i_EoM_ren}
\eea
We have denoted the modified low-energy constants by $a_i^\prime$ in
order to distinguish them from the old ones. Only the low-energy
constant $a_3^\prime$ has changed in comparison with the values given
in Eq.~(\ref{a_i_ren}). Note, however, that $a_1$ and $a_8$ have
disappeared from the list of independent low-energy constants.

The low-energy constants in Eqs~(\ref{a_i_ren}) and
(\ref{a_i_EoM_ren}) have the following general form:
\bea
a_i & = & \Delta_i \pole + a_i^r(\mu) \, , \nonumber \\
a_i^\prime & = & \Delta_i^\prime \pole + a_i^{\prime \, r}(\mu)
\, , 
\eea
i.e.\ they contain a pole term proportional to $\pole$ and a scale
dependent part. We denote the coupling constants $a_i^r(\mu)$ and
$a_i^{\prime \, r}(\mu)$ as renormalized low-energy constants.


\section{Differential operators in the standard model}
\label{app:diffop_HH}

The explicit results for the differential operators $\Dfull$ and $P^T
P$ which appear in Eq.~(\ref{1loopgf_1}) in
Sec.~\ref{sec:SM_gauge_inv} are given below. In the following, upper
case Latin indices $A, B, \ldots $ run from $1$ to $4$, lower case
Latin indices $a, b, \ldots $ run from $1$ to $3$, and Greek indices
$\alpha, \beta, \ldots $ label the components $1,2$.

The components of the differential operator $\tildeD + P P^T +
\delta_P$ in Eq.~(\ref{defD}) are given by
\begin{eqnarray}
d       & = & -\Box + 2 m^2 + 3 m^2 (R^2 - 1) +
                        {1\over 4} \Y_\mu^a \Y_\mu^a  - \widehat h 
\, , \label{firstcomp_defD} \\
\delta^b  & = & - \Y_\rho^a \hvarD_\rho^{ab}
        - {1\over 2} (\hvarD_\rho \Y_\rho)^b \, , \\
{\delta^T}^a & = &  \Y_\rho^a \p_\rho +
        {1\over 2} (\hvarD_\rho \Y_\rho)^a \, , \\
D^{ab}       & = &   - (\hvarD_\rho \hvarD_\rho)^{ab} + \delta^{ab}
        \left( m^2 (R^2 - 1)  - \widehat h \right) + M_W^2 R^2
        \delta^{ab} + {1\over 4} \Y_\rho^a \Y_\rho^b \, , \\
\delta_\nu^B & = & M_W R  \tY_\mu^A \PTT_{\mu\nu}^{AB} \, , 
         \\
\delta_\mu^{T, A} & = & M_W \PTT_{\mu\nu}^{AB} R
\tY_\nu^B \, , \\
\Delta_\nu^{aB} &= & f^{aBc} M_W R \Y_\nu^c
        + 2 M_W (\p_\nu R) \delta^{aB} -s M_Z\delta^{4B}
        \left( 2 \delta^{a3} (\p_\mu R)
        + R T^{ac}\W_\mu^c \right) \PT_{\mu\nu}  \, , \\
\Delta_\mu^{T, Ab} & = &  - f^{Abc} M_W R \Y_\mu^c
        + 2 M_W (\p_\mu R) \delta^{Ab} + s M_Z \delta^{A4}
        \PT_{\mu\nu}\left( R \W_\nu^c T^{cb}
        - 2 (\p_\nu R) \delta^{3b} \right) \, ,   \\
D_{\mu\nu}^{AB} & = &
        -\delta_{\mu\nu}(\tvarD_\rho \tvarD_\rho)^{AB} + 2 f^{ABc}
        \W^c_{\mu\nu} 
+ (\widetilde M^2)^{AB}\PT_{\mu\nu} + M_W^2 \delta^{AB}\PL_{\mu\nu}
\nonumber \\
        & & \mbox{}
        + \PTT_{\mu\alpha}^{AC} (\widetilde
        M^2)^{CD}(R^2-1)\PTT_{\alpha\nu}^{DB}
        + \delta^{A4} \delta^{B4} \PT_{\mu\rho}
        \widehat J_{\rho\sigma} \PT_{\sigma\nu} \, , 
\label{lastcomp_defD}
\eea
where we introduced the quantities
\bea
\hvarD_\mu^{ab} & = & \varD_\mu^{ab} + {1\over 2} \varepsilon^{abc}
\Y_\mu^c \, ,  \\
\varD_\mu^{ab}  & = & \p_\mu \delta^{ab} - \varepsilon^{abc}
\W_\mu^c \, , \\
\tvarD_\mu^{AB} & = & \delta^{AB} \p_\mu - f^{ABc} \W_\mu^c \, , \\
f^{ABc}         & = & \left\{ \begin{array}{ccl}
                                \varepsilon^{abc} & , &  A=a, B=b, \\
                                0 & , & A=4 \ \mbox{and / or} \ B=4, 
                             \end{array} \right.   \\
\tY_\mu^A               & = & \left( \begin{array}{c}
                                \Y_\mu^a \\
                                - {s\over c} \Y_\mu^3 \end{array}
                              \right) \, , \\
\PTT_{\mu\nu} &=& {\rm diag}\left(
\delta_{\mu\nu},\delta_{\mu\nu},\delta_{\mu\nu},\PT_{\mu\nu}\right) \, 
, \\
\PT_{\mu\nu} &=& \delta_{\mu\nu} - \PL_{\mu\nu} \, , \quad 
\PL_{\mu\nu} = {\p_\mu \p_\nu \over \Box} \, , \\ 
\widetilde M^2 & = & \left( \begin{array}{rrrr}
 M_W^2   &      0 &0 &0 \\
       0  & M_W^2 &0 &0 \\
0 &0 & c^2 M_Z^2 & -c s M_Z^2 \\
0 &0 & -c s M_Z^2 & s^2 M_Z^2 \\
\end{array} \right) \, , \\
\widehat J_{\mu\nu}                 & = & \gps v_{dj}
        \left( \delta_{\mu\nu} J_{\kappa}^\alpha J_{\kappa}^\alpha
        - J_{\mu}^\alpha J_{\nu}^\alpha \right) \ .
\eea

In the basis $(f, \eta^a, \tq_\mu^A)$ the differential operator $P$
which creates zero modes can be written as follows:
\begin{equation} \label{zeromodes}
            \left( \begin{array}{c}
                0 \\
                M_W R \delta^{aB} \\
                \tvarD_\mu^{AB}
              \end{array} \right) \alpha^B
            \equiv P \alpha \, , 
\end{equation}
where $\alpha^B$ are four arbitrary scalar functions. From this
expression we obtain the following results for the
differential operators $P P^T$ and $P^T P$ which appear in
Eq.~(\ref{1loopgf_1}):
\begin{eqnarray}
P P^T & = & \left(\begin{array}{ccc}
0 & 0 & 0 \\
0 & M_W^2 R^2 \delta^{ab}& - M_W R \tvarD_\nu^{aB} \\
0 &  M_W \tvarD_\mu^{Ab} R & - (\tvarD_\mu\tvarD_\nu)^{AB}
\end{array}\right) \, , \label{def_PPT} \\
P^T P & = & \left( \begin{array}{cc}
-\varD^{ac}_\mu\varD_{\mu}^{cb} + M_W^2 R^2\delta^{ab} & 0 \\
0 & - \Box
\end{array}\right) \ .  \label{def_PTP}
\end{eqnarray}
Furthermore, the operator $\delta_P$ is defined by
\begin{equation} \label{def_deltaP}
  \delta_P = {\rm diag}\left( 0, 0,
  \delta^{A4}\delta^{4B}M_W^2\PL_{\mu\nu} \right) \ .
\end{equation}

Since we perform a saddle-point approximation in the path integral,
the fields which appear in the list of differential operators
in Eqs.~(\ref{firstcomp_defD})--(\ref{lastcomp_defD}) obey the equations
of motion (\ref{eomRcomp})--(\ref{eomUcompZ}). We have used this fact
to simplify the expressions of those operators which correspond
to the fluctuations $\eta^a$ of the Goldstone bosons.
Furthermore, it is important to ensure that the full differential
operator $\Dfull$ is Hermitian, i.e.\ satisfies the relation
$(y,[\Dfull] y^\prime) = (y^\prime,[\Dfull] y)$ for arbitrary
fluctuation vectors $y,y^\prime$.


\end{document}